\documentclass[showpacs,aps,twocolumn,floats,superscriptaddress]{revtex4-1} 

\usepackage[colorlinks, breaklinks=true, linkcolor=blue, citecolor=blue, linktocpage=true]{hyperref}
\usepackage{color}
\usepackage{graphicx}
\usepackage{dcolumn}
\usepackage{bm}
\usepackage{amsmath}
\usepackage{amssymb}
\usepackage{txfonts}

\newcommand {\bkt} [1] {\langle #1 \rangle}

\begin{document}

\title{Quantum Kinetic Theory of the Chiral Anomaly}

\author{Akihiko Sekine}
\affiliation{Department of Physics, The University of Texas at Austin, Austin, Texas 78712, USA}
\author{Dimitrie Culcer}
\affiliation{School of Physics and Australian Research Council Centre of Excellence in Low-Energy Electronics Technologies, UNSW Node, The University of New South Wales, Sydney 2052, Australia}
\author{Allan H. MacDonald}
\affiliation{Department of Physics, The University of Texas at Austin, Austin, Texas 78712, USA}

\date{\today}

\begin{abstract}
We present a general quantum kinetic theory of low-field magnetotransport in weakly disordered crystals that accounts fully for the interplay between electric-field induced interband coherence, Bloch-state scattering, and an 
external magnetic field.  The quantum kinetic equation we derive for the Bloch-state density matrix  
naturally incorporates the momentum-space Berry phase effects whose influence on 
Bloch-state wavepacket dynamics is normally incorporated into transport theory in an {\it ad hoc} manner.  
The Berry phase correction to the momentum-space density of states in the presence 
of an external magnetic field implied by semiclassical wavepacket dynamics
is captured by our theory as an intrinsic density-matrix response to a magnetic field.
We propose a simple and general procedure for expanding the linear response 
of the Bloch-state density matrix to an electric field in 
powers of magnetic field.  As an illustration, we apply our theory to magnetotransport
 in Weyl semimetals.  We show that the chiral anomaly 
(positive magnetoconductivity quadratic in magnetic field)
that appears when separate Fermi surface pockets surround distinct Weyl points
survives only when intervalley scattering is very weak compared to intravalley scattering.
\\
\\
\\
\end{abstract}

\maketitle

\section{Introduction}
The electrical transport properties of weakly disordered quantum degenerate 
crystalline conductors are normally controlled by occupation probabilities of 
states close to the Fermi surface that change in response to electric fields or temperature gradients.
Transport properties then depend on the shapes of all
Fermi surfaces, on the distributions of group velocities,
and the details of the processes that scatter electrons 
between Bloch states on those surfaces.
It has recently become more widely appreciated
that intrinsic effects, independent of disorder
and related more to Bloch-state wave functions than energies, are sometimes important.
In special cases these effects are nearly completely 
characterized by momentum-space Berry curvatures \cite{Xiao2010},
which can be nonzero only in systems that 
have broken time-reversal symmetry, or broken inversion symmetry, or both. 
One widely recognized and important example of a transport effect that is often dominated by Berry curvature physics 
is the anomalous Hall effect \cite{Nagaosa2010}, i.e., the Hall effect in the absence of a magnetic 
field in a crystal in which time-reversal symmetry is broken by magnetic order.
This paper is motivated by the need for a practical theory suitable for application to 
real crystals that can provide a general description of magnetotransport 
in systems in which Bloch-state wave function properties play an important role.   

Large momentum-space Berry curvatures are often associated with nontrivial band topology.
For example the anomalous Hall conductivity of a two-dimensional (2D) insulator \cite{Thouless1982,Haldane1988}
is equal to the quantum unit of conductance, $e^2/h$, times the
sum of the Chern numbers of all occupied bands.  The Chern number of a band is a 
topological index that is simply the integral of its Berry curvature over the full 2D Brillouin zone
divided by $2 \pi$, and must be an integer.
Indeed, recent interest in the topological classification of crystalline matter started 
with theories of the quantum Hall effect,
and was later extended to time-reversal invariant topological insulators \cite{Kane2005,Kane2005a,Bernevig2006,Hasan2010,Qi2011,Ando2013}.
The classification of topological phases has now been extended to 3D gapless metallic 
systems referred to as Weyl semimetals \cite{Volovik2003,Murakami2007,Wan2011} or Dirac semimetals \cite{Wang2012,Young2012,Wang2013,Yang2014}.
Although Weyl semimetals cannot be characterized by bulk topological invariants, nondegenerate band touching points (Weyl points) are topologically stable and can be regarded as monopole momentum-space
sources of phase flux \cite{Fang2003,AHM2004}.
Dirac points, which can be viewed as a superposition of two degenerate Weyl points, are not generically stable, 
but can be stabilized by crystalline symmetries \cite{Yang2014}.
The experimental identification of Dirac 
semimetals \cite{Liu2014,Liu2014a,Neupane2014,Borisenko2014} and Weyl semimetals \cite{Lv2015,Yang2015,Xu2015,Lu2015}
has motivated a growing effort, with both theoretical and experimental components,
aimed at identifying and exploring novel phenomena in these materials.
Three dimensional topological semimetals host a variety of transport and magnetotransport properties
that are dependent on their unusual Bloch-state wave function properties, some of which we highlight 
in the following paragraphs, and therefore require
the type of transport theory discussed in this paper for a proper theoretical description.

The topologically nontrivial electronic band structures of Weyl and Dirac 
semimetals can lead to an approximate 
realization of the chiral anomaly in condensed matter physics.
In quantum field theory the chiral anomaly, which occurs only in systems with 
odd space dimensions, refers to the violation of axial current conservation,
$\partial_\mu J_5^\mu\neq 0$ as a combined response to electric and magnetic fields.
In the case of Weyl semimetals these response properties are conveniently  
described in terms of a $\theta$ term added to 
the system's electromagnetic action \cite{Wilczek1987}:
\begin{align}
S_\theta=\frac{e^2}{4\pi^2 \hbar c}\int dt\, d^3 r\, \theta(\bm{r},t)\bm{E}\cdot\bm{B},
\label{S_theta_realtime}
\end{align}
where $\theta(\bm{r},t)$ is a coupling coefficient, and $\bm{E}$ and $\bm{B}$ are external electric
and magnetic fields.
The presence of a $\theta$ term implies a charge current
$\bm{j}(\bm{r},t)=\delta S_\theta/\delta \bm{A}=(e^2/4\pi^2 \hbar c)[\nabla\theta(\bm{r},t)\times\bm{E}+\dot{\theta}(\bm{r},t)\bm{B}]$ \cite{Wilczek1987}.
Here, $\bm{A}$ is the electromagnetic field's vector potential.
The electric-field induced current is the anomalous Hall effect, since it is perpendicular to the electric field, and
the magnetic-field induced current is referred to as the chiral magnetic effect \cite{Fukushima2008,Zyuzin2012,Son2012} which we discuss below.  

Weyl semimetals with two Weyl points are characterized by the following approximate 
expression $\theta(\bm{r},t)=2(\bm{b}\cdot\bm{r}-b_0 t)$ \cite{Zyuzin2012,Son2012,Grushin2012,Wang2013a,Goswami2013,Burkov2015}.
Here, $\bm{b}$ is the vector connecting the distinct Weyl points in momentum space, and $b_0$ is the difference between distinct local chemical potentials, if one is
somehow established in different regions of momentum space.
Such a chemical potential difference is of course absent in equilibrium;
the possible existence of an equilibrium chiral magnetic effect has been considered 
theoretically \cite{Vazifeh2013,Chang2015,Buividovich2015,Yamamoto2015,Ma2015,Zhong2016,Zubkov2016},
but can be ruled out in crystalline solids as discussed in Ref.~\cite{Vazifeh2013}.
A related dynamical realization of the chiral magnetic effect was recently 
proposed for axionic insulators \cite{Sekine2016,Sekine2016a}.
A chemical potential difference can 
be generated by the combined influence of electric and magnetic fields as explained below, 
and when present is responsible for a related 
negative magnetoresistance (positive magnetoconductance) 
that is quadratic in magnetic field \cite{Son2013,Burkov2014,Burkov2015a,Spivak2016}.
A positive magnetoconductance stands 
in striking contrast to the familiar negative magnetoconductance due to the Lorentz force.
Remarkably, negative magnetoresistance has recently been observed in the low magnetic field regime in the Dirac 
semimetals Na$_3$Bi \cite{Xiong2015} and 
Cd$_3$As$_2$ \cite{Li2015,Li2016}, and ZrTe$_5$ \cite{Li2016a}, and in the 
Weyl semimetals TaAs \cite{Huang2015} and TaP \cite{Arnold2016}.  
The peculiar positive magnetoconductance behavior occurs only
for parallel electric and magnetic fields, which suggests that 
it is related in some way to an $\bm{E}\cdot\bm{B}$ contribution to the electromagnetic action.
This very specific effect is a particularly attractive target of 
the magnetotransport theory developed here, which 
is able to realistically address its partial realization in real materials.

In this paper, we develop a general quantum kinetic theory of low-field 
magnetotransport in weakly disordered crystals 
that accounts fully for the electric-field induced interband coherence responsible
for Berry phase contributions to wavepacket dynamics and to the 
momentum-space density-of-states, and at the same 
time to account for the interplay between Bloch-state scattering and the 
presence of an external magnetic field.  
We take the effect of magnetic fields into account
using a semiclassical approximation that we expect to be accurate 
when the magnetic field is weak enough that Landau quantization can be neglected.
Our main result is a quantum kinetic equation which includes driving terms associated 
with both external electric and magnetic fields, 
and which we expect to be valuable for understanding the interesting magneto-transport anomalies 
present in any conductor that has band crossings at energies close to the 
Fermi energy.

Our paper is organized as follows.  In Sec.~\ref{Sec-Overview} we provide a 
brief overview and physical explanation of our main results, neglecting all technical details.  
In Sec.~\ref{Sec-Kinetic-Equation} we derive a general quantum kinetic equation that accounts fully
for momentum-dependent Bloch-state wave functions  and the associated interband coherence response
in the presence of external electric and magnetic fields.
In Sec.~\ref{Sec-Magnetotransport} we describe a general scheme to apply our 
theory to magnetotransport phenomena, by performing a systematic low-field expansion.  
In Sec.~\ref{Sec-Scattering} we explain in detail when the intervalley scattering rate emerges as an 
important parameter in transport properties of many-valley electronic systems, and when it does not.
In Sec.~\ref{Sec-Chiral-Anomaly}, as an application of our formalism, we explain how our formalism quite generally captures the pumping between valleys that occurs in systems with 
Fermi surface pockets that enclose Weyl points with a net chirality. 
In Sec.~\ref{Sec-Weyl-Model} we apply our quantum kinetic equation to a  
commonly employed toy model of a two-node Weyl semimetal, 
highlighting the complicated interplay between disorder, free-evolution, and 
driving terms in the kinetic equation.  
In Sec.~\ref{Sec-NMR} we focus 
on the positive magnetoconductance quadratic in magnetic field induced by the 
chiral anomaly in Weyl and Dirac metals.
Finally in Sec.~\ref{Sec-Discussion} we discuss our results and comment on some other potentially 
interesting applications of our transport theory.

\section{Chiral Magnetotransport Anomaly \label{Sec-Overview}} 

The anomalous positive magnetoconductance 
of Weyl semimetals was predicted theoretically by observing that when 
Berry-phase density-of-states and anomalous-velocity corrections are included, 
the theory of Bloch state wavepacket dynamics predicts
that electrons will be pumped between Fermi surface pockets when $\bm{E}\cdot\bm{B} \ne 0$.
Under disorder-free semiclassical dynamics, the number of electrons in a given valley
changes at the rate \cite{Son2012,Son2013}
\begin{equation} 
\frac{\partial N_i}{\partial t} =  Q_i \, \frac{e^2}{4\pi^2\hbar^2c} \bm{E}\cdot\bm{B},
\label{CA-conventional}
\end{equation} 
where $i$ is a valley label and 
\begin{equation} 
Q_i= \int \frac{d^3 k}{2 \pi \hbar}\frac{\partial f_{0}(\varepsilon_{\bm{k}}^m)}{\partial \varepsilon^m_{\bm{k}}}
\bm{v}^m_{\bm{k}} \cdot \bm{\Omega}^m_{\bm{k}}
\end{equation} 
is the chirality of the associated band.
Here, $\bm{\Omega}^m_{\bm{k}}$ is the 
momentum-space Berry curvature of 
band $m$ (a band that possesses a Fermi surface), $\bm{v}^m_{\bm{k}}$ is its 
Bloch-state group velocity, and $f_0(\varepsilon_{\bm{k}}^m)$ is the Fermi-Dirac distribution function whose 
derivative with respect to energy $\varepsilon^m_{\bm{k}}$ provides a $\delta$-function at the Fermi energy $E_F$.
It is easy to show that the chirality of each valley is an integer
equal to the sum of the chiralities \cite{AHM2004} of all 
the Weyl (band crossing) points enclosed by its Fermi surface, and that the 
sum of chirality over all Fermi surfaces vanishes.  

\begin{figure*}[!t]
\centering
\includegraphics[width=1.6\columnwidth]{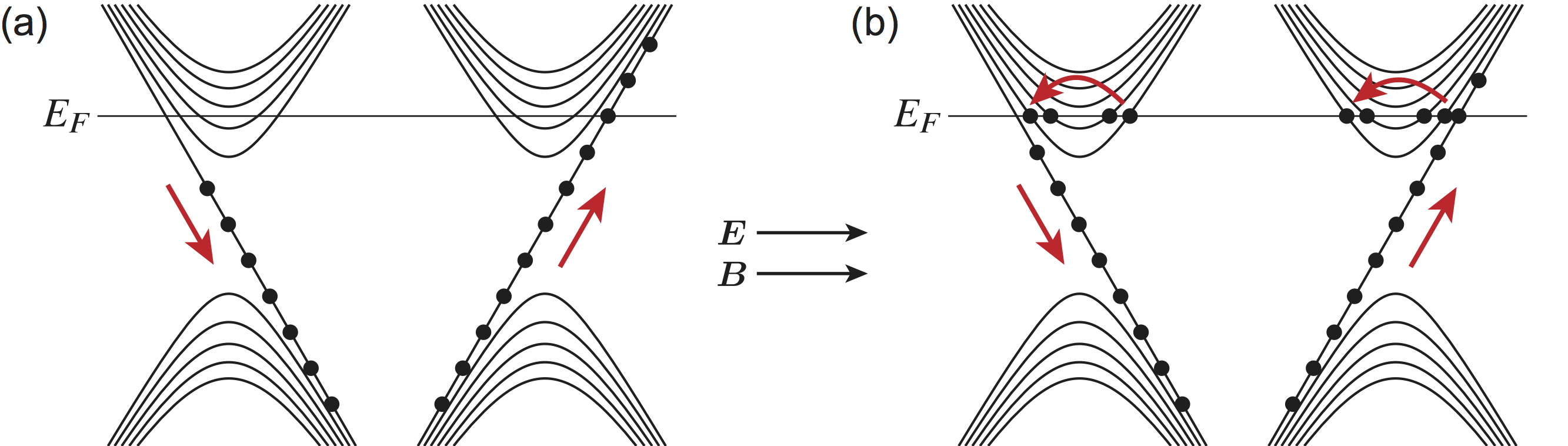}
\caption{(a) Pumping of valley population in a Weyl semimetal
by parallel electric and magnetic fields oriented along the $z$-direction in 
the absence of disorder.  
When the electronic structure of a Weyl semimetal 
is approximated by Dirac cones with masses
that depend on $k_z$, a magnetic field induces an anomalous ($N=0$)
Landau level branch that has only one sign of velocity $v_z$ in a given valley.
It follows that in each valley the density of states is increased for states with one sign of velocity and decreased 
for states with the other sign of velocity, and that the total current summed over a valley is already  
nonzero in equilibrium.  When an electric field drives states through  
momentum space (diagonal red arrows), the total 
number of states in a valley varies.
(b) Scattering within valleys 
(curved red arrows) can relax the current in each valley to its equilibrium 
value, but cannot establish a steady state because the number of states in each valley
still changes at a constant rate.  A steady state can be established only when intervalley 
scattering processes are present.  
}\label{Fig1}
\end{figure*}
The goal of our quantum kinetic theory is to capture all effects that are captured as Berry 
phase corrections in semiclassical wave-packet dynamics, and at the same time to account for 
disorder in a systematic fashion.  The chiral-anomaly-related magnetotransport anomalies of 
Weyl semi-metals are therefore an excellent test case for the theory.  We find that the 
chiral magnetotransport anomaly survives disorder in the limit that intervalley scattering 
is very weak compared to intravalley scattering.  In the following we 
take the direction of the parallel electric and magnetic fields to be the $z$-direction.

In a solid, magnetic and electric fields perturb electrons in qualitatively different ways since
equilibrium can be established in the presence of the former, but not in the presence of the later.
When disorder-free semiclassical wavepacket dynamics is used to derive Eq.~(\ref{CA-conventional}),
the origin of the term on the right hand side proportional to $\Omega_z$ is the change in the 
equilibrium electronic density of states induced by the magnetic field, whereas the 
terms proportional to $\Omega_x$ and $\Omega_y$ originate from the anomalous 
velocity induced by the electric field.
We show below that our quantum kinetic theory is able to completely account for both 
effects.

The robustness of the intervalley pumping effect in the presence of disorder is 
most easily understood using the Landau level picture illustrated in  Fig.~\ref{Fig1},
in which the influence of the magnetic field is accounted for by solving the equilibrium
quantum problem.  Note that at nonzero magnetic field each valley carries a nonzero current in equilibrium
because the number of Landau levels with positive and negative velocities along the 
field direction in an individual valley differ, but this does not lead to important observable effects.  
The intervalley pumping effect central to the chiral anomaly is present only when the electric field is added.    
In the semiclassical theory, the $\Omega_z$ contribution
to intervalley pumping accounts for the influence of Landau quantization 
on the equilibrium density of states, whereas the $\Omega_x$ and $\Omega_y$ corrections account 
for the changing character of Landau level wave functions as they are swept through 
momentum space by the electric field and appear as anomalous velocity contribution to the Lorentz force.
Although it is not possible to establish equilibrium in the presence of an 
electric field, in the presence of disorder scattering it is normally possible to establish a steady state with time-independent 
values of currents and other observables that can be measured experimentally.
The goal of the transport theory we 
present is to describe such a transport steady state and thereby 
to make connection with experimental observables while allowing for
the possibility that momentum space Berry curvature effects survive disorder.  
For the chiral anomaly,  for example, it is important to establish 
the connection between free Bloch-state dynamics and the transport 
steady state, and to distinguish the roles of intervalley and intravalley scattering. 
We will show that the chiral magnetotransport anomaly
is observable only when intervalley scattering at the Fermi energy is very weak compared to
intravalley scattering.  For a microscopic toy model of Weyl semimetals
we compute the positive quadratic magnetoconductivity induced by the chiral 
anomaly $\sigma_{zz}^{\mathrm{CA}}(B_z^2)$ in a fully systematic way.
In the low temperature limit we find that
the contribution to the magnetoconductivity from a single isotropic Weyl cone is 
\begin{align}
\sigma_{zz}^{\mathrm{CA}}(B_z^2)=\mathcal{C} \frac{e^2}{4\pi^2\hbar c^2}\frac{(eB_z)^2v_F^3}{\mu^2}\tau,
\end{align}
where $\mu$ is the chemical potential, $v_F$ is the Fermi velocity, $\tau$ is the intervalley scattering time,
and $\mathcal{C} \sim 1 $ is a coefficient that depends on electronic structure and disorder details. 
In the case of isotropic Weyl points we find that $\mathcal{C}=1$ in the limit of extremely weak intervalley scattering, independent of the details of the disorder scattering.

\section{Quantum Kinetic Equation \label{Sec-Kinetic-Equation}}
In this section, we derive a quantum kinetic equation for Bloch electrons 
in the presence of electric and magnetic fields.
To this end, we start with some general considerations related to Bloch Hamiltonians and basis function choices.
Throughout this paper, we work in the basis of the disorder-free Hamiltonian eigenstates, which we refer to as the eigenstate basis:
\begin{align}
H_0 |m,\bm{k}\rangle=\varepsilon^m_{\bm{k}} |m,\bm{k}\rangle,
\end{align}
where $H_0$ is the crystal Hamiltonian, $\varepsilon^m_{\bm{k}}$ is an eigenvalue of $H_0$, 
$\bm{k}$ is a momentum in the crystal's Brillouin-zone, and $m$ is a band index.  
The eigenstates of a crystal Hamiltonian are Bloch states, products of plane-waves with 
wave-vector $\bm{k}$ and periodic functions that are eigenfunctions 
of the $\bm{k}$-dependent $\bm{k} \cdot \bm{p}$ Hamiltonian.
A Bloch state does not, of course, have to be an eigenstate of the crystal 
Hamiltonian.  Our formalism assumes that a good approximation to the 
Hamiltonian matrix of the perfect crystal is known in a representation of $\bm{k}$-independent
periodic functions.  In the following we assume that the Hamiltonian is known in a 
representation of Wannier functions, but the formalism could be applied with little change 
if we used a representation of $\bm{k} \cdot \bm{p}$ eigenstates at a particular Brillouin-zone 
point of interest.

Given this starting point, we write
\begin{align}
H_0=\sum_{\bm{k},m}\varepsilon^m_{\bm{k}} |m,\bm{k}\rangle\langle m,\bm{k}|=\sum_{\bm{L}\bm{L}' \alpha\alpha'}H^{\alpha\alpha'}_{\bm{L}\bm{L}'}|\alpha,\bm{L}\rangle\langle \alpha',\bm{L}'|
\end{align}
with $H^{\alpha\alpha'}_{\bm{L}\bm{L}'}=\langle \alpha,\bm{L}|H_0|\alpha',\bm{L}'\rangle$.
Here $|\alpha,\bm{L}\rangle$ is a Wannier function associated with 
orbital $\alpha$ and real-space lattice vector $\bm{L}$.  For a particular orbital the Bloch and Wannier functions 
are related by
\begin{align}
|\alpha,\bm{L}\rangle &= \frac{1}{\sqrt{N}}\sum_{\bm{k}}e^{-i\bm{k}\cdot\bm{L}}|\alpha,\bm{k}\rangle, \nonumber\\
|\alpha,\bm{k}\rangle &= \frac{1}{\sqrt{N}}\sum_{\bm{k}}e^{+i\bm{k}\cdot\bm{L}}|\alpha,\bm{L}\rangle,
\end{align}
where $N$ is the number of Bravais lattice sites in the crystal. 
$|\alpha,\bm{k}\rangle$ is a Bloch state with wave function
\begin{equation} 
\langle {\bm{r}}|\alpha,\bm{k}\rangle=e^{i\bm{k}\cdot\bm{r}}u^\alpha_{\bm{k}}(\bm{r}),
\end{equation}
where $\bm{r}$ is position and $u^\alpha_{\bm{k}}(\bm{r})$ is the cell-periodic part of the 
Bloch wave function.
Our transport theory is formulated in terms of the single-particle density matrix. 
The density-matrix operator can be expressed in either the eigenstate or Wannier representations using 
\begin{align}
\rho=\sum_{\bm{k}\bm{k}' mm'}\rho^{mm'}_{\bm{k}\bm{k}'}|m,\bm{k}\rangle\langle m',\bm{k}'|=\sum_{\bm{L}\bm{L}' \alpha\alpha'}\rho^{\alpha\alpha'}_{\bm{L}\bm{L}'}|\alpha,\bm{L}\rangle\langle \alpha',\bm{L}'|,
\end{align}
where $\rho^{mm'}_{\bm{k}\bm{k}'}=\langle m,\bm{k}|\rho |m',\bm{k}'\rangle$ and $\rho^{\alpha\alpha'}_{\bm{L}\bm{L}'}=\langle \alpha,\bm{L}|\rho|\alpha',\bm{L}'\rangle$.
Since we focus on circumstances in which translational symmetry is not broken, 
the nonequilibrium expectation value of the 
density matrix operator $\rho$ will always be diagonal in Bloch state wavevector.
The eigenstate basis and the Wannier basis are related to each other through the 
unitary transformation related to band eigenvectors in the $\alpha$-representation:
$|m,\bm{k}\rangle=\sum_{\alpha}z_\alpha^{m}|\alpha,\bm{k}\rangle$ with
 $z_\alpha^{m}=\langle \alpha,\bm{k}|m,\bm{k}\rangle$.

\subsection{Introducing disorder}
The main result of this paper is a generic quantum kinetic equation that 
accounts for disorder and electric and magnetic fields.
Throughout this paper we take disorder into account within the Born approximation, implicitly assuming 
therefore that disorder is weak. 
We follow a procedure developed in an earlier paper \cite{Culcer2016}, 
generalizing it to allow for magnetic fields.

To establish some notation we first consider the case without external electric and magnetic fields.
We start with the quantum Liouville equation
\begin{align}
\frac{\partial \rho}{\partial t}+\frac{i}{\hbar}[H,\rho]=0,
\label{QLE}
\end{align}
where $\rho$ and $H$ are the density-matrix operator and the Hamiltonian of the system, respectively.
In the absence of external fields, the total Hamiltonian of the system is $H=H_0+U$
where $U$ is the disorder potential.  
We decompose the density matrix $\rho$ into two parts, writing
$\rho=\langle\rho\rangle+g$, where $\langle\rho\rangle$ is the density matrix averaged over 
disorder configurations, and $g$ is the deviation from this average.
The quantum Liouville equation can then be decomposed into 
coupled equations for $\langle\rho\rangle$ and $g$:
\begin{align}
\frac{\partial \langle\rho\rangle}{\partial t}+\frac{i}{\hbar}[H_0,\langle\rho\rangle]+\frac{i}{\hbar}\langle[U,g]\rangle=0,
\label{Eq-for-rho}
\end{align}
\begin{align}
\frac{\partial g}{\partial t}+\frac{i}{\hbar}[H_0,g]+\frac{i}{\hbar}[U,g]-\frac{i}{\hbar}\langle[U,g]\rangle&=-\frac{i}{\hbar}[U,\langle\rho\rangle].
\label{Eq-for-g}
\end{align}
In the Born approximation we can ignore the last two terms on the left hand side of Eq.~(\ref{Eq-for-g}).
The equation for $g$, Eq.~(\ref{Eq-for-g}), can then be solved straightforwardly.
By substituting the solution for $g$ into Eq.~(\ref{Eq-for-rho}), we obtain
\begin{align}
\frac{\partial \langle\rho\rangle}{\partial t}+\frac{i}{\hbar}[H_0,\langle\rho\rangle]+K(\langle\rho\rangle)=0,
\label{QLE-Born}
\end{align}
where the scattering term $K(\langle\rho\rangle)$ is given by \cite{Culcer2016}
\begin{align}
K(\langle\rho\rangle)=\frac{1}{\hbar^2}\int_0^\infty dt'\, \left\langle \left[U, [e^{-iH_0t'/\hbar}U e^{iH_0t'/\hbar}, \langle\rho(t)\rangle]\right]\right\rangle,
\label{scattering-K}
\end{align}
where $\langle\rho(t)\rangle=e^{-iH_0 t/\hbar}\langle\rho\rangle e^{iH_0 t/\hbar}$.

In the following, we do not exhibit
the explicit time dependence of $\langle\rho(t)\rangle$ in order to simplify the notation.
We separate the density matrix into band-diagonal and band-off-diagonal parts
because these two types of components behave quite differently using the notation
$\langle\rho\rangle=\langle n\rangle+\langle S\rangle$, where $\langle n\rangle$ is diagonal in 
band index and 
$\langle S\rangle$ is off-diagonal.  The scattering kernel can be separated into four parts which map $\langle n\rangle$ and 
$\langle S\rangle$ to band-diagonal and band off-diagonal contributions to $\partial \langle\rho(t)\rangle/\partial t$.  
This separation is discussed at length later in connection with magnetotransport.  
Focusing on the elastic-scattering case 
and using the notation of Ref.~\cite{Culcer2016} 
we find that the band-diagonal to band-diagonal part of $K(\langle\rho\rangle)$ from $\langle n\rangle$ is 
\begin{align}
[I(\langle n\rangle)]^{mm}_{\bm{k}} =\frac{2\pi}{\hbar} \sum_{m'\bm{k}'} & \langle U^{mm'}_{\bm{k}\bm{k}'}U^{m'm}_{\bm{k}'\bm{k}}\rangle(n^{m}_{\bm{k}} - n^{m'}_{\bm{k}'})\delta(\varepsilon^m_{\bm{k}} - \varepsilon^{m'}_{\bm{k}'}),
\label{scattering-I}
\end{align}
with $m$ and $m'$ being band indices.
This is exactly Fermi's golden rule.
Similarly, the band-diagonal to band-off-diagonal part of $K(\langle\rho\rangle)$ from $\langle n\rangle$  is
\begin{align}
[J(\langle n\rangle)]^{mm''}_{\bm{k}} = \frac{\pi}{\hbar} \sum_{m'\bm{k}'} \langle U^{mm'}_{\bm{k}\bm{k}'}U^{m'm''}_{\bm{k}'\bm{k}}\rangle \left[ (n^{m}_{\bm{k}} - n^{m'}_{\bm{k}'}) \right. \nonumber\\
\left.\times \delta(\varepsilon^m_{\bm{k}} - \varepsilon^{m'}_{\bm{k}'}) +(n^{m''}_{\bm{k}} - n^{m'}_{\bm{k}'}) \delta(\varepsilon^{m''}_{\bm{k}} - \varepsilon^{m'}_{\bm{k}'})\right], 
\label{Anomalous-driving-term}
\end{align}
where  $m \ne m''$.
The full expression for $K(\langle\rho\rangle)$, including the contributions from the off-diagonal density matrix $\langle S\rangle$, can be found in Ref.~\cite{Culcer2016}.

\subsection{Introducing electric and magnetic fields}
Next, we take the effects of magnetic fields into account using a semiclassical approximation 
that we expect to be accurate when the weak magnetic fields
condition $\omega_c\tau \ll 1$ is satisfied and Landau quantization can be neglected.
Here, $\omega_c$ is the cyclotron frequency and $\tau$ is the transport relaxation time
discussed at greater length below.
In order to derive the kinetic equation for systems in magnetic fields, we apply the 
Wigner transformation to the quantum Liouville equation (\ref{QLE}).
We consider a generic single-particle $D$-dimensional Bloch Hamiltonian $H_0(\bm{p})$ with 
momentum operator $\bm{p}=-i\hbar\nabla$.
In the presence of a vector potential $\bm{A}(\bm{r},t)$, minimal 
coupling results in $\bm{p}\rightarrow \bm{p}+e\bm{A}$ and adopt the 
notation that $e>0$ is 
the magnitude of the electron charge.

The Wigner distribution function in the presence of a vector potential $\bm{A}$ is defined by \cite{Vasko-book}
\begin{align}
\langle\rho\rangle_{\bm{p}}^{mn}(\bm{r})=\int d^D\bm{R}\, e^{-(i/\hbar)\bm{P}\cdot\bm{R}} \langle m,\bm{r}_+|\rho| n,\bm{r}_-\rangle,
\label{Wigner-distribution}
\end{align}
where $\bm{P}=\bm{p}-e\bm{A}$, $\bm{r}_\pm=\bm{r}\pm\bm{R}/2$, and $| n,\bm{r}\rangle=\sum_{m,\bm{k}} |m,\bm{k}\rangle\langle m,\bm{k}| n,\bm{r}\rangle=\sum_{\bm{k}}e^{i\bm{k}\cdot\bm{r}}|n,\bm{k}\rangle$ is the Fourier transform of $|n,\bm{k}\rangle$. 
Note that the sign in front of $e\bm{A}$ in $\bm{P}$ is different from the one in minimal coupling \cite{Vasko-book}.
We project the quantum Liouville equation (\ref{QLE}) onto the $\{|m,\bm{r}\rangle\}$ space.
Let us consider the term
\begin{align}
&\langle m,\bm{r}_+|[H_0,\rho]| n,\bm{r}_-\rangle \nonumber\\
&=\int d^3r'\sum_{m'} \left[\langle m,\bm{r}_+|H_0| m',\bm{r}'\rangle \langle m',\bm{r}'|\rho| n,\bm{r}_-\rangle\right. \nonumber\\
& \left. - \langle m,\bm{r}_+|\rho| m',\bm{r}'\rangle \langle m',\bm{r}'|H_0| n,\bm{r}_-\rangle\right].
\label{multiband-commutation}
\end{align}
Since the Hamiltonian is diagonal in real space, we have $\langle m,\bm{r}_1|H_0| m',\bm{r}_2\rangle=(\langle m',\bm{r}_2|H_0| m,\bm{r}_1\rangle)^*=\sum_{\bm{q},\bm{q}'}e^{-i\bm{q}\cdot\bm{r}_1} e^{i\bm{q}'\cdot\bm{r}_2} \langle m,\bm{q}|H_0(\bm{p}_1+e\bm{A}_1)|m',\bm{q}'\rangle$ with $\bm{p}_1=-i\hbar\partial_{\bm{r}_1}$.
We expand $\bm{p}_\pm$ and $\bm{A}_\pm$ up to linear order in $\bm{R}$ as
$\bm{p}_\pm=-i\hbar\left(\frac{1}{2}\nabla\pm\nabla_{\bm{R}}\right)$ and $\bm{A}_\pm=\bm{A}\pm\frac{1}{2}\left(\bm{R}\cdot\nabla\right)\bm{A}$, where $\nabla_{\bm{R}}\equiv\partial/\partial \bm{R}$.
Finally we obtain
\begin{align}
&\langle m,\bm{q}|H_0(\bm{p}_++\bm{A}_+)|m',\bm{q}' \nonumber\rangle\\
&=\langle m,\bm{q}|H_0(\bm{q})|m',\bm{q}'\rangle-\frac{1}{2}i\hbar\, \langle m,\bm{q}|\nabla_{\bm{q}} H_0(\bm{q})\cdot\nabla |m',\bm{q}'\rangle \nonumber\\
& +\frac{1}{2}e\, \langle m,\bm{q}|\nabla_{\bm{q}} H_0(\bm{q})\cdot\left[\left(\bm{R}\cdot\nabla\right)\bm{A}\right] |m',\bm{q}'\rangle
\label{matrix-element-H_r+}
\end{align}
with $\bm{q}=-i\hbar\nabla_{\bm{R}}+e\bm{A}=\bm{p}+e\bm{A}$.
In reaching Eq.~(\ref{matrix-element-H_r+}) we have observed that it follows
from the definition of the Wigner distribution function (\ref{Wigner-distribution}) that 
 $\bm{R}=i\hbar\nabla_{\bm{p}}$ and $\bm{p}=-i\hbar\nabla_{\bm{R}}$ .

We perform the Wigner transformation on the quantum Liouville equation (\ref{QLE}) as
\begin{align}
\int d^D\bm{R}\, e^{-(i/\hbar)\bm{P}\cdot\bm{R}} \langle m,\bm{r}_+|\left\{\frac{\partial \rho}{\partial t}+\frac{i}{\hbar}[H_0,\rho]\right\} | n,\bm{r}_-\rangle=0.
\end{align}
Note that the scattering term $K(\langle\rho\rangle)$ is not changed after the Wigner transformation.
We use the following identities
\begin{align}
&e^{-(i/\hbar)\bm{P}\cdot\bm{R}}\nabla=\left\{\nabla-(ie/\hbar)[\nabla(\bm{A}\cdot\bm{R})]\right\} e^{-(i/\hbar)\bm{P}\cdot\bm{R}}, \nonumber\\
&e^{-(i/\hbar)\bm{P}\cdot\bm{R}}\partial/\partial t=\left[\partial/\partial t+(ie/\hbar)\bm{E}\cdot\bm{R}\right]e^{-(i/\hbar)\bm{P}\cdot\bm{R}},
\label{nabla-partial_t}
\end{align}
where $\bm{E}=-\partial \bm{A}/\partial t$ is the electric field.
Note that we can obtain the same electric-field dependent term as in Eq.~(\ref{nabla-partial_t}) by taking the 
electric field into account using a scalar potential $\phi(\bm{r})=-\bm{E}\cdot\bm{r}$, i.e., by 
adding $H_E=-e\phi(\bm{r})=e\bm{E}\cdot\bm{r}$ to the Hamiltonian as $H_0 \to H_0+H_E$ in Eq.~(\ref{multiband-commutation}) \cite{Culcer2016}.
By multiplying Eq.~(\ref{matrix-element-H_r+}) by $e^{-(i/\hbar)\bm{P}\cdot\bm{R}}$ from the left 
we see that the second term in the right hand side of Eq.~(\ref{matrix-element-H_r+}) generates an 
additional term proportional to $\nabla(\bm{A}\cdot\bm{R})$.
Finally, the vector potential can be replaced by the magnetic field 
using the identity
\begin{align}
\nabla(\bm{A}\cdot\bm{R})-\left(\bm{R}\cdot\nabla\right)\bm{A}=\bm{R}\times\bm{B},
\end{align}
where $\bm{B}=\nabla\times\bm{A}$ is the magnetic field.
Notice that the matrix element of $\nabla_{\bm{q}} H_0(\bm{q})$ in Eq.~(\ref{matrix-element-H_r+}) can be written as
\begin{align}
&\langle m,\bm{q}|\nabla_{\bm{q}} H_0(\bm{q})|m',\bm{q}'\rangle=\delta(\bm{q}-\bm{q}')\times \nonumber\\
&\left[\nabla_{\bm{q}} \mathcal{H}_{0\bm{q}}^{mm'}+\langle u^m_{\bm{q}}|\nabla_{\bm{q}}u^{n'}_{\bm{q}}\rangle\mathcal{H}_{0\bm{q}}^{n'm'}-\mathcal{H}_{0\bm{q}}^{mn'}\langle u^{n'}_{\bm{q}}|\nabla_{\bm{q}}u^{m'}_{\bm{q}}\rangle\right],
\end{align}
where $\mathcal{H}_{0\bm{q}}^{mm'}=\langle m,\bm{q}|H_0(\bm{q})|m',\bm{q}\rangle=\delta_{mm'}\varepsilon^m_{\bm{q}}$ and $|u^{m}_{\bm{q}}\rangle$ is the periodic part of the Bloch function.
The magnetic-field dependent term in Eq.~(\ref{matrix-element-H_r+}) becomes $-\frac{1}{2}e\sum_{n',\bm{q}''}\langle m,\bm{q}|\nabla_{\bm{q}} H_0(\bm{q})|n',\bm{q}''\rangle\cdot\langle n',\bm{q}''|\bm{R}|m',\bm{q}'\rangle\times\bm{B}$.
Here, the term $\langle n',\bm{q}''|\bm{R}|m',\bm{q}'\rangle$ (with $\bm{R}=i\hbar\nabla_{\bm{q}}$) acts on $\langle m',\bm{r}'|\rho| n,\bm{r}_-\rangle$ in Eq.~(\ref{multiband-commutation}), which finally results in
\begin{align}
&\langle n',\bm{q}''|\bm{R}|m',\bm{q}'\rangle \langle\rho\rangle_{\bm{q}}^{m'n}=i\hbar\, \delta(\bm{q}''-\bm{q}')\delta(\bm{q}'-\bm{q})\times \nonumber\\
&\left[\nabla_{\bm{q}} \langle\rho\rangle_{\bm{q}}^{n'n}+\langle u^{n'}_{\bm{q}}|\nabla_{\bm{q}}u^{m'}_{\bm{q}}\rangle \langle\rho\rangle_{\bm{q}}^{m'n}-\langle\rho\rangle_{\bm{q}}^{n'm'}\langle u^{m'}_{\bm{q}}|\nabla_{\bm{q}}u^{n}_{\bm{q}}\rangle\right],
\end{align}
where the third term on the right hand side comes from the Hermiticity of the equation and 
consistency with the single-band limit.

Combining Eq.~(\ref{QLE-Born}) with the analysis above, we arrive at the final 
form for the quantum kinetic equation in the presence of disorder, an electric field $\bm{E}$,
and a magnetic field $\bm{B}$: 
\begin{align}
\frac{\partial \langle \rho\rangle}{\partial t}+\frac{i}{\hbar}[\mathcal{H}_0,\langle \rho\rangle]+\frac{1}{2\hbar}\left\{\frac{D\mathcal{H}_0}{D\bm{k}} \cdot \nabla\langle \rho\rangle\right\}+K(\langle \rho\rangle) \nonumber\\
=D_E(\langle \rho\rangle)+D_B(\langle \rho\rangle),
\label{full-kinetic-equation}
\end{align}
where $\bm{k}=\bm{q}/\hbar$ is the  crystal wavevector, and the angle brackets in $\langle\rho\rangle$ imply
an impurity average.
Here and below $\{\bm{a}\cdot\bm{b}\}\equiv\bm{a}\cdot\bm{b}+\bm{b}\cdot\bm{a}$ (with $\bm{a}$ and $\bm{b}$ being vectors) denotes a symmetrized operator product.
In Eq.~(\ref{full-kinetic-equation})
$D_E(\langle \rho\rangle)$ and $D_B(\langle \rho\rangle)$ are the electric and magnetic driving terms:
\begin{align}
D_{E}(\langle \rho\rangle)=\frac{e\bm{E}}{\hbar}\cdot\frac{D\langle \rho\rangle}{D\bm{k}}
\label{driving-term-E}
\end{align}
and
\begin{align}
D_{B}(\langle \rho\rangle)=\frac{e}{2\hbar^2}\left\{\left(\frac{D \mathcal{H}_0}{D\bm{k}}\times\bm{B}\right)\cdot\frac{D\langle \rho\rangle}{D\bm{k}}\right\}.
\label{driving-term-B}
\end{align}
Eqs.~(\ref{driving-term-E}) and (\ref{driving-term-B}) have been simplified by introducing the covariant 
derivative notation defined by 
\begin{align}
\frac{D\langle \rho\rangle}{D\bm{k}}=\nabla_{\bm{k}}\langle \rho\rangle-i[\bm{\mathcal{R}}_{\bm{k}},\langle \rho\rangle],
\label{Covariant-derivative}
\end{align}
where $\bm{\mathcal{R}}_{\bm{k}}=\sum_{a=x,y,z}\mathcal{R}_{\bm{k}}^a\bm{e}_a$ with $[\mathcal{R}_{\bm{k}}^a]^{mn}=i\langle u^m_{\bm{k}}|\partial_{k_a}u^n_{\bm{k}}\rangle$ being the generalized Berry connection.
From another point of view, the covariant derivative is simply the derivative evaluated in the $\bm{k}$-independent 
Wannier state representation.

Note that the covariant $\bm{k}$-derivative acting on the density matrix
accounts for changes in the density matrix due to the 
fact that its band-eigenstate representation elements are $\bm{k}$-dependent and also due to
the fact that the band-eigenstates themselves are $\bm{k}$-dependent.  The latter contributions 
capture the momentum-space Berry curvature contributions to semi-classical wavepacket dynamics, 
but now in a formalism that accounts consistently for the scattering terms that are necessary to establish the 
transport steady state.  The covariant derivative involving $\mathcal{H}_0$ 
in Eq.~(\ref{driving-term-B}) is defined in the same way,
simply replacing $\langle \rho\rangle$ by $\mathcal{H}_0$ in Eq.~(\ref{Covariant-derivative}).
For our formulation of magnetotransport, it will be important that $D_{B}$ is linear in the density matrix.
As we see below the $\bm{k}$-dependence of the eigenstates in this derivative play 
the essential role in capturing the chiral anomaly.  

The covariant derivatives reduce to simple derivatives in a spin-independent 
single-band system, for example in a parabolic band system
with $\mathcal{H}_0({\bm{k}})=\hbar^2\bm{k}^2/2m$.
In the same limit, Eq.~(\ref{full-kinetic-equation}) reduces
to the usual semiclassical Boltzmann equation:
\begin{align}
\left[\frac{\partial}{\partial t}+\bm{v}_{\bm{k}}\cdot\nabla-\frac{e}{\hbar}\left(\bm{E}+\bm{v}_{\bm{k}}\times\bm{B}\right)\cdot\nabla_{\bm{k}}\right]f_{\bm{k}}=-I(f_{\bm{k}}),
\label{semiclassical-Boltzmann}
\end{align}
where we have defined the velocity $\bm{v}_{\bm{k}}=(1/\hbar)\nabla_{\bm{k}} \mathcal{H}_0(\bm{k})$ and 
$I(f_{\bm{k}})$ is the single-band version of Eq.~(\ref{scattering-I}).
The quantum kinetic equation (\ref{full-kinetic-equation}) we have derived can
therefore be understood as a generalization of the simple
Boltzmann equation (\ref{semiclassical-Boltzmann})
in which the velocity and distribution function scalars are replaced by matrices,
the simple derivatives $\nabla_{\bm{k}}$ are replaced by covariant 
derivatives $D/D\bm{k}$, and scalar products are replaced by symmetrized 
matrix products $\frac{1}{2}\{\ \ \cdot\ \ \}$.
Equation (\ref{full-kinetic-equation}) is the principal result of this paper.

\section{Magnetotransport Theory \label{Sec-Magnetotransport}}
In this section, we describe a general scheme to apply the 
quantum kinetic equation (\ref{full-kinetic-equation}) to  phenomena in systems 
in which the momentum-space Berry connection plays
an important role.
We start by obtaining the expression for the density matrix in linear response to an electric field in the absence of a magnetic field.
Then we incorporate the effect of a magnetic field by performing a systematic low-magnetic-field expansion.
Throughout this paper, we assume that the system is uniform, i.e., that $\nabla\langle\rho\rangle=0$.

\subsection{Transport at zero magnetic field \label{Sec-rho_E}}
We first briefly summarize the $\bm{B} = 0$ limit of our transport theory.  
In linear response, we write the electron density matrix $\langle\rho\rangle$ as $\langle\rho\rangle=\langle\rho_0\rangle+\langle\rho_E\rangle$, where $\langle\rho_0\rangle$ is the equilibrium density matrix and $\langle\rho_E\rangle$ is the correction to $\langle\rho_0\rangle$ in linear order in an electric field $\bm{E}$.
With this notation we need to solve the kinetic equation in the form
\begin{align}
\frac{\partial \langle\rho_E\rangle}{\partial t}+\frac{i}{\hbar}[\mathcal{H}_0,\langle\rho_E\rangle]+K(\langle\rho_E\rangle)=D_E(\langle \rho_0\rangle).
\label{KE-for-rho_E}
\end{align}
We divide the electron density matrix response $\langle\rho_E\rangle$ into a diagonal 
part $\langle n_E\rangle$ and an off-diagonal part $\langle S_E\rangle$, writing
$\langle\rho_E\rangle=\langle n_E\rangle+\langle S_E\rangle$.
Note that the equilibrium density matrix $\langle\rho_0\rangle$ is diagonal in the band index.

When only band-diagonal to band-diagonal terms are included in the 
scattering kernel, it is easy to solve for the steady-state value of $\langle n_E\rangle$.
The kinetic equation (\ref{KE-for-rho_E}) in this limit is 
\begin{align}
[I(\langle n_E\rangle)]^{mm}_{\bm{k}}=[D_E(\langle \rho_0\rangle)]^{mm}_{\bm{k}}=e\bm{E}\cdot \bm{v}^m_{\bm{k}}\frac{\partial f_0(\varepsilon^m_{\bm{k}})}{\partial \varepsilon^m_{\bm{k}}},
\label{rho_E-diagonal}
\end{align}
where $\bm{v}^m_{\bm{k}}=(1/\hbar)\nabla_{\bm{k}}\varepsilon^m_{\bm{k}}$ and $\langle \rho_0\rangle^{mm}=f_0(\varepsilon^m_{\bm{k}})$ is the Fermi-Dirac distribution function.
The equation for $n_{E}$ is therefore a familiar linear integral equation which is 
discussed at greater length below and yields  
$\langle n_E\rangle^{m}_{\bm{k}}=e\tau^m_{\mathrm{tr}\bm{k}} \bm{E}\cdot \bm{v}^m_{\bm{k}}\partial f_0(\varepsilon^m_{\bm{k}})/\partial \varepsilon^m_{\bm{k}}$, where $\tau^m_{\mathrm{tr}\bm{k}}$ is the transport lifetime which 
is often nearly constant across the Fermi surface.  

Next we consider the solution for the off-diagonal part of the density matrix $\langle S_E\rangle$,
which is independent of weak disorder.  
From Eq.~(\ref{KE-for-rho_E}) the kinetic equation for $\langle S_E\rangle$ is given by
\begin{align}
\frac{\partial \langle S_E\rangle}{\partial t}+\frac{i}{\hbar}[\mathcal{H}_0,\langle S_E\rangle]=D'_E(\langle \rho_0\rangle)-J(\langle n_E\rangle),
\end{align}
where $D'_E(\langle \rho_0\rangle)$ is the off-diagonal part of the intrinsic driving term:
\begin{align}
[D_E(\langle \rho_0\rangle)]^{mm'}_{\bm{k}}=i\frac{e\bm{E}}{\hbar}\cdot\bm{\mathcal{R}}_{\bm{k}}^{mm'}\bigl[f_0(\varepsilon^{m}_{\bm{k}})-f_0(\varepsilon^{m'}_{\bm{k}})\bigr].
\label{D_E-intrinsic}
\end{align}
The off-diagonal part of the driving term is responsible for the Berry phase contribution to the Hall conductivity of systems with broken time-reversal symmetry in the absence of a magnetic field.
As we will see, it also plays an essential role in the anomalous magnetoconductivity response.
The solution to this equation is \cite{Culcer2016}
\begin{align}
\langle S_E\rangle=\int_0^\infty dt'\, e^{-i\mathcal{H}_0 t'/\hbar} [D_E(\langle \rho_0\rangle)-J(\langle n_E\rangle)] e^{i\mathcal{H}_0 t'/\hbar},
\end{align}
where we have not explicitly exhibited
the time dependences of $\langle \rho_0(t-t')\rangle$ and $\langle n_E(t-t')\rangle$.
It can be further expanded in the eigenstate basis by inserting an infinitesimal $e^{-\eta t'}$ and taking the limit $\eta\rightarrow 0$ to obtain
\begin{align}
\langle S_E\rangle_{\bm{k}}^{mm'}=-i\hbar\frac{[D_E(\langle \rho_0\rangle)]^{mm'}_{\bm{k}} - [J(\langle n_E\rangle)]^{mm'}_{\bm{k}}}{\varepsilon_{\bm{k}}^m-\varepsilon_{\bm{k}}^{m'}}.
\label{rho_E-off-diagonal}
\end{align}
Here we have written only the principal value part and omitted $\delta$-function terms which can 
be important when bands touch, giving rise for example to the Zitterbewegung contribution to the minimum 
conductivity \cite{Culcer2016} in graphene.
We note that, as shown in Ref.~\cite{Culcer2016}, the contribution from $J(\langle n_E\rangle)$ corresponds to the vertex correction in the ladder-diagram approximation of perturbation theory.

\subsection{Equilibrium response to a magnetic field}
Next we look closely at the magnetic driving term (\ref{driving-term-B})
by examining linear response to magnetic field in the absence of an electric field.
In this $\bm{E}=0$ limit, we expect that the steady state is actually an equilibrium state, and not one 
in which energy transfer from the external field is balanced by dissipation.
We write the electron density matrix 
as $\langle\rho\rangle=\langle\rho_0\rangle+\langle\rho_B\rangle$, where $\langle\rho_0\rangle$ is the
$\bm{B}=0$ density matrix, and $\langle\rho_B\rangle$ is the correction to $\langle\rho_0\rangle$
at linear order in the magnetic field $\bm{B}$.
This equilibrium density matrix $\langle\rho_B\rangle$ satisfies
\begin{align}
\frac{\partial \langle\rho_B\rangle}{\partial t} + \frac{i}{\hbar}[\mathcal{H}_0, \langle\rho_B\rangle]+K(\langle\rho_B\rangle)= D_B(\langle\rho_0\rangle+\langle\rho_B\rangle).
\label{Kinetic-equation-for-B}
\end{align}
To first order in magnetic field the right-hand side can be replaced by 
$D_B(\langle \rho_0\rangle)$, where $\langle \rho_0\rangle$ is the Fermi-Dirac 
distribution function of Bloch states in the absence of a magnetic field. 
For $\bm{B}=(0,0,B_z)$ we have
\begin{align}
D_B(\langle \rho_{0}\rangle)=\frac{eB_z}{2\hbar^2}\left[\left\{\frac{D\mathcal{H}_0}{Dk_y},\frac{D\langle \rho_{0}\rangle}{Dk_x}\right\}-\left\{\frac{D\mathcal{H}_0}{Dk_x},\frac{D\langle \rho_{0}\rangle}{Dk_y}\right\}\right],
\end{align}
where $\{\ \ ,\ \ \}$ indicates a matrix anticommutator.
We will show in Sec.~\ref{Sec-CME} that $D_B\left(\langle \rho_{0}\rangle\right)$ 
describes valley-dependent electrical equilibrium currents induced 
by a magnetic field (i.e., the chiral magnetic effect) that are apparent in the Landau level representation illustrated 
in Fig.~\ref{Fig1}.  These currents cancel when summed over valleys, but will not 
cancel when the chemical potentials of the two valleys differ.

We separate the electron density matrix response
$\langle\rho_B\rangle$ into its diagonal $\langle \xi_B\rangle$ and
off-diagonal $\langle S_B\rangle$ parts: 
$\langle\rho_B\rangle=\langle \xi_B\rangle+\langle S_B\rangle$.
By writing the Hamiltonian and the $\bm{B}=0$ density matrix operators 
as $\mathcal{H}_0=\sum_{m,\bm{k}}\varepsilon_{\bm{k}}^m |m,\bm{k}\rangle\langle m,\bm{k}|$ and 
$\langle\rho_0\rangle=\sum_{m,\bm{k}}f_0(\varepsilon_{\bm{k}}^m) |m,\bm{k}\rangle\langle m,\bm{k}|$, 
where $\varepsilon_{\bm{k}}^m$ is an energy eigenvalue of band $m$ with momentum $\bm{k}$ and $f_0(\varepsilon_{\bm{k}}^m)$ is the Fermi-Dirac distribution function,
it can be shown quite generally that only band-off-diagonal components in 
$D_B\left(\langle \rho_{0}\rangle\right)$ are nonzero,
and hence that for weak disorder that $K (\langle \rho_B\rangle)=0$.
(See Appendix~\ref{Appendix-D_B-1} for a detailed discussion.)
This means that unlike an electric field [Eq.~(\ref{rho_E-diagonal})],
a magnetic field does not generate a dissipative current when applied to a conductor,
as we know.
However both fields do alter the band off-diagonal part of the density 
matrix.  We find that the solution for the kinetic equation~(\ref{Kinetic-equation-for-B}) is
\begin{align}
\langle S_B\rangle_{\bm{k}}^{mm'}=-i\hbar\frac{[D_B(\langle \rho_0\rangle)]^{mm'}_{\bm{k}}}{\varepsilon_{\bm{k}}^m-\varepsilon_{\bm{k}}^{m'}},
\label{rho_B-formal-expression}
\end{align}
where $m\neq m'$.
See Eq.~(\ref{rho_E-off-diagonal}) for the corresponding electric field equation.

There is another contribution to the equilibrium density matrix linear in magnetic field in the absence of an electric field.
To calculate this, it is convenient to introduce an operator $P$ for an arbitrary density matrix $\langle\rho\rangle$ as
\begin{equation} 
P\langle\rho\rangle \equiv \frac{i}{\hbar}[\mathcal{H}_0,\langle \rho\rangle].
\label{definition-of-P}
\end{equation}
Note that, in the eigenstate representation, the matrix $P$ is purely diagonal in both wavevector and in density-matrix element at a given wavevector, and that it is nonzero only for off-diagonal density-matrix elements.
The notation $P$ for the Bloch-state evolution term $\partial \langle\rho\rangle/\partial t$ in the absence of fields and  
disorder is intended to suggest its role as the many-band generalization of the spin-precession 
terms in two-band models which have only a spin-$1/2$ degree of freedom for each momentum.
Using Eq.~(\ref{definition-of-P}) we can rewrite the kinetic equation~(\ref{Kinetic-equation-for-B}) as
\begin{equation} 
P \langle\rho_{B}\rangle = D_B\left(\langle \rho_{0}\rangle\right),
\end{equation}
which can be viewed as a quantum operator.

As mentioned above, $[D_B(\langle \rho_0\rangle)]^{mm}_{\bm{k}}=0$.
However, it can also be shown quite generally that the band-diagonal components of $P^{-1}D_B(\langle \rho_0\rangle)$ are not zero, i.e., that $[P^{-1}D_B(\langle \rho_0\rangle)]^{mm}_{\bm{k}}\neq 0$.
Here, the operator $P^{-1}$ acting on a driving term $D$ is defined by $[P^{-1}D]^{mm'}=-i\hbar D^{mm'}_{\bm k}/(\varepsilon^{m}_{\bm k} - \varepsilon^{m'}_{\bm k})$.
See Appendix~\ref{Appendix-D_B-1} for a detailed derivation.
We find that in the weak disorder ($ W \tau_{\mathrm{tr}} \gg \hbar$ where $W$ is the scale of the typical 
energetic separation between Bloch bands) 
\begin{align}
\langle \xi_B\rangle_{\bm{k}}^{mm}\equiv[P^{-1}D_B(\langle \rho_0\rangle)]^{mm}_{\bm{k}}=\frac{e}{\hbar}\, f_0(\varepsilon_{\bm{k}}^m)\, \bm{B}\cdot\bm{\Omega}^m_{\bm{k}},
\label{intrinsic-diagonal-matrix-B}
\end{align}
where $\Omega_{\bm{k},a}^m=\epsilon^{abc}\, i\langle \partial_{k_b} u_{\bm{k}}^m|\partial_{k_c} u_{\bm{k}}^m\rangle$ is the Berry curvature of band $m$.
Note that the magnetic-field induced change in the 
diagonal density matrix $\langle \xi_B\rangle$ is quite distinct in character from the 
changes induced by an electric field $\langle n_E\rangle$~(\ref{rho_E-diagonal}) because 
$\langle \xi_B\rangle$ is intrinsic (independent of disorder effects) and not related to 
dissipation. 
Equation~(\ref{intrinsic-diagonal-matrix-B}) captures the same physics as the Berry phase correction to the density of states 
implied by semiclassical wavepacket dynamics, $(2\pi)^{-D}[1+(e/\hbar)\bm{B}\cdot\bm{\Omega}^m_{\bm{k}}]$ \cite{Xiao2005},
and can be viewed as an alternate derivation of that effect.
In our quantum kinetic formalism, however, it is the equilibrium electron density matrix that is modified by a magnetic field, 
while the number of states per momentum space volume is unchanged.
From Eq.~(\ref{intrinsic-diagonal-matrix-B}) we can immediately obtain
the St\v{r}eda formula for the quantum Hall effect with a magnetic field along the $z$ direction \cite{Streda1982}:
\begin{align}
\sigma_{xy}=-e\frac{\partial}{\partial B_z}\mathrm{Tr}[\langle\rho_B\rangle]=-\frac{e^2}{\hbar}\sum_m\int\frac{d^D\bm{k}}{(2\pi)^D}f_0(\varepsilon_{\bm{k}}^m)\Omega_{\bm{k},z}^m.
\end{align}

\begingroup
\renewcommand{\arraystretch}{1.6}
\begin{table*}[!t]
\caption{Schematic comparison of our formalism with semiclassical wavepacket dynamics in the presence of a magnetic field $\bm{B}$.
Semiclassical wavepacket dynamics implies the Berry phase correction to the density of states, $(2\pi)^{-D}[1+(e/\hbar)\bm{B}\cdot\bm{\Omega}^m_{\bm{k}}]$, while the electron distribution function $\mathcal{F}_{\bm{k}}^m$ remains unchanged \cite{Xiao2005}.
In our quantum kinetic formalism, however, it is the equilibrium electron density matrix that is modified by a magnetic field, while the momentum-space density of states remains unchanged [see Eqs.~(\ref{intrinsic-diagonal-matrix-B}) and (\ref{intrinsic-diagonal-matrix-B-general})].
Arrows in this table indicate the changes in physical quantities due to the presence of a magnetic field.
}
\begin{ruledtabular}
\begin{tabular}{ccc}
Formalism & Momentum-space density of states & Momentum-space distribution function\\
\hline
Semiclassical wavepacket dynamics \cite{Xiao2005} & $(2\pi)^{-D}\to (2\pi)^{-D}\left[1+(e/\hbar)\bm{B}\cdot\bm{\Omega}^m_{\bm{k}}\right]$ & $\mathcal{F}_{\bm{k}}^m\to \mathcal{F}_{\bm{k}}^m$\\
Our quantum kinetic formalism & $(2\pi)^{-D}\to (2\pi)^{-D}$ & $\mathcal{F}_{\bm{k}}^m\to \mathcal{F}_{\bm{k}}^m\left[1+(e/\hbar)\bm{B}\cdot\bm{\Omega}^m_{\bm{k}}\right]$
\end{tabular}
\end{ruledtabular}\label{Table1}
\end{table*}
\endgroup
Finally we note that the same result is obtained when $D_B$ acts on any band diagonal density matrix 
$\langle n\rangle=\sum_{m,\bm{k}}\mathcal{F}_{\bm{k}}^m |m,\bm{k}\rangle\langle m,\bm{k}|$.
The calculation details do not appeal to special properties of the equilibrium density matrix.   
In the general case we find that 
\begin{align}
\langle \xi_B\rangle_{\bm{k}}^{mm}=[P^{-1}D_B(\langle n\rangle)]^{mm}_{\bm{k}}=\frac{e}{\hbar}\, \mathcal{F}_{\bm{k}}^m\, \bm{B}\cdot\bm{\Omega}^m_{\bm{k}}.
\label{intrinsic-diagonal-matrix-B-general}
\end{align}
This property will play an important role in 
evaluating contributions to the 
magnetoconductivity (\ref{Magnetoconductivity-general}).
A schematic comparison between our formalism and semiclassical wavepacket dynamics is 
given in Table \ref{Table1}.

\subsection{Magnetotransport}
Now we consider a density matrix in the presence of electric and magnetic fields.
We write the electron density matrix 
as $\langle\rho\rangle=\langle\rho_0\rangle+\langle\rho\rangle_F$, where $\langle\rho_0\rangle$ is the density matrix in the absence of fields, and $\langle\rho\rangle_F$ is the field-induced density matrix.
Then we can rewrite the steady-state uniform system limit of Eq.~(\ref{full-kinetic-equation}) in the form
\begin{equation} 
(\mathcal{L} - D_E - D_B)\langle\rho\rangle_F = (D_E+D_B)\langle \rho_0\rangle,
\end{equation}
where we have defined an operator $\mathcal{L}\equiv P + K$ and have used the fact that $\mathcal{L}\langle \rho_0\rangle=0$.
It follows that
\begin{align}
\langle\rho\rangle_F&=\bigl[1-\mathcal{L}^{-1}(D_E+D_{B})\bigr]^{-1} \mathcal{L}^{-1}(D_E+D_B)\langle \rho_0\rangle \nonumber\\
&=\sum_{N\ge 0} \bigl[\mathcal{L}^{-1}(D_E+D_B)\bigr]^N \mathcal{L}^{-1}(D_E+D_B)\langle \rho_0\rangle.
\label{rho-expansion-general}
\end{align}
We then view the four terms $P$, $K$, $D_E$, and $D_B$ as matrices that act on vectors formed by all 
eigenstate-representation density-matrix components at all wavevectors, both band diagonal and 
band off-diagonal, ordered in any convenient way.
Note also that the matrix $P$ is purely diagonal in both wavevector and in density-matrix
element at a given wavevector, and that it is nonzero only for off-diagonal density-matrix elements.

Equation~(\ref{rho-expansion-general}) can be used to derive a low-magnetic-field expansion for the linear response 
of the density-matrix, and hence of any single-particle observable, to an electric field $\bm{E}$.
From Eq.~(\ref{rho-expansion-general}) we have
\begin{align}
\langle\rho\rangle_F=\sum_{N,N'\ge 0} (\mathcal{L}^{-1}D_B)^N \mathcal{L}^{-1}D_{E} (\mathcal{L}^{-1}D_B)^{N'}\langle \rho_0\rangle.
\end{align}
Here, the $N=N'=0$ term is given by Eqs.~(\ref{rho_E-diagonal}) and (\ref{rho_E-off-diagonal}).
It is convenient to define a density matrix induced solely by the magnetic field as $\langle \rho_B\rangle\equiv\sum_{N\ge 1}(\mathcal{L}^{-1}D_B)^{N}\langle \rho_0\rangle$.
Note that this expression is the generalized solution for Eq.~(\ref{Kinetic-equation-for-B}), i.e., for $(\mathcal{L} - D_{B}) \langle\rho_B\rangle = D_B(\langle\rho_0\rangle)$.
With this expression for $\langle\rho_B\rangle$ we have a simple expression for the low-magnetic-field expansion of the density matrix as
\begin{align}
\langle\rho\rangle_F=\sum_{N\ge 0} (\mathcal{L}^{-1}D_B)^N \mathcal{L}^{-1}D_{E}(\langle \rho_0\rangle+\langle \rho_B\rangle).
\label{rho-expansion}
\end{align}
At each order in $B$, contributions to $\langle\rho\rangle_F$ can quite generally 
be organized by their order in an expansion in powers of scattering strength $\lambda$
by letting $K \to \lambda K$ and identifying terms with a particular power of $\lambda$.  
The various low-field expansion terms are generated by repeated action of $D_{B}$ and $\mathcal{L}^{-1}$.
In Sec.~\ref{Sec-Scattering} we will take a close look at the properties of $\mathcal{L}^{-1}$,
pointing out that even in systems with weak disorder, the terms that have the highest power of $\lambda^{-1}$ 
do not necessarily dominate.  For states near the Fermi level, the low-field expansion is an 
expansion in powers of $\omega_c \tau$, where $\omega_c$ is the cyclotron frequency and 
$\tau$ is the intravalley scattering time discussed below.  

We are now in a position to obtain the 
magnetoconductivity of a system, i.e., the electrical conductivity in the presence of a magnetic field.
Since we are assuming that the magnetic field $\bm{B}$ is very weak, we may set $\langle \rho_B\rangle=\langle\xi_B\rangle$ in Eq.~(\ref{rho-expansion}), which means that the correction to the Fermi-Dirac distribution function due to $\bm{B}$ is given by the Berry phase correction $\langle\xi_B\rangle$ (\ref{intrinsic-diagonal-matrix-B}).
Then with the use of Eq.~(\ref{rho-expansion}) the total conductivity $\sigma_{\mu\nu}$ is obtained from the 
definition $\sigma_{\mu\nu}=\mathrm{Tr}[(-e) v_\mu\langle\rho\rangle_F]/E_\nu=\sum_{N\ge 0} \sigma_{\mu\nu}(B^N)$ ($\mu,\nu=x,y,z$).
Here, $\sigma_{\mu\nu}(B^N)$ ($N\ge 1$) is the magnetoconductivity 
contribution proportional to $B^N$, which is given by
\begin{align}
\sigma_{\mu\nu}(B^N)=&\mathrm{Tr}\left[(-e) v_\mu (\mathcal{L}^{-1}D_B)^N \mathcal{L}^{-1}D_E(\langle\rho_0\rangle)\right]/E_\nu \nonumber\\
&+\mathrm{Tr}\left[(-e) v_\mu (\mathcal{L}^{-1}D_B)^{N-1} \mathcal{L}^{-1}D_E(\langle\xi_B\rangle)\right]/E_\nu.
\label{Magnetoconductivity-general}
\end{align}
Here, $v_{\mu}$ is the $\mu$ component of the 
velocity operator matrix in the eigenstate representation (i.e., $v_\mu=D \mathcal{H}_0/D k_\mu$), and $\mathrm{Tr}$ indicates both a matrix trace and an
integration with respect to $\bm{k}$ over the Brillouin zone.  
Note that in Eq.~(\ref{Magnetoconductivity-general}) we have allowed 
for an arbitrary angle between the directions of the electric and magnetic fields.
In Sec.~\ref{Sec-NMR} we will apply Eq.~(\ref{Magnetoconductivity-general}) to the positive magnetoconductivity proportional to $B^2$, which arises as a consequence of the chiral anomaly, in Dirac and Weyl semimetals.

\section{Fermi Surface Response in Multi-valley Systems \label{Sec-Scattering}}
Anomalous transport is often related to band crossings at energies that are close to the 
Fermi energy and give rise to large Berry curvatures.  
These are often required by symmetry to have replicas at different Brillouin-zone points,
for example at points related by time-reversal operations.  For that reason, 
anomalous transport properties of the type that the present formalism is intended to
describe often occur in systems with more than one Fermi surface pocket.
In what follows, a {\it pocket} refers to a closed Fermi surface segment.
We will follow the common practice in the literature on Weyl and Dirac semimetals by borrowing 
terminology from semiconductor physics and referring to these clearly separated regions of 
momentum space that contribute importantly to the physical properties of interest as valleys. 
It is common that scattering between valleys related to each other by time-reversal,
or by some approximate crystal symmetry, is often weaker than scattering within valleys,
giving rise to long intervalley relaxation times.  In this section, we explain 
when these long relaxation times are physically relevant.

\subsection{Properties of the operator $\mathcal{L}^{-1}$}
The scattering term $K(\langle\rho\rangle)$ [Eq.~(\ref{scattering-K})] may be viewed as a linear operator that
acts on the density matrix and is the 
sum of four terms categorized by the following
separation: processes that map diagonal $\bkt{\rho}$ to diagonal $\partial  \langle\rho\rangle/\partial t $ (defined as $K^{dd}$),
processes that map diagonal $\bkt{\rho}$ to 
off-diagonal $\partial  \langle\rho\rangle/\partial t$ ($K^{od}$), processes that map off-diagonal $\bkt{\rho}$ to 
diagonal $\partial  \langle\rho\rangle/\partial t $ ($K^{do}$) and processes
that map off-diagonal $\bkt{\rho}$ to 
off-diagonal $\partial  \langle\rho\rangle/\partial t $ ($K^{oo}$).
For example, with this notation we can rewrite $I(\langle n\rangle)$ and $J(\langle n\rangle)$ as
\begin{align}
[I(\langle n\rangle)]_{\bm{k}}^{mm}&=\sum_{m',\bm{k}'} K^{dd}[m{\bm k};m'{\bm k}'] n^{m'}_{\bm{k}'}, \nonumber\\
[J(\langle n\rangle)]_{\bm{k}}^{mm'}&=\sum_{m'',\bm{k}'} K^{od}[mm'{\bm k};m''{\bm k}'] n^{m''}_{\bm{k}'}
\label{K-matrix-representaion}
\end{align}
with $m\neq m'$.
Explicit expressions for $K^{dd}$ and $K^{od}$ can be read off from Eqs.~(\ref{scattering-I}) and 
(\ref{Anomalous-driving-term}).
Those for $K^{do}$ and $K^{oo}$ can be found in Ref.~\cite{Culcer2016}.

When block-diagonalized to separate band-diagonal and band-off-diagonal components of the 
density matrix, with this notation the Liouville operator  
\begin{equation} 
\mathcal{L}= P + K =  
\begin{bmatrix}
K^{dd} & K^{do} \\ K^{od} & P+K^{oo}.
\end{bmatrix}.
\end{equation}
The disorder ($K$) contributions to $\mathcal{L}$ are proportional to the disorder strength
parameter $\lambda$, whereas $P$ is intrinsic and independent of disorder strength.
Our expressions for magnetoconductivity are formulated in terms of repeated action of the 
operator
\begin{equation} 
\mathcal{L}^{-1} =
\begin{bmatrix}
(\mathcal{L}^{-1})^{dd} & (\mathcal{L}^{-1})^{do} \\
(\mathcal{L}^{-1})^{od} & (\mathcal{L}^{-1})^{oo}.  
\end{bmatrix}.
\end{equation}
The leading terms in the disorder strength expansion of $\mathcal{L}$ are $\sim \lambda^{-1}$.
As discussed in Ref.~\cite{Culcer2016} it follows that up to the order of $\lambda^{0}$ 
\begin{equation} 
\mathcal{L}^{-1} =
\begin{bmatrix}
(K^{dd})^{-1} & -(K^{dd})^{-1} K^{do} P^{-1} \\
-P^{-1}K^{od}(K^{dd})^{-1} & P^{-1} 
\end{bmatrix}.
\end{equation} 
Note that the ${\it dd}$ block in $\mathcal{L}^{-1}$ is proportional to 
disorder strength $\lambda^{-1}$ and diverges in the limit of weak disorder, 
while all other blocks are proportional to $\lambda^{0}$.
We see from Eqs.~(\ref{rho_E-off-diagonal}) and (\ref{rho_B-formal-expression}) that the 
matrix $P^{-1}$ is diagonal in the density-matrix-component wavevector space.  When it acts on a driving term $D$,
it simply multiplies $D$ by a numerical factor:
\begin{equation}
P^{-1}D=-i\hbar\frac{D^{mm'}_{\bm k}}{\varepsilon^{m}_{\bm k} - \varepsilon^{m'}_{\bm k}}.
\label{Operator-P}
\end{equation}
A similar expression applies when $P^{-1}$ acts on any
Liouville-space density-matrix vector with nonzero off-diagonal components.

We conclude from these considerations that 
the leading order term in the disorder strength expansion of the 
contribution to the response of the {\it diagonal} part of the density matrix $\langle n_N\rangle$ 
at $N$-th order in magnetic field strength $B$ is obtained by setting
$\mathcal{L}^{-1}$ in Eq.~(\ref{rho-expansion}) to 
\begin{equation}
\mathcal{L}^{-1} \to
\begin{bmatrix}
(K^{dd})^{-1} & 0 \\
0 & 0
\end{bmatrix}.
\end{equation}
Then it follows that 
\begin{align}
\langle n_N\rangle=(K^{dd})^{-1}D^d_B(\langle\rho_{N-1}\rangle),
\label{n_B-Nth-order}
\end{align}
where $D^d_B$ is the diagonal part of the magnetic driving term $D_B$, and 
$\langle\rho_{N-1}\rangle=(\mathcal{L}^{-1}D_B)^{N-1} \mathcal{L}^{-1}D_{E}(\langle \rho_0\rangle)$ is the density matrix at $(N-1)$-th order in magnetic field.
On the other hand, the leading order term in the disorder strength expansion of the 
contribution to the response of the {\it off-diagonal} part of the density matrix $\langle S_N\rangle$ is obtained by setting
\begin{equation}
\mathcal{L}^{-1} \to
\begin{bmatrix}
0 & 0 \\
-P^{-1}K^{od}(K^{dd})^{-1} & P^{-1} 
\end{bmatrix}.
\end{equation}
Then it follows that 
\begin{align}
\langle S_N\rangle&=P^{-1}D^o_B(\langle\rho_{N-1}\rangle)-P^{-1}K^{od}(K^{dd})^{-1}D^d_B(\langle\rho_{N-1}\rangle) \nonumber\\
&=P^{-1}\left[D^o_B(\langle\rho_{N-1}\rangle)-K^{od}\langle n_N\rangle\right],
\label{S_B-Nth-order}
\end{align}
where $D^o_B$ is the off-diagonal part of the magnetic driving term $D_B$.
Notice from Eq.~(\ref{K-matrix-representaion}) that $K^{od}\langle n_N\rangle=J(\langle n_N\rangle)$.
Therefore, we see that Eq.~(\ref{S_B-Nth-order}) is similar to the $B=0$ equation
$\langle S_{E}\rangle=P^{-1}[D^o_E(\langle\rho_{0}\rangle)-K^{od}\langle n_{E}\rangle]$ [see Eq.~(\ref{rho_E-off-diagonal})]. 
Here, recall that, as shown in Ref.~\cite{Culcer2016}, the contribution from $J(\langle n_E\rangle)=K^{od}\langle n_{E}\rangle$ corresponds to the vertex correction in the ladder-diagram approximation of perturbation theory.
The same argument is applied to the case of magnetotransport.
Thus, we conclude that the contribution from $K^{od}\langle n_N\rangle$ to $\langle S_N\rangle$ corresponds to 
a vertex correction in the ladder-diagram approximation of perturbative  calculation.
Finally, we note that $\langle S_N\rangle$ is accompanied by a diagonal density matrix
\begin{align}
\langle \xi_N\rangle=P^{-1}D^o_B(\langle n_{N-1}\rangle).
\label{xi_B-Nth-order}
\end{align}
This is the Berry phase correction to the diagonal part of the density matrix, Eq.~(\ref{intrinsic-diagonal-matrix-B-general}).
$\langle n_N\rangle$ will be dominant in the case where both $\langle n_N\rangle$ and $\langle \xi_N\rangle$ are present, since $(K^{dd})^{-1}$ is of the order of $\lambda^{-1}$ while $P^{-1}$ is of the order of $\lambda^0$.
However, as will be shown in Sec.~\ref{Subsec-quadratic-MC}, $\langle \xi_N\rangle$ can make a large contribution to the magnetoconductivity in systems with large momentum-space Berry curvatures.

\subsection{Scattering times and the total number of electrons in a given valley}
We have delayed to this point a discussion of some important properties of $K^{dd}$ and $(K^{dd})^{-1}$ that 
are related to particle number conservation and play an important role in anomalous magnetotransport.
The components of these matrices are labelled by band and 
wavevector labels.  Comparing with Eq.~(\ref{scattering-I}) we see that 
\begin{align}
K^{dd}[m{\bm k};m'{\bm k}']
=& \frac{2\pi}{\hbar} 
\biggl[ \delta_{mm'} \delta_{\bm{k}\bm{k}'} \sum_{n,\bm{p}} \langle |U^{mn}_{\bm{k}\bm{p}}|^2 \rangle
 \delta(\varepsilon^m_{\bm{k}} - \varepsilon^{n}_{\bm{p}}) \nonumber\\
 & - \langle |U^{mm'}_{\bm{k}\bm{k}'}|^2 \rangle
 \delta(\varepsilon^m_{\bm{k}} - \varepsilon^{m'}_{\bm{k}'}) \biggr] .
\label{K^{dd}matrix}
\end{align}
It follows that $\sum_{m'\bm{k}'} K^{dd}[m{\bm k};m'{\bm k}']=0$,
i.e.,that the response vector corresponding to total particle number is an eigenvector of 
$K^{dd}$ with eigenvalue $0$.  Scattering does not relax total particle number.
Since the leading diagonal response $\langle n_E\rangle$ is given as
$\langle n_E\rangle=(K^{dd})^{-1} D_E^{d}$ with $D_E^d$ the diagonal part of $D_E$ [see Eq.~(\ref{rho_E-diagonal})], it might seem that the response is divergent.  However it is easy to see that 
the total particle number component of the driving term, 
i.e.,the driving term $D_E^{d}$ summed over all band and wavevector labels is also zero.
In fact a stronger statement is valid, namely that the sum of the driving term vanishes when summed over 
all wavevectors associated with each individual Fermi surface pocket surrounding each relevant valley: 
\begin{equation}
\frac{\partial N_m}{\partial t}=\sum_{\bm{k}}[D_E(\langle\rho_0\rangle)]^{mm}_{\bm{k}}
 = \sum_{\bm{k}} \frac{e \bm{E}}{\hbar}\cdot \frac{\partial f_0(\varepsilon^m_{\bm{k}})}{\partial \bm{k}} =0,
\label{totaldensitydriving}
\end{equation}
where $N_m$ is the total particle number in band $m$ of a given valley.
The electric field moves electrons through momentum space; it does not create or destroy particles.
In the matrix equations we have been using to express response to electric 
fields, the total particle number component of the band-diagonal response has implicitly been projected out.

The eigenvalues of $(K^{dd})^{-1}$ have units of time and characterize the various time scales on which 
contributions to nonequilibrium populations on the Fermi surface relax.
For example, the time scale for relaxation of the nonequilibrium 
distribution induced by an electric field in a single-valley system, the transport lifetime $\tau_{\mathrm{tr}}$,
is given approximately by
\begin{align} 
\frac{1}{\tau^m_{\mathrm{tr}}}=\frac{2\pi}{\hbar D^m(\mu)}\sum_{\bm{k},\bm{k}'}\, \langle |U^{mm}_{\bm{k}\bm{k}'}|^2 
\rangle(1-\cos\theta_{\bm{k}\bm{k}'}) \delta(\mu - \varepsilon^m_{\bm{k}}) \delta(\mu - \varepsilon^m_{\bm{k}'}),
\end{align}
where $D^m(\mu)=\sum_{\bm{k}}\delta(\mu - \varepsilon^m_{\bm{k}})$ is the density of states at the Fermi energy $\mu$ for valley $m$, and $\theta_{\bm{k}\bm{k}'}$ is the angle between the Bloch state group velocities in valley $m$
at $\bm{k}$ and $\bm{k}'$ (i.e., $\cos\theta_{\bm{k}\bm{k}'}=\bm{k}\cdot\bm{k}'/|\bm{k}||\bm{k}'|$).
In multi-valley systems with weak intervalley scattering it is clear 
that deviations from equilibrium in the total number of electrons in a valley will relax especially slowly.
As we have explained above, these are not driven directly by an electric field, but we will show later that 
they can be driven by the combined action of electric and magnetic fields.

In order to precisely define 
intervalley relaxation times explicitly it is necessary to separate the Bloch state Hilbert space 
into valleys in some precise way.  The best way to do this depends on the system being studied.
For the sake of definiteness, below we use band labels $m$ to distinguish valleys.
When $K^{dd}$ is expressed in the representation in which it is diagonal in the absence 
of intervalley scattering, and intervalley scattering is parametrically weaker than intravalley scattering,
its smallest eigenvalues (longest relaxation times) are associated with an $M \times M$ block 
with entries 
\begin{equation} 
K^{dd}_{mm'} = \delta_{mm'}  \sum_{m''} \frac{1}{\tau^{mm''}_{\mathrm{inter}}} - \frac{1}{\tau^{mm'}_{\mathrm{inter}}}
\end{equation} 
where $(m,m') = 1, \ldots, M$ and 
\begin{align} 
\frac{1}{\tau^{mm'}_{\mathrm{inter}}}=\frac{2\pi}{\hbar D^m(\mu)}\sum_{\bm{k},\bm{k}'}\, \langle |U^{mm'}_{\bm{k}\bm{k}'}|^2\rangle 
 \delta(\mu - \varepsilon^{m}_{\bm{k}}) \delta(\mu - \varepsilon^{m'}_{\bm{k}'}).
\end{align}
Here, $M$ is the number of Fermi surface pockets, and for simplicity we have assumed that all  
pockets have the same Fermi-level density of states because they are related by symmetry.  
Note that this {\it total valley population} block of the scattering kernel still has one 
zero eigenvalue, corresponding to total population summed over all valleys.  
In the $M=2$ case the nonzero eigenvalue has value 
$2/\tau^{1,2}_{\mathrm{inter}}$, which is the time scale for relaxing differences in population 
between the two valleys.  

We are now in a position to explain when the magnetoresponse of semimetals with 
atomically smooth disorder is anomalous, and when it is not, by making the following observations:
i) When the Fermi surface of a conductor consists of a few small valleys that 
are well separated in momentum space and disorder is smooth, scattering between valleys 
is parametrically weaker because it requires large momentum transfers;
ii) When intervalley scattering is negligible, the diagonal-component response 
function $(K^{dd})^{-1}$ is divergent not only in the total particle number channel, 
but also in the valley-projected particle number channel;  iii) because  
the electric driving term vanishes separately for each valley, very weak intervalley
scattering does not lead to anomalous response at $\bm{B}=0$; but iv) in Weyl and Dirac semimetals 
$D_B \mathcal{L}^{-1} D_E(\rho_0) $ does drive the valley-projected total particle number.
The total particle number in a given valley is {\it not} conserved when
the magnetic driving term summed over a given valley is {\it not} zero when it acts on the 
electric-field disturbed density matrix:
\begin{align}
\frac{\partial N}{\partial t}=\sum_{m,\bm{k}}\left[D_B(\langle\rho_E\rangle)\right]^{mm}_{\bm{k}}\neq 0.
\label{D_B-nonzero}
\end{align}
Here, $\langle\rho_E\rangle=\mathcal{L}^{-1} D_{E}(\langle \rho_0\rangle)$ represents a density matrix linear in electric field.
As a consequence of the total particle number nonconservation in a given valley, the intervalley 
scattering time $\tau_{\mathrm{inter}}$ appears as an eigenvalue of $(K^{dd})^{-1}$.

So far we have focused on systems where valleys are not degenerate, i.e., are separated in momentum space, such as Weyl semimetals.
Similar considerations apply to 
systems where valleys are degenerate, such as Dirac semimetals.
3D Dirac semimetals such as Cd$_3$As$_2$ and Na$_3$Bi have two Dirac points which are protected by crystalline symmetry.
The effective Hamiltonian around a Dirac point in such Dirac semimetals can be written in the block-diagonal form \cite{Yang2014}
\begin{align}
\mathcal{H}_{\mathrm{Dirac}}(\bm{q})=
\begin{bmatrix}
\mathcal{H}_{\mathrm{AA}} & \mathcal{H}_{\mathrm{AB}}\\
\mathcal{H}_{\mathrm{BA}} & \mathcal{H}_{\mathrm{BB}}
\end{bmatrix}
=
\begin{bmatrix}
\mathcal{H}_+(\bm{q}) & 0\\
0 & \mathcal{H}_-(\bm{q})
\end{bmatrix},
\label{Dirac-blockdiagonal}
\end{align}
where $A$ and $B$ denote two states after a unitary transformation, and $\mathcal{H}_\pm(\bm{q})$ is a $2\times2$ Weyl Hamiltonian with chirality $\pm1$.
As is seen from Eq.~(\ref{Dirac-blockdiagonal}), two Weyl Hamiltonians that belong to different states $A$ and $B$ are degenerate around the Dirac point.
In other words, two different valleys are degenerate.
In general, atomically smooth disorder does not allow scattering processes between two different states $A$ and $B$.
Namely, intervalley scattering is much weaker than intravalley scattering, i.e., we have $\tau_{\mathrm{inter}}\gg\tau_{\mathrm{intra}}$.
Thus we can apply the same argument as in the case of nondegenerate valleys to the case of degenerate valleys: 
the total particle number in a given valley is not conserved if the magnetic driving term summed over a given valley is not zero [see Eq.~(\ref{D_B-nonzero})].

We will demonstrate in Sec.~\ref{Sec-NMR} that the particle number nonconservation in a given valley characterized by Eq.~(\ref{D_B-nonzero}) occurs in the process of calculating the magnetoconductivity proportional to $B^2$ in Dirac and Weyl semimetals, which is identified as a consequence of the chiral anomaly.

\section{Chiral Anomaly in Three-Dimensional Semimetals \label{Sec-Chiral-Anomaly}}

In Weyl semimetals the term ``chiral anomaly'' is sometimes used to refer to the property that the 
total number of electrons in a given valley is not conserved in the presence of simultaneous 
electric and magnetic fields, and sometimes to the enhanced magnetoconductivity that this lack 
of separate particle number conservation produces.  In this section we focus on pumping of charge 
between valleys without referring to a specific microscopic model, which leads to observable effects only when 
intervalley scattering times are much 
longer than intravalley scattering times.  For long intervalley scattering times, anomalous pumping leads 
to a difference between the effective chemical potentials of different valleys and thus to a 
enhanced current via the magnetoelectric effect, as explained in great deal for a Weyl metal toy model 
in the following sections.
We set $\hbar= 1$ in the rest of this paper.

We study a general model with the Hamiltonian $\mathcal{H}_0=\sum_{m,\bm{k}}\varepsilon_{\bm{k}}^m |m,\bm{k}\rangle\langle m,\bm{k}|$ and the equilibrium density matrix $\langle\rho_0\rangle=\sum_{m,\bm{k}}f_0(\varepsilon_{\bm{k}}^m) |m,\bm{k}\rangle\langle m,\bm{k}|$, where $\varepsilon_{\bm{k}}^m$ is an energy eigenvalue and $f_0(\varepsilon_{\bm{k}}^m)$ is the Fermi-Dirac distribution function.
For concreteness, we choose $\bm{E}=(0,0,E_z)$ and $\bm{B}=(0,0,B_z)$.
Let us consider the following quantity that is linear in both electric and magnetic fields,
\begin{align}
\frac{\partial N}{\partial t}=\mathrm{Tr}[\mathcal{L}\langle\rho\rangle]\equiv \mathrm{Tr}\left[D_{B}\mathcal{L}^{-1} D_{E}(\langle \rho_0\rangle) + D_{E}\mathcal{L}^{-1} D_{B}(\langle \rho_0\rangle)\right]
\label{Rate-of-pumping}
\end{align}
evaluated for a particular Fermi surface pocket associated with a particular valley.  
$\partial N/\partial t$ is the rate of change of the density matrix that is balanced by 
the scattering and free evolution contributions to $\mathcal{L}\langle\rho\rangle$ 
to yield the part of the density matrix that is linear in both electric and magnetic fields;
the steady state density matrix is therefore obtained by acting on $\partial N/\partial t$ with 
$\mathcal{L}^{-1}$.

Let us consider the first term in the right-hand side of Eq.~(\ref{Rate-of-pumping}), i.e., $\mathrm{Tr}[D_{B}\mathcal{L}^{-1} D_{E}(\langle \rho_0\rangle)]$.
As we have explained earlier, $\langle\rho_E\rangle=\mathcal{L}^{-1} D_{E}(\langle \rho_0\rangle)$,
the linear-response density matrix in the absence of a magnetic field, 
contains both band-diagonal $\langle n_E\rangle$ and band off-diagonal 
$\langle S_E\rangle$ contributions: 
$\langle\rho_E\rangle=\langle n_E\rangle+\langle S_E\rangle$.
It follows that
$\mathrm{Tr}\, [D_B \mathcal{L}^{-1} D_{E}(\langle \rho_0\rangle)]=\mathrm{Tr}\, [D_B(\langle n_E\rangle)]+\mathrm{Tr}\, [D_B(\langle S_E\rangle)]$, since $D_B$ is linear in the density matrix.
First we evaluate the diagonal element $[D_B(\langle n_E\rangle)]^{mm}_{\bm{k}}$ with $\langle n_E\rangle=eE_z\sum_{m,\bm{k}}[\partial_{k_z} f_0(\varepsilon_{\bm{k}}^m)] |m,\bm{k}\rangle\langle m,\bm{k}|$.
In this case we can use the result of $[D_B(\langle \rho_0\rangle)]^{mm}_{\bm{k}}$ by replacing $f_0$ by $eE_z\partial_{k_z} f_0$.
Namely, from Eq.~(\ref{diagonal-element-D_B-rho_0}) we get
\begin{align}
[D_B(\langle n_E\rangle)]^{mm}_{\bm{k}}=e^2E_zB_z\left(\frac{\partial \varepsilon_{\bm{k}}^m}{\partial k_y}\frac{\partial}{\partial k_x}-\frac{\partial \varepsilon_{\bm{k}}^m}{\partial k_x}\frac{\partial}{\partial k_y}\right)\frac{\partial f_0(\varepsilon_{\bm{k}}^m)}{\partial k_z}.
\end{align}
It is obvious that this is an odd function of $k_x$, $k_y$, and $k_z$, which means that $\mathrm{Tr}\, [D_B(\langle n_E\rangle)]=0$.
Then we see that one of the two contributions to the nonconservation of the total electron number in a given valley induced by the chiral anomaly is associated with the action of the $D_B$ operator on the off-diagonal 
density matrix $\langle  S_E\rangle$ as
\begin{align}
\mathrm{Tr}\left[D_{B}\mathcal{L}^{-1} D_{E}(\langle \rho_0\rangle)\right]
=\int_{\mathrm{FS}} \frac{d^3k}{(2\pi)^3}\, \sum_m\left[D_B(\langle  S_E\rangle)\right]^{mm}_{\bm{k}},
\label{dN/dt-1}
\end{align}
where FS represents the integration on the Fermi surface of the valley.
After a calculation we obtain the following general expression for the Fermi surface contribution
which can be expressed in terms of band velocities and Berry curvatures alone (see Appendix~\ref{dN/dt-analytical} for a detailed derivation):
\begin{align}
\sum_m \left[D_B(\langle  S_E\rangle)\right]^{mm}_{\bm{k}}=e^2E_zB_z\sum_m\left[\frac{\partial f_0(\varepsilon^m_{\bm{k}})}{\partial k_x}\Omega_{\bm{k},x}^m+\frac{\partial f_0(\varepsilon^m_{\bm{k}})}{\partial k_y}\Omega_{\bm{k},y}^m\right],
\end{align}
where $\Omega_{\bm{k},a}^m=\epsilon^{abc}\, i\langle \partial_{k_b} u_{\bm{k}}^m|\partial_{k_c} u_{\bm{k}}^m\rangle$ is the Berry curvature of band $m$.
Thus the right-hand side of Eq.~(\ref{dN/dt-1}) is written as
\begin{equation} 
\frac{e^2E_zB_z}{4\pi^2}\int \frac{d^3 k}{2 \pi}\frac{\partial f_{0}(\varepsilon_{\bm{k}}^m)}{\partial \varepsilon^m_{\bm{k}}}\big[ v^m_{\bm{k},x} \Omega^m_{\bm{k},x} + v^m_{\bm{k},y} \Omega^m_{\bm{k},y} \big],
\label{dN/dt-2}
\end{equation} 
where $\bm{v}^m_{\bm{k}}$ is the Bloch state group velocity, and we have assumed that only the band $m$ intersects the Fermi surface, i.e., $\partial f_0(\varepsilon^n_{\bm{k}})/\partial \varepsilon^n_{\bm{k}}=\delta_{mn}\partial f_0(\varepsilon^m_{\bm{k}})/\partial \varepsilon^m_{\bm{k}}$ , which can in general apply to multi-valley systems.

Next let us consider the second term in the right-hand side of Eq.~(\ref{Rate-of-pumping}), i.e., $\mathrm{Tr}[D_{E}\mathcal{L}^{-1} D_{B}(\langle \rho_0\rangle)]$.
As we have explained earlier, $\langle\rho_B\rangle=\mathcal{L}^{-1} D_{B}(\langle \rho_0\rangle)$,
the linear-response density matrix in the absence of an electric field, 
contains both band-diagonal $\langle \xi_B\rangle$ and band off-diagonal 
$\langle S_B\rangle$ contributions: 
$\langle\rho_B\rangle=\langle \xi_B\rangle+\langle S_B\rangle$.
In this case the calculation is easier than that of $\mathrm{Tr}[D_{B}(\langle\rho_E\rangle)]$.
After a calculation we find that only the contribution from the Berry phase correction $\langle \xi_B\rangle$ survives:
\begin{align}
\mathrm{Tr}\left[D_{E}\mathcal{L}^{-1} D_{B}(\langle \rho_0\rangle)\right]
&=\int_{\mathrm{FS}} \frac{d^3k}{(2\pi)^3}\, \sum_m\left[D_E(\langle\xi_B\rangle)\right]^{mm}_{\bm{k}} \nonumber\\
&=\frac{e^2E_zB_z}{4\pi^2}\int \frac{d^3 k}{2 \pi}\frac{\partial f_{0}(\varepsilon_{\bm{k}}^m)}{\partial \varepsilon^m_{\bm{k}}}v^m_{\bm{k},z} \Omega^m_{\bm{k},z}.
\label{dN/dt-3}
\end{align}

Combining Eqs.~(\ref{dN/dt-2}) and (\ref{dN/dt-3}) we arrive at the final expression for the rate of pumping of electrons between valleys due to the chiral anomaly
\begin{align}
\left.\frac{\partial N}{\partial t}\right|_{\mathrm{CA}}
&=\mathrm{Tr}\left[D_{B}(\langle S_E\rangle)\right]+\mathrm{Tr}\left[D_{E}(\langle \xi_B\rangle)\right] \nonumber\\
&=\frac{e^2E_zB_z}{4\pi^2}\int \frac{d^3 k}{2 \pi}\frac{\partial f_{0}(\varepsilon_{\bm{k}}^m)}{\partial \varepsilon^m_{\bm{k}}}\bm{v}^m_{\bm{k}} \cdot \bm{\Omega}^m_{\bm{k}},
\label{dN/dt-total}
\end{align}
which is in complete agreement with the expression obtained by disorder-free semiclassical dynamics [Eq.~(\ref{CA-conventional})] \cite{Son2012,Son2013}.
When disorder-free semiclassical wavepacket dynamics is used to derive Eq.~(\ref{CA-conventional}),
the origin of the term on the right hand side proportional to $\Omega_z$ is quite different from the
terms proportional to $\Omega_x$ and $\Omega_y$.  The later are due to the well known 
anomalous velocity of electrons in a magnetic field, which contributes to the Lorentz force and hence to
the rate of motion of states through momentum space.
The anomalous velocity effect is an interband wave-function polarization effect,
akin to the dielectric screening response of insulators.
The $\Omega_z$ contribution, on the other hand, is due to the correction of the density of states 
in momentum space in the presence of a magnetic field.
The density-of-states correction is the semiclassical manifestation of the relationship between 
Berry curvatures (Berry phases) and Landau quantization.
We have shown in Eq.~(\ref{dN/dt-total}) that our quantum kinetic theory is able to completely account for both effects.

Here, let us consider the case of Weyl semimetals.
As depicted in Fig.~\ref{Fig1}, a magnetic field induces an anomalous ($N = 0$) Landau level branch that has only one sign of velocity $v_z$ in a given valley.
It follows that in each valley the density of states is increased for states with one sign of velocity and decreased for states with the other sign of velocity, and that the total current summed over a valley is already nonzero in equilibrium.
When an electric field drives states through momentum space, the total number of states in a valley varies.
Scattering within valleys can relax the current in each valley to its equilibrium value, but cannot establish a steady state because the number of states in each valley still changes at a constant rate.
A steady state can be established only when scattering processes between valleys are present.
When Bloch-state scattering between valleys is much weaker than scattering within valleys the anomalous $\bm{E}\cdot\bm{B}$ pumping leads to differences between the chemical potentials of different valleys.

\section{Chiral Anomaly in Weyl Semimetals \label{Sec-Weyl-Model}}
So far we have i) formulated a general theory of low-field  that accounts for 
the wavevector-dependence of Bloch states in multiband systems, ii) shown that the Berry-phase 
related density of states and anomalous velocities of semiclassical wavepacket dynamics are 
fully captured by our theory, and iii) demonstrated that Bloch state scattering, which is 
consistently described in our transport theory, can alter theoretical predictions made by introducing 
Berry phase effects in semiclassical wavepacket dynamics into the transport theory in an {\it ad hoc} manner.    
In this section, we present an analysis of the magnetotransport properties 
of a specific toy model of a Weyl semimetal which illustrates in more detail 
the interplay between intrinsic dynamics, scattering, and external field driving terms 
in the kinetic equation.

\subsection{Theoretical model}
We start with the $4\times4$ continuum toy model
Hamiltonian for two-node Weyl semimetals with broken time-reversal symmetry \cite{Burkov2011,Vazifeh2013,Sekine2014,Burkov2015}
\begin{align}
\mathcal{H}_0(\bm{k})=v_F\tau_z\bm{k}\cdot\bm{\sigma}+\Delta\tau_x+b\sigma_z,
\label{WSM-H_eff}
\end{align}
where the Pauli matrices $\tau_i$ and $\sigma_i$ are Weyl-node and spin degrees of freedom, respectively, $v_F$ is the Dirac velocity, and $\Delta$ is the mass of 3D Dirac fermions.
The term $b\sigma_z$ can be regarded as a magnetic interaction of the $z$ component of spin such as $s$-$d$ coupling and Zeeman coupling.
Therefore, it can be said that the above Hamiltonian represents a 3D (topological or normal) insulator doped with magnetic impurities.
As will be shown just below, a Weyl semimetal is realized when $|b/\Delta|>1$.
The time-reversal $\mathcal{T}$ and spatial inversion (parity) $\mathcal{P}$ operators are given by $\mathcal{T}=-i\sigma_y\mathcal{C}$ with $\mathcal{C}$ the complex conjugation operator and $\mathcal{P}=\tau_x$.
Here, we have the following relations: $\mathcal{T}^{-1}=i\sigma_y\mathcal{C}$, $\mathcal{P}^{-1}=\tau_x$, $\mathcal{T}^2=-1$, and $\mathcal{P}^2=1$.
Then it is easily seen that time-reversal symmetry of the system is broken such that $\mathcal{T}\mathcal{H}_0(\bm{k})\mathcal{T}^{-1}\neq \mathcal{H}_0(-\bm{k})$, but inversion symmetry of the system is preserved such that $\mathcal{P}\mathcal{H}_0(\bm{k})\mathcal{P}^{-1}=\mathcal{H}_0(-\bm{k})$.

Performing a canonical transformation such that $\sigma_{x,y}\rightarrow\tau_z\sigma_{x,y}$ and $\tau_{x,y}\rightarrow\sigma_z\tau_{x,y}$, Eq.~(\ref{WSM-H_eff}) can be rewritten as $\tilde{\mathcal{H}}_0(\bm{k})=v_F(k_x\sigma_x+k_y\sigma_y)+(b+v_F\tau_zk_z+\Delta\tau_x)\sigma_z$.
In the representation of eigenstates of $b+v_F\tau_zk_z+\Delta\tau_x$, Eq.~(\ref{WSM-H_eff}) 
is block-diagonal with two $2\times2$ Hamiltonians given by \cite{Burkov2015}
\begin{align}
\mathcal{H}_{\pm}(\bm{k})=v_F(k_x\sigma_x+k_y\sigma_y)+m_\pm(k_z)\sigma_z
\label{H_pm}
\end{align}
with $m_\pm(k_z)=b\pm\sqrt{v_F^2 k_z^2+\Delta^2}$.
The two Weyl nodes are located at $W_\pm=(0,0,\pm k_0)$ 
with $k_0=\sqrt{b^2-\Delta^2}/v_F$.  Near the Weyl nodes,
we have $m_-(q_z)\approx \mp v_F^2(k_0/b)q_z$ with momentum $\bm{q}=\bm{k}-W_\pm$ ($\bm{q}^2\ll 1$).
The eigenvectors of $\mathcal{H}_t(\bm{k})$ $(t=\pm)$ with 
eigenvalues $\varepsilon^\pm_{t\bm{k}}=\pm\varepsilon_{t\bm{k}}=\pm\sqrt{v_F^2(k_x^2+k_y^2)+m_t^2}$ are 
given by
\begin{align}
|u_{t\bm{k}}^\pm\rangle=\frac{1}{\sqrt{2}}
\begin{bmatrix}
\sqrt{1\pm\frac{m_t(k_z)}{\varepsilon_{t\bm{k}}}}\\
\pm e^{i\theta}\sqrt{1\mp\frac{m_t(k_z)}{\varepsilon_{t\bm{k}}}}
\end{bmatrix},
\label{WSM-eigenstates}
\end{align}
where $e^{i\theta}=(k_x+ik_y)/k_\perp$ with $k_\perp=\sqrt{k_x^2+k_y^2}$.
We see that $\mathcal{H}_-(\bm{k})$ describes a subsystem with  
two Weyl nodes, while $\mathcal{H}_+(\bm{k})$ has a fully gapped spectrum.
For convenience, we omit the subscripts $t=\pm$ at 
intermediate steps of the calculations below and restore them in the final results.

The generalized Berry connection in the eigenstate representation is given by $[\mathcal{R}_{\bm{k},\alpha}]^{mn}=i\langle u_{\bm{k}}^m|\partial_{k_\alpha}u_{\bm{k}}^n\rangle$ with $\alpha=x,y,z$ and $m,n=\pm$.
The individual components are given explicitly by
\begin{align}
\mathcal{R}_{\bm{k},x}
=&\frac{1}{2 k_\perp}\sin\theta-\tilde{\sigma}_z\frac{1}{2 k_\perp}\frac{m}{\varepsilon_{\bm{k}}}\sin\theta -\tilde{\sigma}_y\frac{v_F m}{2\varepsilon_{\bm{k}}^2}\cos\theta\nonumber\\
& -\tilde{\sigma}_x\frac{v_F}{2\varepsilon_{\bm{k}}}\sin\theta, \nonumber\\
\mathcal{R}_{\bm{k},y}
=&-\frac{1}{2 k_\perp}\cos\theta+\tilde{\sigma}_z\frac{1}{2 k_\perp}\frac{m}{\varepsilon_{\bm{k}}}\cos\theta -\tilde{\sigma}_y\frac{v_F m}{2\varepsilon_{\bm{k}}^2}\sin\theta \nonumber\\
& +\tilde{\sigma}_x\frac{v_F}{2\varepsilon_{\bm{k}}}\cos\theta, \nonumber\\
\mathcal{R}_{\bm{k},z}
=&\tilde{\sigma}_y\frac{v_F k_\perp}{2\varepsilon_{\bm{k}}^2}\frac{\partial m}{\partial k_z},
\label{Berry-connection}
\end{align}
where $\tilde{\sigma}_\alpha$ are the Pauli matrices in the eigenstate basis.
Also, the individual components of the Berry curvature, $\Omega^\pm_{\bm{k},a}=\epsilon^{abc}\, i\langle \partial_{k_b} u_{\bm{k}}^\pm|\partial_{k_c} u_{\bm{k}}^\pm\rangle$, are given by
\begin{align}
&\Omega^\pm_{\bm{k},x}=\mp\frac{\partial m}{\partial k_z}\frac{v_F^2 k_x}{2\varepsilon^2_{\bm{k}}},\ \ \Omega^\pm_{\bm{k},y}=\mp\frac{\partial m}{\partial k_z}\frac{v_F^2 k_y}{2\varepsilon^3_{\bm{k}}},\ \ \Omega^\pm_{\bm{k},z}=\mp\frac{v_F^2 m}{2\varepsilon^3_{\bm{k}}}.
\label{Berry-curvature}
\end{align}

\begin{figure}[!t]
\centering
\includegraphics[width=0.8\columnwidth]{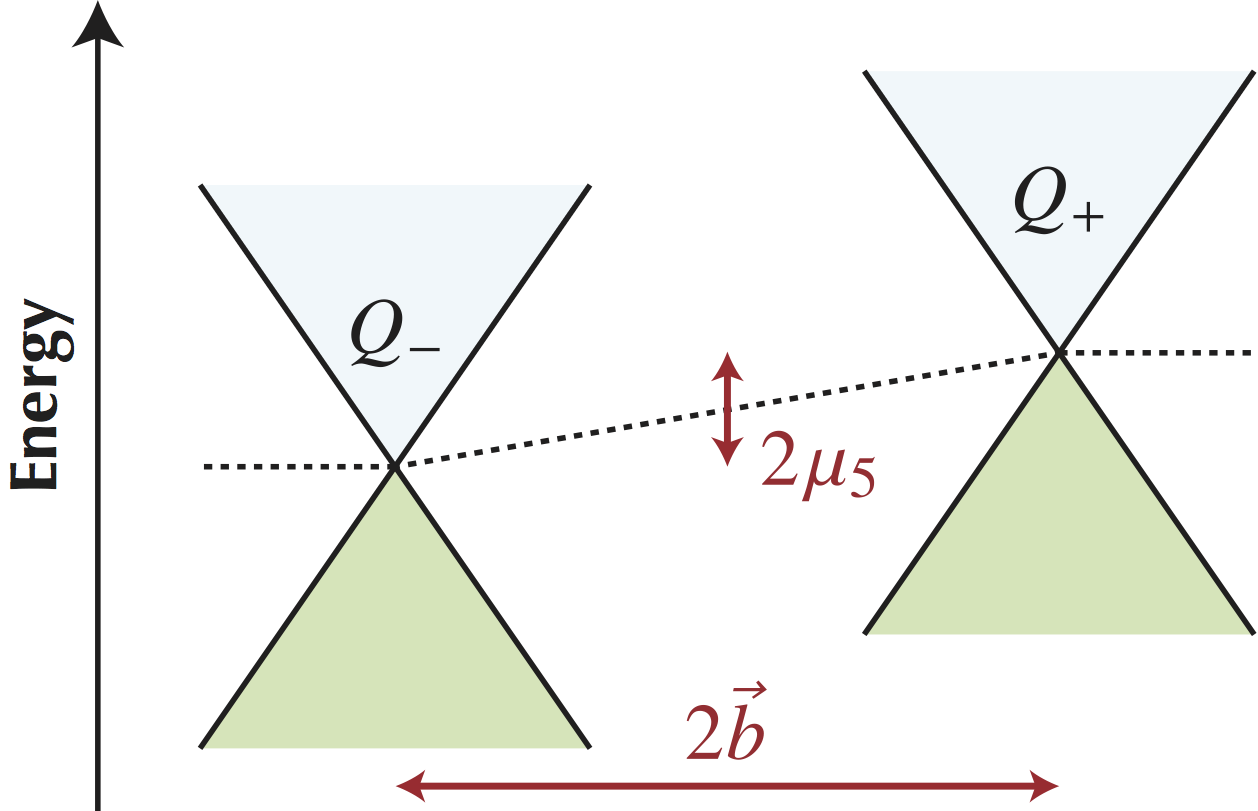}
\caption{Schematic illustration of a Weyl semimetal with 
two Weyl points with chiralities $Q_\pm=\pm1$.
$2\bm{b}$ and $2\mu_5$ are the momentum-space distance
 and the chemical potential difference between the Weyl nodes, respectively.
The dashed line indicates the Fermi level of the system, which can have different 
values in different valleys when the system is driven from equilibrium, for example, by 
the combined effect of electric and magnetic fields.
}\label{Fig2}
\end{figure}
Here, we briefly review the derivation of the $\theta$ term from the microscopic four-band model (\ref{WSM-H_eff}).
In order to describe a more generic Weyl semimetal, we add the term $\mu_5\tau_z$ to the Hamiltonian, which generates a chemical potential difference $2\mu_5$ between the two Weyl nodes, and replace $b\sigma_z$ by $\bm{b}\cdot\bm{\sigma}$.
Note that this term breaks inversion symmetry.
Figure \ref{Fig2} shows a schematic diagram of a two-node Weyl semimetal.
We set $\Delta=0$ for simplicity.
The action of the system in the presence of an external electromagnetic potential $A_\mu=(A_0,-\bm{A})$ is given by
\begin{align}
S&=\int dt\, d^3r\, \psi^\dag\left\{i(\partial_t-ieA_0)-[\mathcal{H}_0(\bm{k}+e\bm{A})-\mu_5\tau_z]\right\}\psi \nonumber\\
&=\int dt\, d^3r\, \bar{\psi}i\gamma^\mu(\partial_\mu-ieA_\mu-ib_\mu\gamma^5)\psi,
\end{align}
where $e>0$, $\psi$ is a four-component spinor, $\bar{\gamma}=\psi^\dag\gamma^0$, $\gamma^0=\tau_x$, $\gamma^j=\tau_x\tau_z\sigma_j=-i\tau_y\sigma_j$, $\gamma^5=i\gamma^0\gamma^1\gamma^2\gamma^3=\tau_z$, and $b_\mu=(\mu_5, -\bm{b})$.
After applying the Fujikawa method \cite{Fujikawa}, i.e., applying an infinitesimal gauge transformation such that $\psi\rightarrow e^{-id\phi\theta(\bm{r},t)\gamma^5/2}\psi$ with $\theta(\bm{r},t)=-2x^\mu b_\mu=2(\bm{b}\cdot\bm{r}-\mu_5 t)$ and $\phi\in [0,1]$ for infinite times, the action of the system becomes \cite{Zyuzin2012,Burkov2015}
\begin{align}
S=\int dt\, d^3r\, \bar{\psi}[i\gamma^\mu(\partial_\mu-ieA_\mu)]\psi+S_\theta,
\label{S_effective-WSM}
\end{align}
where $S_\theta$ is the $\theta$ term [Eq.~(\ref{S_theta_realtime})] with $\theta(\bm{r},t)=2(\bm{b}\cdot\bm{r}-\mu_5 t)$.
Note that the first term in Eq.~(\ref{S_effective-WSM}) represents the (trivial) action of massless Dirac fermions.

In general, regardless of whether the system is gapless or gapped, the four-current density $j^\nu$ can be obtained from the variation of the $\theta$ term with respect to the four-potential $A_\nu$ as
$j^\nu=\delta S_\theta/\delta A_\nu=-(e^2/4\pi^2)\epsilon^{\mu \nu\rho\lambda}[\partial_\mu \theta(\bm{r},t)]\partial_\rho A_\lambda$ \cite{Wilczek1987}.
The induced current density in the presence of external electric and magnetic fields is given by
\begin{align}
\bm{j}(\bm{r},t)=\frac{e^2}{4\pi^2}\left[\nabla\theta(\bm{r},t)\times\bm{E}+\dot{\theta}(\bm{r},t)\bm{B}\right],
\label{Current-Eq}
\end{align}
where $\dot{\theta}=\partial \theta(\bm{r},t)/\partial t$, $\bm{E}=-\nabla A_0-\partial \bm{A}/\partial t$, and $\bm{B}=\nabla\times\bm{A}$.
In the present case of Weyl semimetals with $\theta(\bm{r},t)=2(\bm{b}\cdot\bm{r}-\mu_5 t)$, we obtain in the ground state a static current of the form
\begin{align}
\bm{j}=\frac{e^2}{2\pi^2}\left(\bm{b}\times\bm{E}-\mu_5\bm{B}\right),
\label{Current-Eq-WSM}
\end{align}
where the electric-field induced and magnetic-field induced terms are the anomalous Hall effect and chiral magnetic effect, respectively \cite{Zyuzin2012,Son2012,Grushin2012,Wang2013a,Goswami2013,Burkov2015,Fukushima2008,Vazifeh2013,Sekine2016}.

The anomalous Hall effect in Weyl semimetals can be understood straightforwardly.
It is well known that the Hall conductivity of 2D massive Dirac fermions (such as those on the surface of 3D topological insulators) of the form (\ref{H_pm}) is given by $\sigma_{xy}^\pm(k_z)=\mathrm{sgn}[m_\pm(k_z)]e^2/2h$, when the chemical potential lies in the energy gaps.
This (half-)quantized value holds valid even in the presence of disorder \cite{Nomura2011,Culcer2011a}.
We see that the 3D Hall conductivity is nonzero only in the region $-k_0\le k_z\le k_0$, which gives $\sigma^{\mathrm{3D}}_{xy}=\int_{-k_0}^{k_0}\, (k_z/2\pi) \sigma_{xy}^\pm(k_z)=k_0 e^2/\pi h$.
This value is exactly the same as the first term in Eq.~(\ref{Current-Eq-WSM}).
On the other hand, the chiral magnetic effect in Weyl semimetals looks like a peculiar phenomenon.
The chiral magnetic effect indicates a direct current generation along a static magnetic field (without electric fields) in the presence of a chemical potential difference between the two nodes, as is seen from the second term in Eq.~(\ref{Current-Eq-WSM}).
If the static chiral magnetic effect exists in real materials, there will be substantial possible applications.
Discussions on the existence of the static chiral magnetic effect in Weyl semimetals have continued theoretically \cite{Vazifeh2013,Chang2015,Buividovich2015,Yamamoto2015,Ma2015,Zhong2016,Zubkov2016}.
However, the existence of the static chiral magnetic effect would be ruled out in crystalline solids (i.e., lattice systems) as discussed in Ref.~\cite{Vazifeh2013}, which is consistent with our understanding that static magnetic fields do not generate equilibrium currents.

Lastly, we refer to these two phenomena from the symmetry viewpoint.
The term $\bm{b}\cdot\bm{\sigma}$ breaks time-reversal symmetry but preserves inversion symmetry: $\mathcal{T}\sigma_i\mathcal{T}^{-1}=-\sigma_i$ and $\mathcal{P}\sigma_i\mathcal{P}^{-1}=\sigma_i$ ($i=x,y,z$).
This indicates that the occurrence of the anomalous Hall effect in Weyl semimetals requires the breaking of time-reversal symmetry, which is consistent with the generic requirement for the realization of the anomalous Hall effect.
On the other hand, the term $\mu_5\tau_z$ breaks inversion symmetry but preserves time-reversal symmetry: $\mathcal{T}\tau_z\mathcal{T}^{-1}=\tau_z$ and $\mathcal{P}\tau_z\mathcal{P}^{-1}=-\tau_z$.
This indicates that the occurrence of the chiral magnetic effect in Weyl semimetals requires the breaking of inversion symmetry.
Note that both the anomalous and chiral magnetic effects are possible (at least theoretically) in systems with broken time-reversal and inversion symmetries.

\subsection{Anomalous Hall effect in Weyl metals \label{Sec-AHE}}
Let us see that the electric driving term (\ref{driving-term-E}) properly describes the anomalous Hall effect in Weyl metals.
As described in Sec.~\ref{Sec-Magnetotransport}, the diagonal part of the electric driving term $D_E$ results in usual Drude conductivity.
Hence, we need the explicit matrix expression for the off-diagonal part of the electron density $\langle S_E\rangle$ to obtain the anomalous Hall conductivity.
As is seen from Eq.~(\ref{rho_E-off-diagonal}), there are contributions from the electric driving term $D_E(\langle \rho_0\rangle)$ and the anomalous driving term $J(\langle n_E\rangle)$ to the off-diagonal part of the density matrix $\langle S_E\rangle$.
Let us denote the resultant total anomalous Hall conductivity of the system as
\begin{align}
\sigma_{xy}=\sigma^{\rm I}_{xy}+\sigma^{\rm II}_{xy},
\end{align}
where $\sigma^{\rm I}_{xy}$ is the Hall conductivity from $D_E(\langle \rho_0\rangle)$ and $\sigma^{\rm II}_{xy}$ is that from $J(\langle n_E\rangle)$.
In the following, we consider the case where an electric field is applied along the $y$ direction as $\bm{E}=E_y\bm{e}_y$, and set $\mu_5=0$.
Also, we add a positive small chemical potential $\mu$ which lies sufficiently close to the Weyl nodes.
Note that the sign of $\mu$ is not essential, since the Hamiltonian (\ref{H_pm}) has particle-hole symmetry.

First, we evaluate the intrinsic contribution $\sigma^{\rm I}_{xy}$, which originates from the Fermi sea of the electronic band.
Using the expression for $D_E(\langle\rho_0\rangle)$ (\ref{D_E-intrinsic}) and the Berry connection (\ref{Berry-connection}), the off-diagonal part of the electron density matrix $\langle S_E\rangle$ is given by
\begin{align}
\langle S_E\rangle&=eE_y\frac{f_0(\varepsilon^+_{\bm{k}})-f_0(\varepsilon^-_{\bm{k}})}{2\varepsilon_{\bm{k}}}
\begin{bmatrix}
0 && \mathcal{R}_{\bm{k},y}^{+-}\\
\mathcal{R}_{\bm{k},y}^{-+} && 0
\end{bmatrix} \nonumber\\
&=eE_y\frac{f_0(\varepsilon^+_{\bm{k}})-f_0(\varepsilon^-_{\bm{k}})}{4\varepsilon_{\bm{k}}^2}v_F\left(\tilde{\sigma}_x\cos\theta-\tilde{\sigma}_y\frac{m}{\varepsilon_{\bm{k}}}\sin\theta\right),
\end{align}
where $f_0(\varepsilon^\pm_{\bm{k}})=1/[e^{(\varepsilon^\pm_{\bm{k}}-\mu)/T}+1]$ is the Fermi-Dirac distribution function.
The velocity operator is given by $v_x=\partial \mathcal{H}_t/\partial k_x=v_F\sigma_x$.
We work in the eigenstate basis, in which the matrix element $[v_x]^{mn}=v_F\langle u_{\bm{k}}^m|\sigma_x |u_{\bm{k}}^n\rangle$ ($m,n=\pm$) is given by
\begin{align}
v_x=v_F\left(\tilde{\sigma}_z\frac{v_F k_\perp}{\varepsilon_{\bm{k}}}\cos\theta+\tilde{\sigma}_y\sin\theta-\tilde{\sigma}_x\frac{m}{\varepsilon_{\bm{k}}}\cos\theta\right).
\label{v_x}
\end{align}
The anomalous Hall conductivity $\sigma^{\rm I}_{xy}$ is calculated from the definition $\sigma^{\rm I}_{xy}=\mathrm{Tr}\left[(-e)v_x\langle S_E\rangle\right]/E_y$.
Combining the above ingredients, we obtain
\begin{align}
\sigma^{\rm I}_{xy}&=2e^2\sum_{t=\pm}\int_{-\infty}^\infty\frac{d^3k}{(2\pi)^3}\, \frac{v_F^2 m_t}{4\varepsilon_{t\bm{k}}^3}\left[f_0(\varepsilon^+_{t\bm{k}})-f_0(\varepsilon^-_{t\bm{k}})\right] \nonumber\\
&=\frac{e^2}{2}\int_{-\infty}^\infty\frac{d^3k}{(2\pi)^3}\left[\frac{v_F^2 m_-}{\varepsilon_{-\bm{k}}^3}f_0(\varepsilon^+_{-\bm{k}})-\sum_{t=\pm}\frac{v_F^2 m_t}{\varepsilon_{t\bm{k}}^3}\right] \nonumber\\
&\approx\frac{e^2}{2}\int_{-\delta}^\delta\frac{d^3q}{(2\pi)^3}\, \frac{\frac{k_0}{b}q_z-\frac{k_0}{b}q_z}{\varepsilon_{-\bm{q}}^3}-\frac{e^2}{4\pi^2}\sum_{t=\pm}\int_0^\infty dk_z\, \frac{m_t}{|m_t|} \nonumber\\
&=-\frac{e^2}{2\pi^2}k_0,
\label{AHE-Intrinsic}
\end{align}
where we have used the fact that $m_-(k_z)\approx\mp v_F^2(k_0/b)q_z$ around the Weyl nodes with $\bm{q}=\bm{k}-W_\pm$ ($\bm{q}^2\ll 1$).
Then we see that the contribution from $\varepsilon^+_{t\bm{k}}$ always vanishes as long as the chemical potential $\mu$ lies sufficiently close to the Weyl nodes.
Also, it should be noted that the quantity $\Omega_{t\bm{k},z}=v_F^2m_t/2\varepsilon_{t\bm{k}}^3$ is the $z$ component of the Berry curvature of the two-band Hamiltonian (\ref{H_pm}).
The anomalous Hall conductivity (\ref{AHE-Intrinsic}) is exactly the same as the value obtained from a field-theoretical approach, the first term in Eq.~(\ref{Current-Eq-WSM}).

Next, we need to evaluate the extrinsic contribution $\sigma^{\rm II}_{xy}$, which originates from the Fermi surface and the presence of disorder.
As described in Sec.~\ref{Sec-Kinetic-Equation}, we treat disorder within the Born approximation.
We consider a short-range (on-site) disorder potential of the form $U(\bm{r})=U_0\sum_i\delta(\bm{r}-\bm{r}_i)$, and assume that the correlation function satisfies $\langle U(\bm{r})U(\bm{r}')\rangle=n_{\mathrm{imp}}U_0^2\,\delta(\bm{r}-\bm{r}')$ with $n_{\mathrm{imp}}$ the impurity density.
Then we obtain
\begin{align}
&\langle U^{++}_{\bm{k}\bm{k}'}U^{+-}_{\bm{k}'\bm{k}}\rangle \nonumber\\
&=\frac{n_{\mathrm{imp}}U_0^2}{2}\left[\frac{k_\perp}{\varepsilon_{\bm{k}}}\frac{m(k'_z)}{\varepsilon_{\bm{k}'}}+\left(i\sin\gamma-\frac{m(k_z)}{\varepsilon_{\bm{k}}}\cos\gamma\right)\frac{k_\perp'}{\varepsilon_{\bm{k}'}}\right],
\end{align}
where $\gamma=\theta'-\theta$ with $e^{i\theta'}=(k'_x+ik'_y)/k'_\perp$.
We also have the relations $\langle U^{+-}_{\bm{k}\bm{k}'}U^{--}_{\bm{k}'\bm{k}}\rangle=-\langle U^{++}_{\bm{k}\bm{k}'}U^{+-}_{\bm{k}'\bm{k}}\rangle$, $\langle U^{--}_{\bm{k}\bm{k}'}U^{-+}_{\bm{k}'\bm{k}}\rangle=-\langle U^{++}_{\bm{k}\bm{k}'}U^{+-}_{\bm{k}'\bm{k}}\rangle^*$, and $\langle U^{-+}_{\bm{k}\bm{k}'}U^{++}_{\bm{k}'\bm{k}}\rangle=\langle U^{++}_{\bm{k}\bm{k}'}U^{+-}_{\bm{k}'\bm{k}}\rangle^*$.
From the expression for $J(\langle n_E\rangle)$ [Eq.~(\ref{Anomalous-driving-term})] with $\langle n_E\rangle=\mathrm{diag}[n^+_{E\bm{k}},n^-_{E\bm{k}}]$, we get
\begin{align}
[J(\langle n_E\rangle)]^{+-}_{\bm{k}}=&\pi\sum_{\bm{k}'}\langle U^{++}_{\bm{k}\bm{k}'}U^{+-}_{\bm{k}'\bm{k}}\rangle\left[n^+_{E\bm{k}}\delta(\varepsilon^+_{\bm{k}}-\varepsilon^+_{\bm{k}'})\right. \nonumber\\
& \left.-n^+_{E\bm{k}'}\delta(\varepsilon^+_{\bm{k}}-\varepsilon^+_{\bm{k}'})\right],
\label{J^+-}
\end{align}
where $n^+_{E\bm{k}}=e\tau_+E_y\partial f_0(\varepsilon^+_{\bm{k}})/\partial k_y$ with $\tau_+\propto 1/n_{\mathrm{imp}}U_0^2$ the scattering rime for the band $\varepsilon^+_{\bm{k}}$.
Here, we have used $\delta(\varepsilon^+_{\bm{k}}-\varepsilon^-_{\bm{k}'})=\delta(\varepsilon^-_{\bm{k}}-\varepsilon^+_{\bm{k}'})=0$ and the fact that $n^-_{E\bm{k}}=n^-_{E\bm{k}'}=0$, because we have considered the case of $\mu>0$.
Similarly, we obtain $[J(\langle n_E\rangle)]^{-+}_{\bm{k}}=\{[J(\langle n_E\rangle)]^{+-}_{\bm{k}}\}^*$.
Then the off-diagonal density matrix resulting from $J(\langle n_E\rangle)$ is given by
\begin{align}
\langle S'_E\rangle=-\tilde{\sigma}_y\frac{\mathrm{Re}\left\{[J(\langle n_E\rangle)]^{+-}_{\bm{k}}\right\}}{2\varepsilon_{\bm{k}}}-\tilde{\sigma}_x\frac{\mathrm{Im}\left\{[J(\langle n_E\rangle)]^{+-}_{\bm{k}}\right\}}{2\varepsilon_{\bm{k}}},
\label{rho-from-J}
\end{align}
where $\mathrm{Re}$ and $\mathrm{Im}$ represent the real and imaginary part, respectively.
The calculation of the conductivity from the definition $\sigma^{\rm II}_{xy}=\mathrm{Tr}[(-e)v_x\langle S'_E\rangle]/E_y$ is somewhat long.
See Appendix~\ref{Appendix-AHE-extrinsic-contribution} for a detailed calculation.
Finally, we find that $\sigma^{\rm II}_{xy}=0$ when the chemical potential $\mu$ lies sufficiently close to the Weyl nodes.

In the end, the total anomalous Hall conductivity of the Weyl metal reads
\begin{align}
\sigma_{xy}=\sigma^{\rm I}_{xy}+\sigma^{\rm II}_{xy}=-\frac{e^2}{2\pi^2}k_0,
\label{AHE-total}
\end{align}
which means that the anomalous Hall conductivity of Weyl metals is purely intrinsic, i.e., is determined by the Berry curvature of the filled band, as long as the chemical potential $\mu$ lies sufficiently close to the Weyl nodes.
This plateau-like behavior of the anomalous Hall conductivity is consistent with a calculation by Burkov \cite{Burkov2014}.
As mentioned in Sec.~\ref{Sec-rho_E} the contribution from $J(\langle n\rangle)$, i.e., $\sigma^{\rm II}_{xy}$, corresponds to the vertex correction in the ladder approximation.
Therefore it can be said that the vertex correction to the anomalous Hall conductivity of Weyl metals described by the Hamiltonian (\ref{WSM-H_eff}) is absent, as long as the chemical potential $\mu$ lies sufficiently close to the Weyl nodes.

\subsection{Chiral magnetic effect in Weyl metals \label{Sec-CME}}
Let us see that the magnetic driving term (\ref{driving-term-B}) properly describes the chiral magnetic effect in Weyl metals.
The chiral magnetic effect, the second term in Eq.~(\ref{Current-Eq-WSM}), is written as
\begin{align}
\bm{j}=\frac{e^2}{4\pi^2}\bm{B}\sum_{\nu} Q_\nu\mu_\nu,
\label{CME}
\end{align}
where $Q_\nu$ and $\mu_\nu$ are the chirality and chiral chemical potential of a Weyl node, respectively.
In two-node Weyl semimetals, we have $Q_{\pm}=\pm 1$ and $2\mu_5=\mu_+-\mu_-$ (see Fig.~\ref{Fig2}).
As is obvious, the chiral magnetic effect does not occur when there is no chemical potential difference between the nodes as $\mu_+=\mu_-$.

\subsubsection{The case of $\mu_+=\mu_-$}
First, it is informative to check the absence of the chiral magnetic effect in our formulation when $\mu_+=\mu_-$.
Since there is no electric fields, we start from the diagonal density matrix $\langle\rho_0\rangle=\mathrm{diag}[f_0(\varepsilon^+_{\bm{k}}),f_0(\varepsilon^-_{\bm{k}})]$ with $f_0(\varepsilon^\pm_{\bm{k}})=1/[e^{(\varepsilon^\pm_{\bm{k}}-\mu)/T}+1]$.
Without loss of generality we can consider the case of $\bm{B}=(0,0,B_z)$ and can set $\mu>0$.
Explicit matrix expression for $D_B(\langle\rho_0\rangle)$ can be found in Appendix~\ref{Appendix-D_B-2}.
Using Eqs.~(\ref{rho_B-formal-expression}) and (\ref{D_B-from-diagonal-density}), the linear-response off-diagonal electron density matrix induced by the magnetic field is obtained as
\begin{align}
\langle S_B\rangle=&\frac{e}{2}B_z\left\{\frac{\partial [f_0(\varepsilon^+_{\bm{k}})+f_0(\varepsilon^-_{\bm{k}})]}{\partial k_x}
\begin{bmatrix}
0 && \mathcal{R}^{+-}_{\bm{k},y}\\
\mathcal{R}^{-+}_{\bm{k},y}\ && 0
\end{bmatrix}\right. \nonumber\\
&\left.-\frac{\partial [f_0(\varepsilon^+_{\bm{k}})+f_0(\varepsilon^-_{\bm{k}})]}{\partial k_y}
\begin{bmatrix}
0 && \mathcal{R}^{+-}_{\bm{k},x}\\
\mathcal{R}^{-+}_{\bm{k},x} && 0
\end{bmatrix}\right\}.
\label{rho_B-CME}
\end{align}
By substituting the Berry connections (\ref{Berry-connection}) into this equation, we get a concrete expression for $\langle S_B\rangle$.
Also, the intrinsic diagonal density matrix induced by the magnetic field (\ref{intrinsic-diagonal-matrix-B}) is given by
\begin{align}
\langle\xi_B\rangle=eB_z
\begin{bmatrix}
f_0(\varepsilon^+_{\bm{k}})\Omega^+_{\bm{k},z} && 0\\
0 && f_0(\varepsilon^-_{\bm{k}})\Omega^-_{\bm{k},z}
\end{bmatrix}.
\label{xi_B-CME}
\end{align}
The velocity operator is written in the eigenstate basis as
\begin{align}
v_z=\frac{\partial m}{\partial k_z}\left(\frac{m}{\varepsilon_{\bm{k}}}\tilde{\sigma}_z+\frac{v_F k_\perp}{\varepsilon_{\bm{k}}}\tilde{\sigma}_x\right).
\label{v_z-matrix}
\end{align}
We are in a position to calculate the electric current.
Since the chemical potential $\mu$ lies sufficiently close to the Weyl nodes, we do not need to take into account the contribution from the $t=+$ band.
Finally, we find that the electric current along the magnetic field $\bm{B}$ vanishes as expected:
\begin{align}
j_z=&\mathrm{Tr}[(-e)v_z\langle S_B\rangle]+\mathrm{Tr}[(-e)v_z\langle \xi_B\rangle] \nonumber\\
=&-\frac{e^2}{2}B_z\int \frac{d^3k}{(2\pi)^3}\, \frac{v_F^4 k_z}{m_--b}\frac{1}{\varepsilon_{-\bm{k}}^2}\sum_{a=x,y}k_a\frac{\partial f_0(\varepsilon^+_{-\bm{k}})}{\partial k_a} \nonumber\\
&+e^2B_z\int \frac{d^3k}{(2\pi)^3}\, \frac{v_F^4 k_z}{m_--b}\frac{m_-^2}{2\varepsilon^4_{-\bm{k}}}\bigl[f_0(\varepsilon^+_{-\bm{k}})+f_0(\varepsilon^-_{-\bm{k}})\bigr] \nonumber\\
=&0,
\end{align}
where we have used the fact that the integrand is an odd function of $k_z$.
This shows that the chiral magnetic effect does not occur when there is no chemical potential difference between the Weyl nodes, i.e., when $\mu_+=\mu_-$.

\subsubsection{The case of $\mu_+\neq\mu_-$}
Next, let us consider the case where there is a chemical potential difference between the two nodes as $2\mu_5=\mu_+-\mu_-$.
As is shown above, the expression for the chiral magnetic effect (\ref{CME}) is derived analytically from the four-band model (\ref{WSM-H_eff}) with nonzero $\mu_5\tau_z$ term.
However, it is difficult to obtain analytically the eigenvalues and eigenstates in the presence of the $\mu_5\tau_z$ term, unlike in the case of the $b\sigma_z$ term where we can obtain the eigenstates analytically as Eq.~(\ref{WSM-eigenstates}). To determine whether the magnetic driving term (\ref{driving-term-B}) can describe the chiral magnetic effect, we consider two-band Hamiltonians around each Weyl node separately and sum each contribution to the electric current. The Hamiltonian around each Weyl node reads
\begin{align}
\tilde{\mathcal{H}}^\nu(\bm{q})=v_F(q_x\sigma_x+q_y\sigma_y+Q_\nu q_z\sigma_z),
\end{align}
which gives the eigenvalues $\varepsilon^\pm_{\bm{q}}=\pm v_F\sqrt{q_x^2+q_y^2+q_z^2}$.
Here, $\nu=\pm$ is the node index and $Q_\pm=\pm 1$ is the chirality of each Weyl node.
The Fermi-Dirac distribution function around each node is given by $f_0^\nu(\varepsilon^\pm_{\bm{q}})=1/[e^{(\varepsilon^\pm_{\bm{q}}-\tilde{\mu}_\nu)/T}+1]$ with $\tilde{\mu}_\nu=\mu+\mu_\nu$, where $\mu$ is a uniform chemical potential and $\mu_\nu$ is a chiral chemical potential such that $\mu_+\neq \mu_-$.
Without loss of generality, we can set $\tilde{\mu}_\nu>0$.
In linear response, the diagonal part $\langle \xi_{B,\nu}\rangle$ and the off-diagonal part $\langle S_{B,\nu}\rangle$ of the density matrix induced by the magnetic field are given respectively by Eqs.~(\ref{xi_B-CME}) and (\ref{rho_B-CME}), with $f_0(\varepsilon^\pm_{\bm{q}})$ replaced by $f_0^\nu(\varepsilon^\pm_{\bm{q}})$.
The velocity operator is given by
\begin{align}
v_{z,\nu}=Q_\nu v_F\left(\frac{Q_\nu v_F q_z}{\varepsilon_{\bm{q}}}\tilde{\sigma}_z+\frac{v_F q_\perp}{\varepsilon_{\bm{q}}}\tilde{\sigma}_x\right).
\label{v_z-each-node}
\end{align}

In the present case, we may approximate the total electric current by the sum of the contributions from each node.
Then the total electric current along the magnetic field $\bm{B}$ is calculated as $j_z=\sum_{\nu=\pm}(j_{z,\nu}^S+j_{z,\nu}^\xi)$, where
$j_{z,\nu}^X\equiv\mathrm{Tr}[(-e)v_{z,\nu}\langle X_{B,\nu}\rangle]$ ($X=S,\xi$).
The contribution from $\langle S_{B,\nu}\rangle$ is calculated to be
\begin{align}
j_{z,\nu}^S&=-\frac{e^2}{2}B_zQ_\nu \int_{\delta\Lambda_\nu}\frac{d^3q}{(2\pi)^3}\, \frac{v_F^3}{\varepsilon_{\bm{q}}^2}\sum_{a=x,y}q_a\frac{\partial f_0^\nu(\varepsilon^+_{\bm{q}})}{\partial q_a} \nonumber\\
&=\frac{e^2 B_z v_F}{8\pi^2}Q_\nu \int_0^\pi d\theta \int_0^{\delta q_\nu}dq\, q\sin^3\theta\delta(q-\tilde{\mu}_\nu/v_F), \nonumber\\
&=\frac{2}{3}\frac{e^2}{4\pi^2}B_zQ_\nu\tilde{\mu}_\nu,
\label{CME-approximate-integral}
\end{align}
where $\delta\Lambda_\nu$ is a small momentum space around each node, $q=\sqrt{q_x^2+q_y^2+q_z^2}$, and we have used $\partial f_0^\nu(\varepsilon^+_{\bm{q}})/\partial \varepsilon^+_{\bm{q}}=-\delta(\varepsilon_{\bm{q}}-\tilde{\mu}_\nu)$.
The contribution from $\langle \xi_{B,\nu}\rangle$ is calculated to be
\begin{align}
j_{z,\nu}^\xi&=e^2B_zQ_\nu \int \frac{d^3q}{(2\pi)^3}\, \frac{v_F^5 q_z^2}{2\varepsilon^4_{\bm{q}}}\bigl[f^\nu_0(\varepsilon^+_{\bm{q}})+f^\nu_0(\varepsilon^-_{\bm{q}})\bigr] \nonumber\\
&=\frac{e^2B_z}{8\pi^2}Q_\nu \int_0^\pi d\theta \left[\int_0^{\tilde{\mu}_\nu}dq+\int_0^{\Lambda}dq\right]\sin\theta\cos^2\theta \nonumber\\
&=\frac{1}{3}\frac{e^2}{4\pi^2}B_zQ_\nu(\tilde{\mu}_\nu+\Lambda),
\label{CME-approximate-integral2}
\end{align}
where $\Lambda$ is the cutoff of the Fermi sea of Weyl cones.

Combining Eqs.~(\ref{CME-approximate-integral}) and (\ref{CME-approximate-integral2}), we arrive at the final expression for the chiral magnetic effect:
\begin{align}
j_z=\frac{e^2}{4\pi^2}B_z\sum_\nu Q_\nu\mu_\nu,
\label{CME-approx}
\end{align}
where we have used $\sum_{\nu=\pm}Q_\nu\tilde{\mu}_\nu=\sum_{\nu=\pm}Q_\nu\mu_\nu$ and $\sum_{\nu=\pm}Q_\nu\Lambda=0$.
This is in full agreement with the result by a field-theoretical approach (\ref{CME}).
Here, we comment on an important observation in this study.
The portion 2/3 comes from the Fermi surface contribution, as seen from Eq.~(\ref{CME-approximate-integral}).
The portion 1/3 comes from the Fermi sea contribution, as see from Eq.~(\ref{CME-approximate-integral2}).
This observation is in contrast to a derivation by semiclassical wavepacket dynamics \cite{Son2013,Spivak2016} in which the chiral magnetic effect comes entirely from the Fermi surface contribution.

\section{Negative magnetoresistance in Weyl and Dirac Metals \label{Sec-NMR}}
\begin{figure*}[!t]
\centering
\includegraphics[width=1.85\columnwidth]{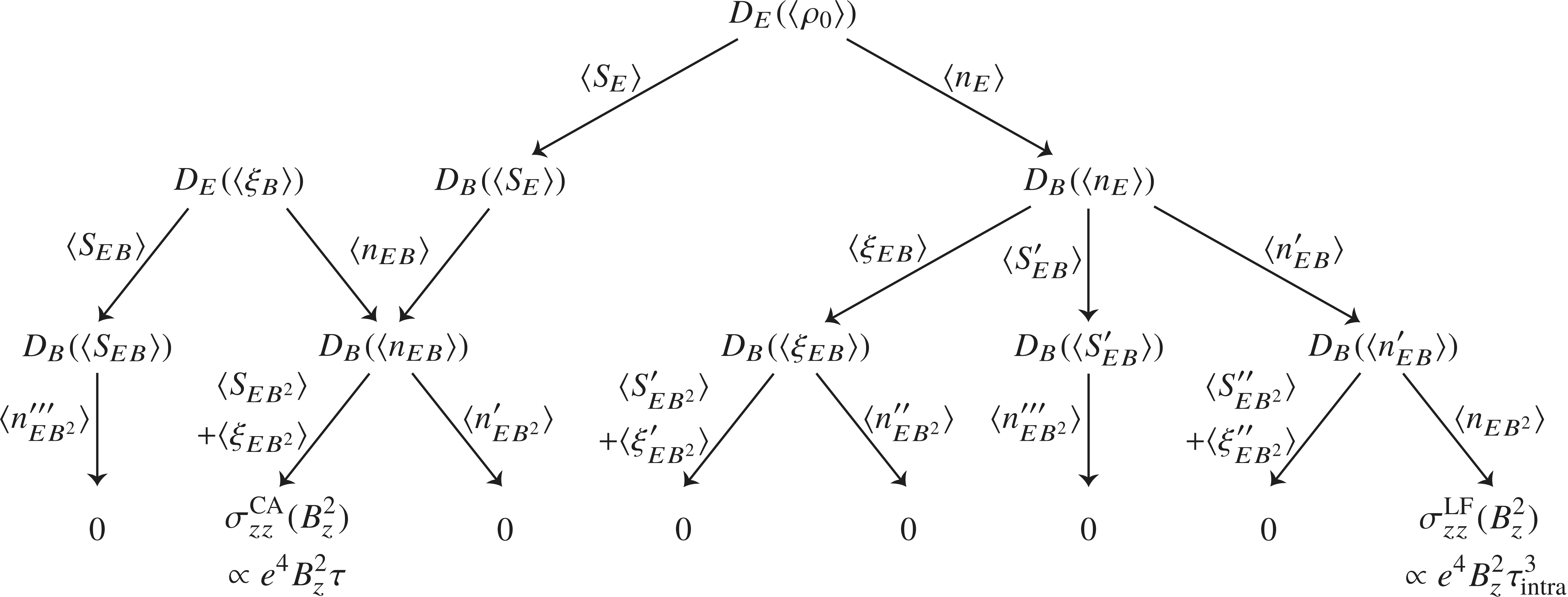}
\caption{Schematic illustration of longitudinal quadratic magnetoconductivity contributions 
 $\sigma_{zz}(B_z^2)$ (\ref{Sigma-zz-formal})  for $\bm{E}\parallel \bm{B}$ in a Weyl metal 
 described by the Hamiltonian (\ref{WSM-H_eff}).
$\langle n\rangle$ and $\langle \xi\rangle$ indicates band-diagonal density matrix components,
and $\langle S\rangle$ indicates band-off-diagonal density matrix components.
See Eqs.~(\ref{n_B-Nth-order}), (\ref{S_B-Nth-order}) and (\ref{xi_B-Nth-order}) respectively for the formal 
expressions for $\langle n\rangle$, $\langle S\rangle$ and $\langle \xi\rangle$ in a magnetic field.
$\tau$ and $\tau_{\mathrm{intra}}$ are the intervalley and intravalley scattering times, respectively.
$\sigma_{zz}(B_z^2)$ is given by the sum of the contribution from the chiral anomaly $\sigma^{\mathrm{CA}}_{zz}(B_z^2)$ and that from the Lorentz force $\sigma^{\mathrm{LF}}_{zz}(B_z^2)$ as $\sigma_{zz}(B_z^2)=\sigma^{\mathrm{CA}}_{zz}(B_z^2)+\sigma^{\mathrm{LF}}_{zz}(B_z^2)$.
The other contributions to the quadratic magnetoconductivity vanish.
Note that the contribution from the Lorentz force vanishes when the energy bands are isotropic in the $(k_x,k_y)$ plane [see the discussion below Eq.~(\ref{D_B-Lorentz})].
}\label{Fig3}
\end{figure*}
So far we have shown that the electric driving term $D_E$ (\ref{driving-term-E}) and magnetic driving term $D_B$ (\ref{driving-term-B}) properly describe the chiral-anomaly induced phenomena in Weyl metals, the anomalous Hall effect and chiral magnetic effect, respectively.
Another phenomenon manifested by the chiral anomaly is a negative magnetoresistance (or equivalently positive magnetoconductance) quadratic in magnetic field for parallel electric and magnetic fields in Weyl and Dirac metals \cite{Son2013,Burkov2014,Burkov2015a,Spivak2016}.
Such an unusual negative magnetoresistance has recently been experimentally observed in Dirac semimetals Na$_3$Bi \cite{Xiong2015}, Cd$_3$As$_2$ \cite{Li2015,Li2016}, and ZrTe$_5$ \cite{Li2016a}, and in the Weyl semimetals TaAs \cite{Huang2015} and TaP \cite{Arnold2016}.
Here, note that the usual magnetoresistance due to Lorentz force is positive.
As denoted in the introduction, the chiral anomaly is described by the $\theta$ term $S_\theta=\int dt\, d^3 r\, (e^2/4\pi^2)\theta(\bm{r},t)\bm{E}\cdot\bm{B}$.
As we can see from this expression for the $\theta$ term, the $\bm{E}\cdot\bm{B}$ contribution becomes largest in the case of parallel electric and magnetic fields.
The positive quadratic magnetoconductivity arising from the chiral anomaly first derived by Son and Spivak reads \cite{Son2013}
\begin{align}
\sigma_{zz}(B_z^2)=\frac{e^2}{4\pi^2}\frac{(eB_z)^2v_F^3}{\mu^2}\tau,
\label{Sigma_zz-Son&Spivak}
\end{align}
where $\mu$ is the chiral potential and $\tau$ is the intervalley scattering time.
Since their derivation is based on semiclassical wavepacket dynamics alone,
it is difficult to take disorder scattering into account, and the microscopic origin of the 
intervalley scattering time in Eq.~(\ref{Sigma_zz-Son&Spivak}) is not precisely clear.

In this section, we apply the  theory we have developed so far to the positive quadratic magnetoconductivity, starting from the microscopic continuum model of Weyl semimetals (\ref{WSM-H_eff}).
It should be mentioned that the positive quadratic magnetoconductivity has been experimentally observed in the low magnetic field regime.
Thus, our semiclassical treatment of magnetic fields can be justified.
In our theory, the formal expression for the $\mu\nu$-component of the quadratic magnetoconductivity is written from Eq.~(\ref{Magnetoconductivity-general}) in the form
\begin{align}
\sigma_{\mu\nu}(B^2)=&\mathrm{Tr}\left[(-e) v_\mu (\mathcal{L}^{-1}D_B)^2 \mathcal{L}^{-1}D_E(\langle\rho_0\rangle)\right]/E_\nu \nonumber\\
&+\mathrm{Tr}\left[(-e) v_\mu \mathcal{L}^{-1}D_B \mathcal{L}^{-1}D_E(\langle\xi_B\rangle)\right]/E_\nu.
\label{Sigma-zz-formal}
\end{align}
Here, note that the angle between the electric and magnetic fields is arbitrary in this formalism.

Our evaluation of Eq.~(\ref{Sigma-zz-formal}) involves a number of subtleties.
It is informed by noting (i) that the expression of $\sigma_{zz}(B_z^2)$ (\ref{Sigma_zz-Son&Spivak}) has a linear dependence on $\tau$, and (ii) that this expression is derived via the chiral magnetic effect \cite{Son2013,Burkov2014,Burkov2015a,Spivak2016,Li2016a}.
We use the fact that a magnetic driving term obtained from an off-diagonal density matrix is purely diagonal, and that a magnetic driving term obtained from a diagonal density matrix has 
diagonal and off-diagonal components that lead to the chiral magnetic effect as we have seen in Sec.~\ref{Sec-CME}
(see Appendix~\ref{Appendix-D_B-2} for explicit expressions for the magnetic driving term).
Indeed, our explicit calculations summarized schematically in Fig.~\ref{Fig3} 
confirmed that among all the possible contributions that are second order in magnetic field,
only magnetoconductances originating from the chiral anomaly and the usual Lorentz force survive.

To obtain the quadratic magnetoconductivity, we divide the calculation of the matrices $(\mathcal{L}^{-1}D_B)^2 \mathcal{L}^{-1}D_E(\langle\rho_0\rangle)$ and $\mathcal{L}^{-1}D_B \mathcal{L}^{-1}D_E(\langle\xi_B\rangle)$ in Eq.~(\ref{Sigma-zz-formal}) into four steps as follows:
\begin{align}
\langle S_E\rangle&=\mathcal{L}^{-1}D_E(\langle\rho_0\rangle)=[D_E(\langle\rho_0\rangle)]_{\bm{k}}^{mm'}/i(\varepsilon^m_{\bm{k}}-\varepsilon^{m'}_{\bm{k}}) \nonumber\\
&\propto eE_z, \nonumber\\
\langle n_{EB}\rangle&=\mathcal{L}^{-1}\left[D_B(\langle S_E\rangle)+D_E(\langle \xi_B\rangle)\right] \nonumber\\
&=\tau \left[D_B(\langle S_E\rangle)+D_E(\langle \xi_B\rangle)\right]_{\bm{k}}^{mm} \nonumber\\
&\propto e^2E_zB_z\tau, \nonumber\\
\langle S_{EB^2}\rangle&=\mathcal{L}^{-1}D_B(\langle n_{EB}\rangle)=[D_B(\langle n_{EB}\rangle)]_{\bm{k}}^{mm'}/i(\varepsilon^m_{\bm{k}}-\varepsilon^{m'}_{\bm{k}}) \nonumber\\
&\propto e^3E_zB_z^2\tau, \nonumber\\
\langle \xi_{EB^2}\rangle&=P^{-1}D_B(\langle n_{EB}\rangle)=e\, \langle n_{EB}\rangle^{mm}\bm{B}\cdot\bm{\Omega}^m_{\bm{k}} \nonumber\\
&\propto e^3E_zB_z^2\tau,
\label{procedures}
\end{align}
where $\langle\rho_0\rangle=\mathrm{diag}[f_0(\varepsilon^+_{\bm{k}}),f_0(\varepsilon^-_{\bm{k}})]$ with $f_0(\varepsilon^\pm_{\bm{k}})=1/[e^{(\pm \varepsilon_{\bm{k}}-\mu)/T}+1]$ being the Fermi-Dirac distribution function, $\langle n\rangle$ and $\langle \xi\rangle$ indicate diagonal density matrices, $\langle S\rangle$ indicates an off-diagonal density matrix, and $\tau$ is the intervalley scattering time.
See Eqs.~(\ref{n_B-Nth-order}), (\ref{S_B-Nth-order}) and (\ref{xi_B-Nth-order}) for the formal expressions for $\langle n\rangle$, $\langle S\rangle$ and $\langle \xi\rangle$ respectively in a magnetic field.
As we shall show below, the appearance of the intervalley scattering time $\tau$ is due to that
the driving terms $D_B(\langle S_E\rangle)$ and $D_E(\langle \xi_B\rangle)$ have nonzero values when 
integrated over a given valley (Weyl cone).
In the following, we consider the low-temperature case where $T\ll \mu$, with $T$ and $\mu>0$ being the temperature and chemical potential of the system, respectively.

\subsection{Quadratic magnetoconductivity for $\vec{E}\parallel \vec{B}$ \label{Subsec-quadratic-MC}}
\subsubsection{Contribution from the chiral anomaly \label{Sec-MC-from-CA}}
Let us consider the case of parallel electric and magnetic fields $\bm{E}=(0,0,E_z)$ and $\bm{B}=(0,0,B_z)$.
We start by obtaining the off-diagonal part of the density matrix induced solely by the electric field, $\langle  S_E\rangle$ in Eq.~(\ref{procedures}).
As we have seen in Sec.~\ref{Sec-AHE}, this off-diagonal density matrix results in an intrinsic effect, i.e., the anomalous Hall effect.
Using the expression for the electric driving term $D_E(\langle\rho_0\rangle)$ (\ref{D_E-intrinsic}) and the Berry connection~(\ref{Berry-connection}), we get
\begin{align}
\langle  S_E\rangle=\tilde{\sigma}_y eE_z[f_0(\varepsilon^+_{\bm{k}})-f_0(\varepsilon^-_{\bm{k}})]\frac{v_F k_\perp}{4\varepsilon_{\bm{k}}^3}\frac{\partial m}{\partial k_z}.
\label{step1-explicit}
\end{align}
At this stage we recall from Eq.~(\ref{rho_E-off-diagonal}) that there exists an extrinsic contribution to the off-diagonal density matrix $\langle  S_E\rangle$ from the anomalous driving term $J(\langle n_E\rangle)$.
Here, $\langle n_E\rangle=\mathrm{diag}[n^+_{E\bm{k}},n^-_{E\bm{k}}]$ with $n^\pm_{E\bm{k}}=e\tau_{\mathrm{intra}} E_z\partial f_0(\varepsilon^\pm_{\bm{k}})/\partial k_z$ ($\tau_{\mathrm{intra}}$ is the intravalley scattering time).
Let us consider the same setup as in the evaluation of the anomalous Hall conductivity in Sec.~\ref{Sec-AHE}.
Namely, we consider the case of a short-range (on-site) disorder potential of the form $U(\bm{r})=U_0\sum_i\delta(\bm{r}-\bm{r}_i)$, and assume that the correlation function satisfies $\langle U(\bm{r})U(\bm{r}')\rangle=n_{\mathrm{imp}}U_0^2\,\delta(\bm{r}-\bm{r}')$ with $n_{\mathrm{imp}}$ the impurity density.
In this case, we can show explicitly from Eqs.~(\ref{Im-J-explicit}) and (\ref{Re-J-explicit}) that
\begin{align}
\mathrm{Re}\left\{[J(\langle n_E\rangle)]_{\bm{k}}^{+-}\right\}=\mathrm{Im}\left\{[J(\langle n_E\rangle)]_{\bm{k}}^{+-}\right\}=0,
\end{align}
and hence $[J(\langle n_E\rangle)]_{\bm{k}}^{+-}=[J(\langle n_E\rangle)]_{\bm{k}}^{-+}=0$, as long as the chemical potential $\mu$ lies sufficiently close to the Weyl nodes.
Hence, the expression for $\langle  S_E\rangle$, Eq.~(\ref{step1-explicit}), remains valid even after the anomalous driving term is taken into account in the case of sufficiently small $\mu$.

Second, we compute the diagonal density matrix $\langle n_{EB}\rangle$ proportional to $E_zB_z$ in Eq.~(\ref{procedures}).
This $\langle n_{EB}\rangle$ is the most important quantity
in our derivation of the magnetoconductivity induced by the chiral anomaly, as is understood from its form similar to $\bm{E}\cdot\bm{B}$.
As described in Appendix~\ref{Appendix-D_B-2}, the magnetic driving term obtained from an off-diagonal matrix is purely diagonal.
Using the expressions for the magnetic driving term $D_B(\langle S_E\rangle)$ (\ref{Expression-of-AB}) and (\ref{D_B-from-offdiagonal-density}) and the Berry connection~(\ref{Berry-connection}), we have
\begin{align}
D_B(\langle S_E\rangle)=e^2 E_z B_z\mathcal{F}_{\bm{k}}\bm{1}
\label{step2-explicit}
\end{align}
with
\begin{align}
\mathcal{F}_{\bm{k}}&=-\frac{v_F^3 k_\perp}{\varepsilon_{\bm{k}}^2}c_{\bm{k}}-\frac{v_F m^2}{\varepsilon_{\bm{k}}^2k_\perp}c_{\bm{k}}-v_F\cos\theta\frac{\partial c_{\bm{k}}}{\partial k_x}-v_F\sin\theta\frac{\partial c_{\bm{k}}}{\partial k_y},
\label{A-B}
\end{align}
where $\bm{1}$ is the $2\times2$ identity matrix and $c_{\bm{k}}=[f_0(\varepsilon^+_{\bm{k}})-f_0(\varepsilon^-_{\bm{k}})](v_F k_\perp/4\varepsilon_{\bm{k}}^3)\partial m/\partial k_z$.
Also, using the expressions for the electric driving term (\ref{rho_E-diagonal}) and (\ref{D_E-intrinsic}) and the Berry phase contribution (\ref{intrinsic-diagonal-matrix-B}), we have
\begin{align}
D_E(\langle \xi_B\rangle)=e^2 E_z B_z\frac{\partial}{\partial k_z}
\begin{bmatrix}
f_0(\varepsilon^+_{\bm{k}})\Omega^+_{\bm{k},z} && 0\\
0 && f_0(\varepsilon^-_{\bm{k}})\Omega^-_{\bm{k},z}
\end{bmatrix}
+\mathcal{L}\, \langle S_{EB}\rangle,
\label{step2-explicit2}
\end{align}
where $\langle S_{EB}\rangle$ is a purely off-diagonal matrix obtained by replacing $\langle\rho_0\rangle$ by $\langle \xi_B\rangle$ in Eq.~(\ref{step1-explicit}).

Now we show that these $D_B(\langle S_E\rangle)$ and $D_E(\langle \xi_B\rangle)$ have a special property which is not 
present when the system is driven by an electric field alone.  
Around a Weyl node with 
momentum $\bm{q}=\bm{k}-W_\pm$ ($\bm{q}^2\ll 1$), $m_-(k_z)$ is approximated as $m_-(q_z)\approx \mp v_F^2(k_0/b)q_z$.
Then we find that $\mathcal{F}_{\bm{q}}$ and $\partial f_0(\varepsilon^{\pm}_{\bm{q}})\Omega^{\pm}_{\bm{q},z}/\partial q_z$ are both even functions of $q_x$, $q_y$, and $q_z$.
Accordingly, we find that the integral of $D_B(\langle  S_E\rangle)+D_E(\langle \xi_B\rangle)$ over the Fermi surface of a given valley has a nonzero value: $\int_{\mathrm{FS}}\frac{d^3q}{(2\pi)^3}\, \sum_m [D_B(\langle  S_E\rangle)+D_E(\langle \xi_B\rangle)]_{\bm{q}}^{mm}\neq 0$.
As discussed in Sec.~\ref{Sec-Scattering}, this is a consequence of the total particle number nonconservation in a given Weyl cone.
Namely we obtain the rate of pumping
\begin{align}
\frac{\partial N}{\partial t}=\frac{e^2 E_z B_z}{4\pi^2}\int_{\mathrm{FS}}\frac{d^3q}{2\pi}\left(2\, \mathcal{F}_{\bm{q}}+\sum_{m=\pm}\frac{\partial f_0(\varepsilon^m_{\bm{q}})}{\partial q_z}\Omega^m_{\bm{q},z}
\right).
\label{Nonzero-integrated-D_B}
\end{align}
In the case of isotropic Weyl fermions [i.e., $m_-(q_z)\equiv v_F q_z$], Eq.~(\ref{A-B}) can be simplified to be
\begin{align}
\mathcal{F}^{\mathrm{iso}}_{\bm{q}}=\sum_{a=x,y}\left[\frac{1}{2}\Omega_{\bm{q},a}^+\frac{\partial}{\partial q_a} + 3q\, (\Omega_{\bm{q},a}^+)^2\right]\bigl[f_0(\varepsilon^+_{\bm{q}})-f_0(\varepsilon^-_{\bm{q}})\bigr],
\end{align}
where $\Omega_{\bm{q},a}^+=-q_a/(2q^3)$ is the Berry curvature and $q=\sqrt{q_x^2+q_y^2+q_z^2}$.
Integrating this over the Fermi surface we have
\begin{align}
&\int_{\mathrm{FS}} \frac{d^3q}{2\pi}\left(2\, \mathcal{F}^{\mathrm{iso}}_{\bm{q}}+\sum_{m=\pm}\frac{\partial f_0(\varepsilon^m_{\bm{q}})}{\partial q_z}\Omega^m_{\bm{q},z}\right) \nonumber\\
&=\sum_{m=\pm}\int \frac{d^3 \bm{q}}{2 \pi}\frac{\partial f_{0}(\varepsilon^m_{\bm{q}})}{\partial \varepsilon^m_{\bm{q}}} \bm{v}^m_{\bm{q}} \cdot \bm{\Omega}^m_{\bm{q}}=1,
\label{dN/dt-isotropic}
\end{align}
which is indeed consistent with the general expression (\ref{dN/dt-total}).

It follows from Eq.~(\ref{n_B-Nth-order}) that the intervalley scattering time $\tau$ appears when $\mathcal{L}^{-1}$ acts on $D_B(\langle S_E\rangle)$ and $D_E(\langle \xi_B\rangle)$.  When intravalley scattering is much stronger than intervalley 
scattering,  $\mathcal{L}^{-1}$ acting on $D_B(\langle S_E\rangle)$ and $D_E(\langle \xi_B\rangle)$ will also change the 
way in which the pumped charge is distributed across the Fermi surface, altering it so that 
it is equally distributed across the Fermi surface and corresponds simply to a change in chemical potential.  
In that limit the remaining steps in the calculation correspond precisely to the magnetoelectric effect 
calculation outlined in the previous section.  In general though some anisotropy remains when 
$\mathcal{L}^{-1}$ acts on the pumped charge.  To illustrate its potential role, we continue the 
calculation here considering the opposite limit in which only intervalley scattering is present.
Then the diagonal density matrix $\langle n_{EB}\rangle$ is obtained as
\begin{align}
\langle n_{EB}\rangle=\mathcal{L}^{-1}\left[D_B(\langle S_E\rangle)+D_E(\langle \xi_B\rangle)\right]=e^2 E_z B_z\tau
\begin{bmatrix}
\tilde{\mathcal{F}}_{\bm{k}}^{++} && 0\\
0 && \tilde{\mathcal{F}}_{\bm{k}}^{--}
\end{bmatrix},
\end{align}
where
\begin{align}
\tilde{\mathcal{F}}_{\bm{k}}^{mm}=\frac{1}{2}\sum_{m'=\pm}\left[\frac{\partial f_0(\varepsilon^{m'}_{\bm{k}})}{\partial k_x}\Omega_{\bm{k},x}^{m'}+\frac{\partial f_0(\varepsilon^{m'}_{\bm{k}})}{\partial k_y}\Omega_{\bm{k},y}^{m'}\right]+\frac{\partial f_0(\varepsilon^m_{\bm{k}})}{\partial k_z}\Omega^m_{\bm{k},z}
\label{F_k-tilde}
\end{align}
is the component which represents the Fermi surface response in Eqs.~(\ref{step2-explicit}) and (\ref{step2-explicit2}).
Note that we have neglected the Fermi sea response in Eqs.~(\ref{step2-explicit}) and (\ref{step2-explicit2}).

Third, we compute the off-diagonal density matrix $\langle S_{EB^2}\rangle$ proportional to $E_zB_z^2$ in Eq.~(\ref{procedures}).
Before doing this, notice from Eq.~(\ref{D_B-diagonal-component}) that there also are the diagonal components in $D_B(\langle n_{EB}\rangle)$ as
$[D_B(\langle n_{EB}\rangle)]^{++}_{\bm{k}}=e^3 E_z B_z^2\tau(\frac{\partial \varepsilon_{\bm{k}}}{\partial k_y}\frac{\partial \tilde{\mathcal{F}}^{++}_{\bm{k}}}{\partial k_x}-\frac{\partial \varepsilon_{\bm{k}}}{\partial k_x}\frac{\partial \tilde{\mathcal{F}}^{++}_{\bm{k}}}{\partial k_y})$ and $[D_B(\langle n_{EB}\rangle)]^{--}_{\bm{k}}=-e^3 E_z B_z^2\tau(\frac{\partial \varepsilon_{\bm{k}}}{\partial k_y}\frac{\partial \tilde{\mathcal{F}}^{--}_{\bm{k}}}{\partial k_x}-\frac{\partial \varepsilon_{\bm{k}}}{\partial k_x}\frac{\partial \tilde{\mathcal{F}}^{--}_{\bm{k}}}{\partial k_y})$, which represent the Lorentz force contribution.
Since these components are odd functions of $k_x$ and $k_y$ around a Weyl node, $D_B(\langle n_{EB}\rangle)$ does not drive the total particle number in a given Weyl cone.
Hence the diagonal density matrix  $\langle n'_{EB^2}\rangle$ obtained from these $[D_B(\langle n_{EB}\rangle)]^{mm}_{\bm{k}}$ contains the intravalley scattering time as $\langle n'_{EB^2}\rangle^{mm}_{\bm{k}}=\tau_{\mathrm{intra}}[D_B(\langle n_{EB}\rangle)]^{mm}_{\bm{k}}=m e^3 E_z B_z^2\tau\tau_{\mathrm{intra}}(\frac{\partial \varepsilon_{\bm{k}}}{\partial k_y}\frac{\partial \tilde{\mathcal{F}}^{mm}_{\bm{k}}}{\partial k_x}-\frac{\partial \varepsilon_{\bm{k}}}{\partial k_x}\frac{\partial \tilde{\mathcal{F}}^{mm}_{\bm{k}}}{\partial k_y})$.
However, we see that $\langle n'_{EB^2}\rangle$ does not contribute to the magnetoconductivity proportional to $B_z^2$, since it is an odd function of $k_x$ and $k_y$ and thus vanishes when integrated over momentum space.
From Eqs.~(\ref{S_B-Nth-order}) and (\ref{D_B-from-diagonal-density}), we obtain the relevant off-diagonal density matrix as
\begin{align}
\langle S_{EB^2}\rangle=&\frac{e^3 E_z B_z^2\tau}{2}\times \nonumber\\
&\left[\frac{\partial (\tilde{\mathcal{F}}^{++}_{\bm{k}}+\tilde{\mathcal{F}}^{--}_{\bm{k}})}{\partial k_x}\left(\tilde{\sigma}_x\frac{v_F}{2\varepsilon_{\bm{k}}}\cos\theta-\tilde{\sigma}_y\frac{v_F m}{2\varepsilon_{\bm{k}}^2}\sin\theta\right)\right. \nonumber\\
&\left.+\frac{\partial (\tilde{\mathcal{F}}^{++}_{\bm{k}}+\tilde{\mathcal{F}}^{--}_{\bm{k}})}{\partial k_y}\left(\tilde{\sigma}_x\frac{v_F}{2\varepsilon_{\bm{k}}}\sin\theta+\tilde{\sigma}_y\frac{v_F m}{2\varepsilon_{\bm{k}}^2}\cos\theta\right)\right].
\label{step3-explicit}
\end{align}
Here, let us recall that there exists an extrinsic contribution to the off-diagonal density matrix $\langle  S_{EB^2}\rangle$ from the anomalous driving term $J(\langle n'_{EB^2}\rangle)$, as in the case of $\langle S_E\rangle$.
We can show explicitly by replacing $\langle n_E\rangle$ in Eqs.~(\ref{Im-J-explicit}) and (\ref{Re-J-explicit}) by $\langle n'_{EB^2}\rangle$ that
\begin{align}
\mathrm{Re}\left\{[J(\langle n'_{EB^2}\rangle)]_{\bm{k}}^{+-}\right\}=\mathrm{Im}\left\{[J(\langle n'_{EB^2}\rangle)]_{\bm{k}}^{+-}\right\}=0,
\end{align}
and hence $[J(\langle n'_{EB^2}\rangle)]_{\bm{k}}^{+-}=[J(\langle n'_{EB^2}\rangle)]_{\bm{k}}^{-+}=0$, as long as the chemical potential $\mu$ lies sufficiently close to the Weyl nodes.
Hence, the expression for $\langle S_{EB^2}\rangle$, Eq.~(\ref{step3-explicit}), remains valid even after the anomalous driving term is taken into account in the case of sufficiently small $\mu$.

Notice from Eqs.~(\ref{intrinsic-diagonal-matrix-B-general}) and (\ref{xi_B-Nth-order}) that $\langle S_{EB^2}\rangle$ is accompanied by an intrinsic Berry phase contribution to the diagonal part of the density matrix proportional to $E_zB_z^2$.
From Eq.~(\ref{intrinsic-diagonal-matrix-B-general}) we readily obtain
\begin{align}
\langle\xi_{EB^2}\rangle=-e^3 E_z B_z^2\tau \frac{v_F^2 m}{2\varepsilon^3_{\bm{k}}}
\begin{bmatrix}
\tilde{\mathcal{F}}_{\bm{k}}^{++} && 0\\
0 && -\tilde{\mathcal{F}}_{\bm{k}}^{--}
\end{bmatrix}.
\label{step3-explicit-xi}
\end{align}

We are now in a position to evaluate the $zz$-component of the magnetoconductivity proportional to $B_z^2$, $\sigma_{zz}^{\mathrm{CA}}(B_z^2)=\mathrm{Tr}\{(-e)v_z[\langle S_{EB^2}\rangle+\langle \xi_{EB^2}\rangle]\}/E_z$.
Since we have considered the scattering of electrons at the Fermi surface, we do not need to take into account the contribution from the $t=+$ band.
From Eqs.~(\ref{v_z-matrix}), (\ref{step3-explicit}) and (\ref{step3-explicit-xi}), an explicit expression for $\sigma_{zz}(B_z^2)$ at low temperatures such that $T\ll \mu$ is obtained as
\begin{align}
\sigma_{zz}^{\mathrm{CA}}(B_z^2)=&-\frac{e^4B_z^2\tau}{2}\int\frac{d^3k}{(2\pi)^3} \frac{\partial m_-}{\partial k_z}\frac{v_F^2}{\varepsilon_{-\bm{k}}^2}\sum_{a=x,y}k_a\frac{\partial (\tilde{\mathcal{F}}^{++}_{\bm{k}}+\tilde{\mathcal{F}}^{--}_{\bm{k}})}{\partial k_a} \nonumber\\
& + \frac{e^4B_z^2\tau}{2}\int\frac{d^3k}{(2\pi)^3} \frac{\partial m_-}{\partial k_z}\frac{v_F^2 m_-^2}{\varepsilon^4_{-\bm{k}}}(\tilde{\mathcal{F}}^{++}_{\bm{k}}+\tilde{\mathcal{F}}^{--}_{\bm{k}})\nonumber\\
\equiv & \frac{e^2}{4\pi^2}\mathcal{C}(b, \Delta, T)\frac{(eB_z)^2v_F^3}{\mu^2}\tau,
\label{sigma-zz-CA}
\end{align}
which takes a positive value.
We numerically find that $\sigma^{\mathrm{CA}}_{zz}(B_z^2)$ is proportional to $1/\mu^2$ and $v_F^3$.
This is consistent with the result by Son and Spivak, Eq.~(\ref{Sigma_zz-Son&Spivak}).
We also find that, as in the case of the chiral magnetic effect (\ref{CME-approx}), the contributions from $\langle S_{EB^2}\rangle$ and $\langle \xi_{EB^2}\rangle$ to $\sigma^{\mathrm{CA}}_{zz}(B_z^2)$ are respectively 2/3 and 1/3 of the total value (\ref{sigma-zz-CA}).

\begin{figure}[!t]
\centering
\includegraphics[width=\columnwidth]{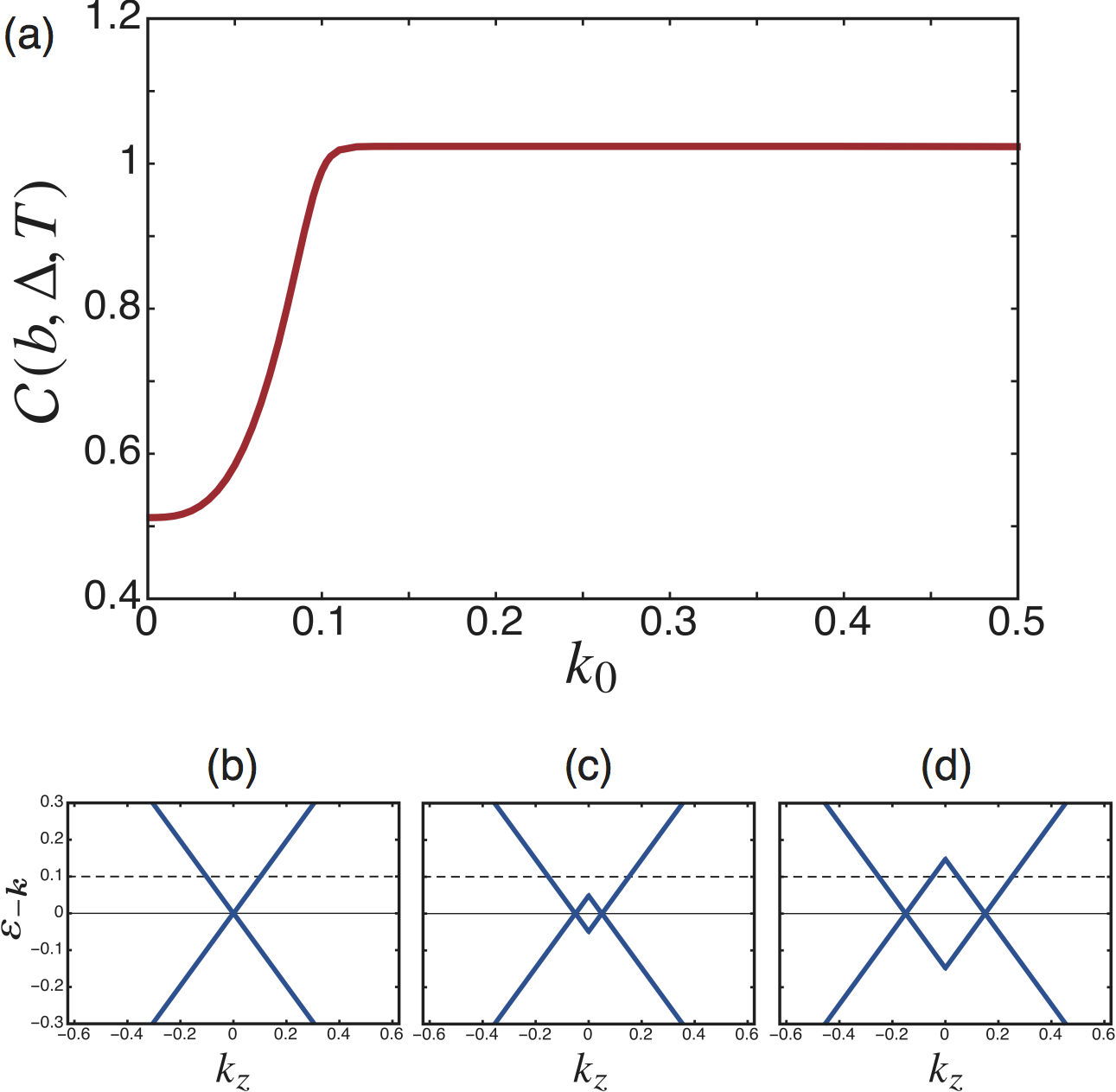}
\caption{(a) $k_0$ dependence of $\mathcal{C}(b, \Delta, T)$ in Eq.~(\ref{sigma-zz-CA}) for $\Delta=0$, $v_F=1\, \mathrm{eV\cdot\AA}$, $T=0.005\, \mathrm{eV}$, and $\mu=0.1\, \mathrm{eV}$.
In our model Hamiltonian for a two-node Weyl metal (\ref{WSM-H_eff}), the distance between the two Weyl nodes in momentum space is given by $2k_0=2\sqrt{b^2-\Delta^2}/v_F$.
Here, $k_0$ is given in units of $\mathrm{eV}/\hbar v_F(\sim\AA^{-1})$.
The value of $\mathcal{C}$ with $b=\Delta=0$ (with large $b$ and $\Delta=0$) approaches $\mathcal{C}=1/2$ ($\mathcal{C}=1$) in the low-temperature limit $T/\mu\to 0$.
Also, the $k_z$ dependences of the energy bands $\pm\varepsilon_{-\bm{k}}$ at $k_x=k_y=0$ for $k_0=0$, $k_0=0.05$, and $k_0=0.15$ in (a) are shown in (b), (c), and (d), respectively.
The dashed lines indicate the Fermi level, $\mu=0.1\, \mathrm{eV}$.
There exists a single isotropic Weyl cone when $k_0=0$, and it splits into two isotropic Weyl cones when $k_0\neq 0$.
}\label{Fig4}
\end{figure}
$\mathcal{C}(b, \Delta, T)$ in Eq.~(\ref{sigma-zz-CA}) is a dimensionless parameter of order $1$.
In Fig.~\ref{Fig4}(a) we show the $k_0$ dependence of $\mathcal{C}(b, \Delta, T)$ with $\Delta=0$, where $2k_0=2\sqrt{b^2-\Delta^2}/v_F$ is the distance between the two Weyl nodes in momentum space.
We find 
that $\mathcal{C}$ is an increasing function of $k_0$ and that is saturates
when $k_0$ is sufficiently large.
As shown in Figs.~\ref{Fig4}(b), (c), and (d), a single Weyl cone splits into two Weyl cones when $k_0\neq 0$.
The two Weyl cones partly overlap at the Fermi level and therefore there is only one Fermi surface in the region where $\mathcal{C}$ is increasing as $k_0$ becomes larger.
The separate contributions from the two Weyl cones are therefore incomplete in this region.
When two Fermi surfaces are present, 
the value of $\mathcal{C}$ saturates to the contributions from the two independent Weyl cones.
In the present case, the value of $k_0$ at which the saturation begins is given by $k_0=\mu/v_F$.
Here, note that we have not considered the $k_0$ dependence of the intervalley scattering time 
$\tau$.
Although it is not easy to take into account such a $k_0$ dependence in $\tau$, it is expected that the value of $\tau$ will become larger as the value of $k_0$ becomes larger, i.e., as the distance between the two Weyl nodes becomes longer.
This is because the strength of intervalley scattering processes will 
decrease as the value of $k_0$ becomes larger, since intervalley scattering processes 
with large $k_0$ require large momentum transfer.
Thus, based on these viewpoints, we conclude that the value of $\sigma_{zz}^{\mathrm{CA}}(B_z^2)$ in Weyl metals becomes larger as the distance between the two Weyl nodes becomes longer. 

Finally, we note that the contribution to $\sigma_{zz}^{\mathrm{CA}}(B_z^2)$ from a {\it single} isotropic Weyl cone in the low temperature limit $T/\mu\to 0$ is obtained by setting $\mathcal{C}(b, \Delta, T)=1/2$ and $b=\Delta=0$ in Eq.~(\ref{sigma-zz-CA}) (see Fig.~\ref{Fig4}), since our Hamiltonian (\ref{H_pm}) with $b=\Delta=0$ (i.e, $k_0=0$) possesses a single isotropic Weyl cone.
Also, the saturated value $\mathcal{C}=1$ obtained at large $k_0$ with $T/\mu\to 0$ is the 
contribution from {\it two} isotropic Weyl cones.
The value from a single isotropic Weyl cone, $\mathcal{C}=1/2$, is smaller than the value obtained by invoking semiclassical wavepacket dynamics and neglecting 
scattering [Eq.~(\ref{Sigma_zz-Son&Spivak})] that corresponds to $\mathcal{C}=1$.
This result can be interpreted as follows.
The rate of pumping into a valley (\ref{dN/dt-isotropic}), $\partial N/\partial t=e^2 E_z B_z/4\pi^2$, is the same as that obtained by invoking semiclassical wavepacket dynamics and neglecting scattering [Eq.~(\ref{CA-conventional})].
When intravalley scattering is dominant (i.e., in the limit of extremely weak intervalley scattering), by following the chiral magnetic effect we will find that $\mathcal{C}=1$.
In general, however, 
the electron distribution on the Fermi surface $\tilde{\mathcal{F}}^{mm}_{\bm{k}}$ (\ref{F_k-tilde}) can be 
anisotropic due to intravalley scattering.
The numerical results in Fig.~\ref{Fig4}, obtained neglecting intravalley scattering, 
demonstrate that the precise value of $\mathcal{C}$ can be depend on the details of disorder.

\subsubsection{The case of Dirac semimetals}
Here, we briefly discuss the case of Dirac semimetals.
To this end let us consider the contribution from a single Dirac cone, i.e., two degenerate Weyl cones.
Time reversal symmetry is preserved in Dirac metals, and therefore we have to set $b=0$.
Also, for the existence of a Dirac point we have to set $\Delta=0$.
As mentioned above, Eq.~(\ref{sigma-zz-CA}) with $b=\Delta=0$ represents the contribution from a single Weyl cone.
Hence, the value of $\sigma_{zz}^{\mathrm{CA}}(B_z^2)$ coming from a single Dirac cone in the low temparature limit $T/\mu\to 0$ is obtained by setting $b=\Delta=0$ (i.e., $\mathcal{C}=1/2$) and multiplying a factor $2$ in Eq.~(\ref{sigma-zz-CA}):
\begin{align}
\sigma_{zz}^{\mathrm{CA}}(B_z^2)_{\mathrm{Dirac}}=\frac{e^2}{4\pi^2}\frac{(eB_z)^2v_F^3}{\mu^2}\tau_{\mathrm{Dirac}},
\end{align} 
where $\tau_{\mathrm{Dirac}}$ is the intervalley scattering time between the two Weyl cones that are degenerate in momentum space.
As discussed in Sec.~\ref{Sec-Scattering} the effective Hamiltonian around a Dirac point in Dirac semimetals which are protected by crystalline symmetry can be written in the block-diagonal form \cite{Yang2014}
\begin{align}
\mathcal{H}_{\mathrm{Dirac}}(\bm{q})=
\begin{bmatrix}
\mathcal{H}_{\mathrm{AA}} & \mathcal{H}_{\mathrm{AB}}\\
\mathcal{H}_{\mathrm{BA}} & \mathcal{H}_{\mathrm{BB}}
\end{bmatrix}
=
\begin{bmatrix}
\mathcal{H}_+(\bm{q}) & 0\\
0 & \mathcal{H}_-(\bm{q})
\end{bmatrix},
\label{Dirac-blockdiagonal2}
\end{align}
where $A$ and $B$ denote two states after a unitary transformation, and $\mathcal{H}_\pm(\bm{q})$ is a $2\times2$ Weyl Hamiltonian with chirality $\pm1$.
In general, atomically smooth disorder does not allow scattering processes between two different states $A$ and $B$.
Therefore, the intervalley scattering time $\tau_{\mathrm{Dirac}}$ is much larger than intravalley scattering time.

\subsubsection{Contribution from the Lorentz force}
Next, we compute the magnetoconductivity proportional to $B_z^2$ due to the Lorentz force.
Such a magnetoconductivity does not originate from the intrinsic (i.e., the Berry curvature) contribution but originates solely from the Fermi surface contribution.
Thus, only the diagonal components in the electric and magnetic driving terms are relevant to obtaining it.
We start by obtaining the diagonal part of the density matrix induced by the electric field.
It is easily seen that the integral of $D_E(\langle \rho_0\rangle)$ over the Fermi surface of a given Weyl cone is zero.
Then, from the discussion in Sec.~\ref{Sec-Scattering}, the diagonal density matrix obtained from $D_E(\langle \rho_0\rangle)$ contains the intravalley scattering time as $\langle n_{E}\rangle=\tau_{\mathrm{intra}}D_E(\langle \rho_0\rangle)$, where $n_{E\bm{k}}^+=eE_z\tau_{\mathrm{intra}}[\partial f_0(\varepsilon^+_{\bm{k}})/\partial k_z]$ and $n_{E\bm{k}}^-=0$ (since we consider the case of $\mu>0$).
From Eq.~(\ref{D_B-diagonal-component}), the diagonal components in $D_B(\langle n_E\rangle)$ are given by
\begin{align}
[D_B(\langle n_E\rangle)]^{++}_{\bm{k}}=eB_z\left(\frac{\partial \varepsilon_{\bm{k}}}{\partial k_y}\frac{\partial n^+_{E\bm{k}}}{\partial k_x}-\frac{\partial \varepsilon_{\bm{k}}}{\partial k_x}\frac{\partial n^+_{E\bm{k}}}{\partial k_y}\right)
\label{D_B-Lorentz}
\end{align}
and $[D_B(\langle n_E\rangle)]^{--}_{\bm{k}}=0$.
We see that Eq.~(\ref{D_B-Lorentz}) indeed represents the Lorentz force term, as $[D_B(\langle n_E\rangle)]^{++}_{\bm{k}}=e(\bm{v}_{\bm{k}}\times\bm{B})\cdot\nabla_{\bm{k}}n^+_{E\bm{k}}$ with $\bm{v}_{\bm{k}}=\nabla_{\bm{k}} \varepsilon_{\bm{k}}$.
Here, note that Eq.~(\ref{D_B-Lorentz}) can be rewritten as
\begin{align}
[D_B(\langle n_E\rangle)]^{++}_{\bm{k}}=e^2E_zB_z\tau_{\mathrm{intra}}\left(\frac{\partial \varepsilon_{\bm{k}}}{\partial k_y}\frac{\partial^2 \varepsilon_{\bm{k}}}{\partial k_x k_z}-\frac{\partial \varepsilon_{\bm{k}}}{\partial k_x}\frac{\partial^2 \varepsilon_{\bm{k}}}{\partial k_y k_z}\right)\frac{\partial f_0(\varepsilon_{\bm{k}})}{\partial \varepsilon_{\bm{k}}},
\label{D_B-Lorentz}
\end{align}
from which we see clearly that $[D_B(\langle n_E\rangle)]^{++}_{\bm{k}}=0$ for the present case of the isotropic energy band $\varepsilon_{\bm{k}}$ in the $(k_x, k_y)$ plane, $\varepsilon_{\bm{k}}=\sqrt{v_F^2k_x^2+v_F^2k_y^2+m_\pm(k_z)^2}$.
Namely, there are no contributions from the Lorentz force in the present case, as is also understood from the fact that the Lorentz force does not act when the velocity is along the magnetic-field direction.
However, when the energy band $\varepsilon_{\bm{k}}$ is anisotropic in the $(k_x, k_y)$ plane, we see that in general $[D_B(\langle n_E\rangle)]^{++}_{\bm{k}}\neq 0$.
See, for example, Ref.~\cite{Pal2010} for a detailed discussion on the conditions for nonzero contributions from the Lorentz force in the case of $\bm{E}\parallel \bm{B}$.

The Weyl cones in real materials will be anisotropic.
In the following, let us assume that there is an anisotropy in $\varepsilon_{\bm{k}}$, for example, by replacing $m_\pm(k_z)$ by $m_\pm(k_x, k_z)=b\pm\sqrt{v_F^2k_z^2+\alpha k_x^2+\Delta^2}$.
In such as case we have $[D_B(\langle n_E\rangle)]^{++}_{\bm{k}}\neq 0$.
Again we find that the integral of $D_B(\langle n_E\rangle)$ over the Fermi surface of a given Weyl cone is zero.
Then we see from Eq.~(\ref{n_B-Nth-order}) the diagonal density matrix obtained from $D_B(\langle n_E\rangle)$ contains the intravalley scattering time as $\langle n'_{EB}\rangle=\tau_{\mathrm{intra}}D_B(\langle n_E\rangle)$.
The same procedure can be applied to obtain the diagonal density matrix proportional to $E_zB_z^2$, and we obtain $\langle n_{EB^2}\rangle=\tau_{\mathrm{intra}}D_B(\langle n_{EB}\rangle)=\mathrm{diag}[(eB_z)^2\tau_{\mathrm{intra}}^2(\frac{\partial \varepsilon_{\bm{k}}}{\partial k_y}\frac{\partial}{\partial k_x}-\frac{\partial \varepsilon_{\bm{k}}}{\partial k_x}\frac{\partial}{\partial k_y})^2 n^+_{E\bm{k}},0]$.
Finally, noting that only the $t=-$ band is relevant, the magnetoconductivity proportional to $B_z^2$ induced by the Lorentz force at low temperatures such that $T\ll \mu$ is calculated to be
\begin{align}
\sigma_{zz}^{\mathrm{LF}}(B_z^2)=& -e^4B_z^2\tau_{\mathrm{intra}}^3\int\frac{d^3k}{(2\pi)^3}\, \frac{\partial m_-}{\partial k_z}\frac{m_-}{\varepsilon_{-\bm{k}}} \nonumber\\
&\times\left(\frac{\partial \varepsilon_{-\bm{k}}}{\partial k_y}\frac{\partial}{\partial k_x}-\frac{\partial \varepsilon_{-\bm{k}}}{\partial k_x}\frac{\partial}{\partial k_y}\right)^2\frac{\partial f_0(\varepsilon_{-\bm{k}})}{\partial k_z} \nonumber\\
\sim & -\sigma_{zz}^0\left(\omega_c\tau_{\mathrm{intra}}\right)^2,
\label{Conductivity-LF}
\end{align}
which takes a negative value.
Here, $\sigma_{zz}^0\approx e^2(\mu^2/v_F)\tau_{\mathrm{intra}}$ is the Drude conductivity, and $\omega_c=eB_z v_F^2/\mu$ is the cyclotron frequency.

\subsubsection{Total magnetoconductivity of the system}
We have seen that the contribution from the Lorentz force is absent when the Weyl cones are isotropic.
In this case the total quadratic magnetoconductivity for parallel electric and magnetic fields is written in the low-field limit as
\begin{align}
\sigma_{zz}(B_z^2)=\sigma_{zz}^{\mathrm{CA}}(B_z^2)>0.
\end{align}

On the other hand, when the Weyl cones are anisotropic, the contribution from the Lorentz force is present.
In this case, from Eqs.~(\ref{sigma-zz-CA}) and (\ref{Conductivity-LF}), the total quadratic magnetoconductivity for parallel electric and magnetic fields is written in the low-field limit as $\sigma_{zz}(B_z^2)=\sigma_{zz}^{\mathrm{CA}}(B_z^2)+\sigma_{zz}^{\mathrm{LF}}(B_z^2)$.
The ratio of these two contributions to the magnetoconductivity is estimated as
\begin{align}
\left|\frac{\sigma_{zz}^{\mathrm{LF}}}{\sigma_{zz}^{\mathrm{CA}}}\right|\sim\frac{\tau_{\mathrm{intra}}}{\tau}(\mu\tau_{\mathrm{intra}})^2.
\end{align}
In general, the intervalley scattering time $\tau$ is much larger than the intravalley scattering time $\tau_{\mathrm{intra}}$, i.e., $\tau_{\mathrm{intra}}/\tau\ll 1$, since the intervalley scatterings require large momentum transfers, i.e., the number of intervalley scattering processes is much smaller than that of intravalley scattering processes.
On the other hand, in usual high-mobility semiconductors the condition $(\mu\tau_{\mathrm{intra}})^2\gg 1$ is satisfied.
When the contribution from the chiral anomaly dominates that from the Lorentz force, we have the total magnetoconductivity in the low-field limit as
\begin{align}
\sigma_{zz}(B_z^2)\approx\sigma_{zz}^{\mathrm{CA}}(B_z^2)>0.
\end{align}

We have confirmed that the magnetoconductivity linear in $B$ vanishes in our formalism.
Namely, the magnetoconductivity that has the lowest $B$-dependence is the one proportional to $B^2$, which is in agreement with experimental results on the magnetoconductivity in the low-field limit in the Dirac semimetal Cd$_3$As$_2$ \cite{Li2016}.
On other hand, note that in this study we have not considered the contribution from the weak antilocalization such that $\sim a\sqrt{B}$ with $a<0$ in the low-field limit which is observed in the Weyl semimetal TaAs \cite{Huang2015}.

\subsection{Quadratic magnetoconductivity for $\vec{E}\perp \vec{B}$}
Let us consider the case of perpendicular electric and magnetic fields as $\bm{E}=(0,0,E_z)$ and $\bm{B}=(B_x,0,0)$.
Even in this case, the intrinsic contribution to the magnetoconductivity proportional to $B_x^2$, which resembles that from the chiral anomaly (\ref{sigma-zz-CA}), takes nonzero value in our formalism.
However, it is not characterized by the intervalley scattering time $\tau$ which appeared in the case of parallel fields, but characterized by the intravalley scattering time $\tau_{\mathrm{intra}}$.
We show that as the result the contribution from the Lorentz force dominates the intrinsic contribution, which is consistent with experimental results.

We compute the intrinsic contribution to the quadratic magnetoconductivity.
The exactly same calculation as in the case of parallel electric and magnetic fields can be applied here.
After a calculation, we obtain the magnetic driving term which is proportional to $E_zB_x$ as
\begin{align}
D_B(\langle S_E\rangle)
&=e^2E_zB_x
\mathcal{G}_{\bm{k}}\bm{1}
\label{D_B-S_E-perp}
\end{align}
with
\begin{align}
\mathcal{G}_{\bm{k}}=v_F\cos\theta\frac{\partial}{\partial k_z}\left\{\bigl[f_0(\varepsilon^+_{\bm{k}})-f_0(\varepsilon^-_{\bm{k}})\bigr]\frac{v_F k_\perp}{4\varepsilon_{\bm{k}}^3}\frac{\partial m}{\partial k_z}\right\}.
\label{G_k}
\end{align}
We also have the electric driving term which is proportional to $E_zB_x$ as
\begin{align}
D_E(\langle \xi_B\rangle)=e^2 E_z B_x\frac{\partial}{\partial k_z}
\begin{bmatrix}
f_0(\varepsilon^+_{\bm{k}})\Omega^+_{\bm{k},x} && 0\\
0 && f_0(\varepsilon^-_{\bm{k}})\Omega^-_{\bm{k},x}
\end{bmatrix}
+\mathcal{L}\, \langle S_{EB}\rangle,
\label{D_E-xi_B-perp}
\end{align}
where $\langle S_{EB}\rangle$ is a purely off-diagonal matrix.
Here, we show that these $D_B(\langle S_E\rangle)$ and $D_E(\langle \xi_B\rangle)$ have the opposite property of the case of parallel electric and magnetic fields.
Around a Weyl node with momentum $\bm{q}=\bm{k}-W_\pm$ ($\bm{q}^2\ll 1$), we find that $\mathcal{G}_{\bm{q}}$ and $\partial f_0(\varepsilon^\pm_{\bm{q}})\Omega^\pm_{\bm{q},x}/\partial q_z$ are odd functions of $q_x$ and $q_z$.
Accordingly, we find that the integral of $D_B(\langle  S_E\rangle)+D_E(\langle \xi_B\rangle)$ over the Fermi surface of a given valley is zero: $\int_{\mathrm{FS}} \frac{d^3q}{(2\pi)^3}\, \sum_m[D_B(\langle  S_E\rangle)+D_E(\langle \xi_B\rangle)]^{mm}_{\bm{q}}= 0$.
As discussed in Sec.~\ref{Sec-Scattering}, this is a consequence of the total particle number conservation in a given Weyl cone.
Namely we obtain
\begin{align}
\frac{\partial N}{\partial t}=\frac{e^2 E_z B_x}{4\pi^2}\int_{\mathrm{FS}} \frac{d^3q}{2\pi}\left(2\, \mathcal{G}_{\bm{q}}+\sum_{m=\pm}\frac{\partial f_0(\varepsilon^m_{\bm{q}})}{\partial q_z}\Omega^m_{\bm{q},x}\right)=0.
\end{align}
Then it follows from Eq.~(\ref{n_B-Nth-order}) that the intravalley scattering time $\tau_{\mathrm{intra}}$ appears when $\mathcal{L}^{-1}$ acts on $D_B(\langle S_E\rangle)$ and $D_E(\langle \xi_B\rangle)$.
In the end, the diagonal density matrix $\langle n_{EB}\rangle$ is obtained as
\begin{align}
\langle n_{EB}\rangle&=\mathcal{L}^{-1}\left[D_B(\langle S_E\rangle)+D_E(\langle \xi_B\rangle)\right] \nonumber\\
&=e^2 E_z B_x\tau_{\mathrm{intra}}
\begin{bmatrix}
\tilde{\mathcal{G}}_{\bm{k}}^{++} && 0\\
0 && \tilde{\mathcal{G}}_{\bm{k}}^{--}
\end{bmatrix},
\end{align}
where
\begin{align}
\tilde{\mathcal{G}}_{\bm{k}}^{mm}=\frac{1}{2}\Omega_{\bm{k},x}^-\frac{\partial}{\partial k_z}[f_0(\varepsilon^+_{\bm{k}})-f_0(\varepsilon^-_{\bm{k}})]+\frac{\partial f_0(\varepsilon^m_{\bm{k}})}{\partial k_z}\Omega^m_{\bm{k},x}
\label{F_k-tilde}
\end{align}
is the component which represents the Fermi surface response in Eqs.~(\ref{D_B-S_E-perp}) and (\ref{D_E-xi_B-perp}).
The off-diagonal part $\langle S_{EB^2}\rangle$ and diagonal part $\langle \xi_{EB^2}\rangle$ of the density matrix proportional to $E_zB_x^2$ can be obtained in a similar way as Eqs.~(\ref{step3-explicit}) and (\ref{step3-explicit-xi}), respectively.

Finally, noting that only the $t=-$ band is relevant, the $zz$-component of the quadratic magnetoconductivity from the intrinsic contribution at low temperatures such that $T\ll \mu$ is obtained as
\begin{align}
\sigma_{zz}^{\mathrm{In}}(B_x^2)=&\frac{e^4B_x^2\tau_{\mathrm{intra}}}{2}\int\frac{d^3k}{(2\pi)^3}\frac{\partial m_-}{\partial k_z}\frac{v_F^2 k_x}{\varepsilon_{-\bm{k}}^2}\frac{\partial (\tilde{\mathcal{G}}^{++}_{\bm{k}}+\tilde{\mathcal{G}}^{--}_{\bm{k}})}{\partial k_z} \nonumber\\
& +\frac{e^4B_x^2\tau_{\mathrm{intra}}}{2}\int\frac{d^3k}{(2\pi)^3}\frac{\partial m_-}{\partial k_z}\frac{v_F^3 k_x m_-}{\varepsilon_{-\bm{k}}^4}(\tilde{\mathcal{G}}^{++}_{\bm{k}}+\tilde{\mathcal{G}}^{--}_{\bm{k}}) \nonumber\\
\equiv&\frac{e^2}{4\pi^2}\mathcal{C}'(b, \Delta, T)\frac{(eB_z)^2v_F^3}{\mu^2}\tau_{\mathrm{intra}},
\label{Conductivity-In}
\end{align}
which takes a positive value with $\mathcal{C}'(b, \Delta, T)\sim\mathcal{O}(0.1)$.
As mentioned above, this form resembles that from the chiral anomaly (\ref{sigma-zz-CA}).
However, the appearance of the intravalley scattering time indicates that the total particle number in a given valley is conserved, while the chiral anomaly requires nonconservation of the total particle number in a given valley.
Thus, the magnetoconductivity (\ref{Conductivity-In}) does not originate from the chiral anomaly.

The quadratic magnetoconductivity due to the Lorentz force in the case of perpendicular electric and magnetic fields can be evaluated in the same way as Eq.~(\ref{Conductivity-LF}).
At low temperatures such that $T\ll \mu$ we have
\begin{align}
\sigma_{zz}^{\mathrm{LF}}(B_x^2)=& -e^4B_x^2\tau_{\mathrm{intra}}^3\int\frac{d^3k}{(2\pi)^3}\, \frac{\partial m_-}{\partial k_z}\frac{m_-}{\varepsilon_{-\bm{k}}} \nonumber\\
&\times\left(\frac{\partial \varepsilon_{-\bm{k}}}{\partial k_z}\frac{\partial}{\partial k_y}-\frac{\partial \varepsilon_{-\bm{k}}}{\partial k_y}\frac{\partial}{\partial k_z}\right)^2\frac{\partial f_0(\varepsilon_{-\bm{k}})}{\partial k_z} \nonumber\\
\approx & -\sigma_{zz}^0\left(\omega_c\tau_{\mathrm{intra}}\right)^2,
\label{Conductivity-LF-perp}
\end{align}
which takes a negative value.
Here, $\sigma_{zz}^0\approx e^2(\mu^2/v_F)\tau_{\mathrm{intra}}$ is the Drude conductivity, and $\omega_c=eB_x v_F^2/\mu$ is the cyclotron frequency.
Note that, in contrast to the case of parallel electric and magnetic fields, $\sigma_{zz}^{\mathrm{LF}}(B_x^2)$ is nonzero even when the Weyl cones are isotropic.

Therefore from Eqs.~(\ref{Conductivity-In}) and (\ref{Conductivity-LF-perp}), the total quadratic magnetoconductivity for perpendicular electric and magnetic fields is written in the low-field limit as $\sigma_{zz}(B_x^2)=\sigma_{zz}^{\mathrm{In}}(B_x^2)+\sigma_{zz}^{\mathrm{LF}}(B_x^2)$.
The ratio of these two contributions to the magnetoconductivity is estimated as
\begin{align}
\left|\sigma_{zz}^{\mathrm{LF}}/\sigma_{zz}^{\mathrm{In}}\right|\sim(\mu\tau_{\mathrm{intra}})^2\gg 1,
\end{align}
where we have used that in usual high-mobility semiconductors the condition $(\mu\tau_{\mathrm{intra}})^2\gg 1$ is satisfied.
Thus, we have the total magnetoconductivity in the low-field limit as
\begin{align}
\sigma_{zz}(B_x^2)\approx\sigma_{zz}^{\mathrm{LF}}(B_x^2)<0,
\end{align}
which is in agreement with experimental results in Dirac and Weyl semimetals \cite{Xiong2015,Li2015,Li2016,Li2016a,Huang2015,Arnold2016}.
As in the case of parallel electric and magnetic fields, we have confirmed that the magnetoconductivity linear in $B$ vanishes in our formalism.

\section{Discussions \label{Sec-Discussion}}
We have derived a multi-band quantum kinetic equation for the Bloch-state density matrix 
of a crystal that includes driving terms 
related to both electric and magnetic fields.   The magnetic driving term $D_B$ is new, as 
far as we are aware, and has a simple form that elegantly generalizes the 
well known Lorentz force driving term of the scalar single-band Boltzmann quantum kinetic equation,
$-e(\nabla_{\bm{k}}\varepsilon_{\bm{k}}\times\bm{B})\cdot\nabla_{\bm{k}}f_{\bm{k}}$,  to
a multi-band matrix form.  We have also shown that our
quantum kinetic equation captures two effects that have been previously identified by examining Bloch-state wavepacket dynamics in multi-band systems, namely an anomalous contribution to velocity 
that is proportional to $\dot{\bm{k}} \times \bm{\Omega}$ and a relative change in the momentum-space 
density of states associated with a particular band that is proportional 
to $ \bm{B} \cdot \bm{\Omega}$, where $\bm{B}$ is an external magnetic field and 
$\bm{\Omega}$ is the momentum-space Berry phase curvature associated with the band. 
In transport the former effect is responsible for the intrinsic anomalous Hall effect of magnetic crystals.  
When disorder scattering is neglected \cite{Son2013}, the two effects combine to yield a 
remarkable condensed matter realization of the chiral anomaly in which charge is pumped between 
Fermi surfaces surrounding different Weyl points.  As an illustration of the physics that can be captured using 
our quantum kinetic equation we have examined how the chiral anomaly is observably manifested in the 
magnetoconductance of Weyl semimetals, discovering a complex interplay between 
electric and magnetic field driving terms, free-particle dynamics, and scattering contributions to the 
quantum kinetic equations.  We have found that the charge pumping effect 
survives only when intervalley scattering is very weak compared to intravalley scattering.  
 
The free-evolution ($P$), scattering ($K$), and electric and magnetic driving 
term ($D_E$ and $D_B$) contributions to the quantum kinetic equation can all be viewed as operators that 
act on the Bloch-state density matrices that characterize the transport steady state.
Only band-off-diagonal elements of this density matrix have time dependence 
in the absence of external fields and disorder, i.e., the projection of 
$P$ onto the band-diagonal part of the Liouville density-matrix vector space is zero.
For this reason, the roles of diagonal and off-diagonal 
density-matrix components in the quantum kinetic equations are quite distinct. 
The magnetic driving term maps the equilibrium density 
matrix $\langle\rho_0\rangle$ (i.e., the Fermi-Dirac distribution function),
which is diagonal in a band-eigenstate representation, to a purely band  
off-diagonal density matrix. 
In other words, $D_B\left(\langle \rho_{0}\rangle\right)$ is a purely band off-diagonal matrix.
On the other hand, the matrix $P^{-1}D_B\left(\langle \rho_{0}\rangle\right)$ has diagonal components 
$\langle\xi_B\rangle$.  Similar diagonal components are obtained whenever 
$P^{-1}D_B$ acts on any band-diagonal density matrix components.  These magnetic-field induced changes in the occupation probabilities of Bloch states capture the 
Berry-phase related density-of-states changes in magnetic field of semiclassical 
wavepacket dynamics.  
As is shown in Sec.~\ref{Sec-CME}, $\langle S_B\rangle$ and $\langle\xi_B\rangle$ 
combine to produce a magnetic-field driven current when the local chemical potentials of
Fermi surface pockets near different Weyl points are different, thus capturing what is 
referred to as the chiral magnetic effect.  
As shown in Sec.~\ref{Subsec-quadratic-MC}, the density matrices $\langle S_{EB^2}\rangle$ and 
$\langle\xi_{EB^2}\rangle$, which result in the positive quadratic magnetoconductivity in Weyl and Dirac metals,
are closely related $\langle S_B\rangle$ and $\langle\xi_B\rangle$.  In the limit of extremely weak 
intervalley scattering, the chemical potential difference of the chiral magnetic effect is 
established by balancing charge pumping between valleys and intervalley relaxation.  
Although our calculations show that the precise numerical value of the positive quadratic magnetoconductivity
can be quite complex and depend subtly on the details of specific systems, it is still true,
as explained in earlier theoretical work \cite{Son2013,Burkov2014,Burkov2015a,Spivak2016,Li2016a},
that the positive quadratic magnetoconductivity is basically
a simple consequence of the chiral magnetic effect.
It is important to emphasize that the anomalous Hall effect and 
the chiral magnetic effect arise essentially from band off-diagonal parts of the electric and 
magnetic driving terms, respectively.
This indicates that transport phenomena induced by the 
chiral anomaly are an important example of interband coherence response
in conductors.

This study is an extension of 
Ref.~\cite{Culcer2016} which developed a quantum kinetic theory for 
crystals that is sensitive to recent developments in electronic structure theory that allow
accurate Wannier representations of the Bloch-state Hamiltonian which are 
convenient for transport theory to be constructed, and also sensitive to recent awareness 
of the importance of momentum-space Berry phase effects in a number of different contexts.  
The aim of the theory is to be able to account for intrinsic effect related to Bloch-state 
wave function properties and extrinsic effects related to disorder scattering to be accounted 
for on a consistent footing in real materials with complicated electronic structure.  
The quantum kinetic theory in Ref.~\cite{Culcer2016}, which is the $\bm{B}=0$ limit of the 
theory presented in this paper, conveniently captures effects that appear 
in Kubo formula formulations of transport theory with disorder treated as a perturbation,
as ladder-diagram vertex corrections, for example the well known
absence of a spin Hall conductivity \cite{Inoue2004} and 
anomalous Hall conductivity \cite{Inoue2006} in certain Rashba models.
An important aspect of the detailed calculations we have presented for the 
simplified Weyl semimetal toy model, is its demonstration that 
the corresponding vertex corrections are {\it absent} in the anomalous Hall conductivity (\ref{AHE-total}) and positive quadratic magnetoconductivity (\ref{sigma-zz-CA}) of Weyl metals, 
as long as the chemical potential lies sufficiently close to the Weyl points.

We briefly discuss other possible applications of our theory.
In this connection, we emphasize that although our theory is formulated as an independent 
particle theory, we really intend it as a theory of independent quasiparticles in a 
mean-field approximation to an interacting electron theory.  For example the 
mean-field quasiparticles can be viewed as the Kohn-Sham quasiparticles of 
density functional theory.   
Because our theory calculates the density-matrix response to electric and 
magnetic fields, it can be used to determine the response of any crystal observables,
including observables like the spin density that can feed back into the 
steady state crystal Hamiltonian.  
All single-particle observables $\mathcal{O}$ maintain their crystal periodicity 
when they respond to spatially constant electric and magnetic fields
and therefore have expectation values of the form
\begin{align}
\langle \mathcal{O}\rangle=\mathrm{Tr}\left[\mathcal{O}\langle\rho\rangle\right],
\end{align}
where $\langle\rho\rangle$ is a density matrix we have calculated.
The evaluation of field-induced spin currents and spin densities, which are 
related to the current-induced torques of spintronics is one practically important
problem to which our quantum kinetic theory can be applied. 
Our theory can be flexibly applied to metals, insulators,
nodal-line semimetals, and to systems with many other types of electronic structure.
It can be applied to toy models that capture the essence of different phenomena or to 
realistic models of specific materials.   
We anticipate, for example, that it will have interesting implications for the properties of 
2D multivalley systems such as graphene and transition metal dichalcogenides.
The kinetic equation approach can also be applicable to interacting systems 
that are described at a mean-field theory level \cite{Culcer2011,Culcer2013,Culcer2016a}.
It will be interesting to investigate from the quantum kinetic theory viewpoint the interplay between nontrivial band topology and interactions in the presence of electric and magnetic fields.

\section{Summary}
In summary, we have developed a general quantum kinetic theory of low-field magnetotransport
 in weakly disordered multiband systems.
Our theory naturally incorporates momentum-space Berry phase effects, 
which are often discussed in the context of semiclassical wavepacket dynamics, into transport theory.
By applying the Wigner transformation to the quantum Liouville equation, we have derived a quantum kinetic equation~(\ref{full-kinetic-equation}), which is the principle result of this study.
From this equation we have obtained a generic expression for the magnetoconductivity that is applicable 
for arbitrary angle between electric and magnetic fields.
We note that the purely band-diagonal contributions in our theory describe all the regular Fermi-surface dominated effects that are familiar from textbooks.
Our theory is able to simply explain and predict when large momentum-space Berry curvatures yield important 
corrections to the standard Boltzmann theory of Fermi-surface dominated magnetotransport.

We have applied our theory to study transport phenomena induced by the 
chiral anomaly in a toy model of two-node Weyl metals (\ref{WSM-H_eff}).
We have shown that the anomalous Hall effect in the model is purely intrinsic, which means that the vertex 
correction in the ladder-diagram approximation is absent.
We have also shown, by constructing the linear response theory to a magnetic field in the absence of electric fields, that the magnetic driving term we have newly introduced describes properly the chiral magnetic effect in a 
continuum model of Weyl metals.
We have obtained an explicit expression for the positive quadratic magnetoconductivity that includes the intervalley scattering time in parallel electric and magnetic fields.
In the process of obtaining this expression, we have shown that the vertex correction in the ladder-diagram approximation is absent.
Our study indicates that the positive quadratic magnetoconductivity is a consequence of the chiral magnetic effect.
On the other hand, in the case of perpendicular electric and magnetic fields, we have obtained a negative quadratic magnetoconductivity due to the Lorentz force that includes the intravalley scattering time, which is in agreement with experimental results.
We have clarified that the chiral anomaly is observable only when intervalley scattering at the Fermi energy is very weak compared to intravalley scattering.

\acknowledgements
Work at Austin was supported by the Department of Energy, Office of Basic Energy Sciences under Contract No.~DE-FG02-ER45958 and by the Welch foundation under Grant No.~TBF1473.
A.S. is supported by the JSPS Overseas Research Fellowship.
D.C. is supported by the Australian Research Council Centre of Excellence in Future Low-Energy Electronics Technologies (Project No.~CE170100039).
A.H.M. acknowledges stimulating conversations with Anton Burkov and Boris Spivak.

\appendix
\begin{widetext}
\section{General properties of the magnetic driving term $D_B$  \label{Appendix-D_B-1}}
In this Appendix, we consider general properties of the magnetic driving term (\ref{driving-term-B}).
We write the Hamiltonian and the density matrix of a system as $\mathcal{H}_0=\sum_{m}\varepsilon_m |m\rangle\langle m|$ and $\langle\rho_0\rangle=\sum_m f_{0m} |m\rangle\langle m|$ with $m$ a band index and $f_{0m}$ a Fermi-Dirac distribution function, where we have omitted the wavevector dependences in these equations to simplify the notation.
For concreteness, we consider the case of $\bm{B}=(0,0,B_z)$.
Then the driving term (\ref{driving-term-B}) is written as
\begin{align}
D_B(\langle \rho_0\rangle)=\frac{e}{2\hbar^2}\left\{\left(\frac{D \mathcal{H}_0}{D\bm{k}}\times\bm{B}\right)\cdot\frac{D\langle \rho_0\rangle}{D\bm{k}}\right\}=\frac{eB_z}{2\hbar^2}\left[\left\{\frac{D\mathcal{H}_0}{Dk_y},\frac{D\langle \rho_0\rangle}{Dk_x}\right\}-\left\{\frac{D\mathcal{H}_0}{Dk_x},\frac{D\langle \rho_0\rangle}{Dk_y}\right\}\right].
\end{align}

First, we consider a diagonal component of $D_B(\langle \rho_0\rangle)$, i.e., $\langle m|D_B(\langle \rho_0\rangle)|m\rangle$.
We immediately get
\begin{align}
\frac{D\mathcal{H}_0}{Dk_y}&=\sum_{m'}\partial_y\varepsilon_{m'}|m'\rangle\langle m'|+\sum_{m'}\varepsilon_{m'}\bigl[|\partial_y m'\rangle\langle m'|+|m'\rangle\langle \partial_y m'|\bigr], \nonumber\\
\frac{D\langle \rho_0\rangle}{Dk_x}&=\sum_{n'}\partial_x f_{0n'}|n'\rangle\langle n'|+\sum_{n'}f_{0n'}\bigl[|\partial_x n'\rangle\langle n'|+|n'\rangle\langle \partial_x n'|\bigr],
\label{appendix-D_B-element1}
\end{align}
where $\partial_a=\partial/\partial k_a$.
Then we have
\begin{align}
\frac{D\mathcal{H}_0}{Dk_y}\frac{D\langle \rho_0\rangle}{Dk_x}=&\sum_{m'}\partial_y\varepsilon_{m'}\partial_x f_{0m'}|m'\rangle\langle m'|+\sum_{m' n'}\partial_y\varepsilon_{m'}f_{0n'}|m'\rangle\langle m'|\partial_x n'\rangle\langle n'|+\sum_{m'}\partial_y\varepsilon_{m'}f_{0m'}|m'\rangle\langle \partial_x m'| \nonumber\\
&+\sum_{m'}\varepsilon_{m'}\partial_x f_{0m'}|\partial_y m'\rangle\langle m'|+\sum_{m' n'}\varepsilon_{m'}f_{0n'}|\partial_y m'\rangle\langle m'|\partial_x n'\rangle\langle n'|+\sum_{m'}\varepsilon_{m'}f_{0m'}|\partial_y m'\rangle\langle \partial_x m'| \nonumber\\
&+\sum_{m' n'}\varepsilon_{m'}\partial_x f_{0n'}|m'\rangle\langle \partial_y m'|n'\rangle\langle n'|+\sum_{m' n'}\varepsilon_{m'}f_{0n'}|m'\rangle\langle \partial_y m'|\partial_x n'\rangle\langle n'|+\sum_{m' n'}\varepsilon_{m'}f_{0n'}|m'\rangle\langle \partial_y m'|n'\rangle\langle \partial_x n'|,
\label{appendix-D_B-element2}
\end{align}
from which a diagonal component is obtained as
\begin{align}
\langle m|\frac{D\mathcal{H}_0}{Dk_y}\frac{D\langle \rho_0\rangle}{Dk_x}|m\rangle=&\partial_y\varepsilon_{m}\partial_x f_{0m}+\sum_{m'}\varepsilon_{m'}f_{0m}\langle m|\partial_y m'\rangle\langle m'|\partial_x m\rangle+\sum_{m'}\varepsilon_{m'}f_{0m'}\langle m|\partial_y m'\rangle\langle \partial_x m'| m\rangle \nonumber\\
&+\varepsilon_{m}f_{0m}\langle \partial_y m|\partial_x m\rangle+\sum_{n'}\varepsilon_{m}f_{0n'}\langle \partial_y m|n'\rangle\langle \partial_x n'| m\rangle,
\end{align}
where we have used $\langle m'|\partial_a n'\rangle+\langle\partial_a  m'|n'\rangle=\partial_a (\delta_{m'n'})=0$.
Similarly we have
\begin{align}
\langle m|\frac{D\langle \rho_0\rangle}{Dk_x}\frac{D\mathcal{H}_0}{Dk_y}|m\rangle=&\partial_y\varepsilon_{m}\partial_x f_{0m}+\sum_{n'}f_{0n'}\varepsilon_{m}\langle m|\partial_x n'\rangle\langle n'|\partial_y m\rangle+\sum_{n'}f_{0n'}\varepsilon_{n'}\langle m|\partial_x n'\rangle\langle \partial_y n'| m\rangle \nonumber\\
&+f_{0m}\varepsilon_{m}\langle \partial_x m|\partial_y m\rangle+\sum_{m'}f_{0m}\varepsilon_{m'}\langle \partial_x m|m'\rangle\langle \partial_y m'| m\rangle.
\end{align}
Then we see that
\begin{align}
\langle m|D_B(\langle \rho_0\rangle)|m\rangle=eB_z(\partial_y\varepsilon_{m}\partial_x f_{0m}-\partial_x\varepsilon_{m}\partial_y f_{0m})=0.
\label{diagonal-element-D_B-rho_0}
\end{align}

Next, we consider a diagonal component of $P^{-1}D_B(\langle \rho_0\rangle)$, i.e., $\langle m|P^{-1}D_B(\langle \rho_0\rangle)|m\rangle$.
Here, the matrix $P$ is defined by $P\langle\rho\rangle \equiv \frac{i}{\hbar}[\mathcal{H}_0,\langle \rho\rangle]$, i.e., $P$
is the precession term that accounts for the time dependence of the density matrix in the absence of fields and  
disorder.
Recall from Eq.~(\ref{rho_E-off-diagonal}) that when the matrix $P^{-1}$ acts on a 
driving term $D$, it simply multiplies $D$ by a numerical factor:
\begin{equation}
P^{-1}D=-i\hbar\frac{D^{mn}_{\bm k}}{\varepsilon^{m}_{\bm k} - \varepsilon^{n}_{\bm k}}.
\end{equation}
Let us take a closer look at the second term in the third line of Eq. (\ref{appendix-D_B-element2}).
Noticing the fact that $\partial_a(\sum_{m'}|m'\rangle\langle m'|)=0$, we can replace $\varepsilon_{m'}$ by $(\varepsilon_{m'}-\varepsilon_{n'})$ in Eq. (\ref{appendix-D_B-element1}).
Then we can rewrite it as
\begin{align}
\frac{D\mathcal{H}_0}{Dk_y}\frac{D\langle \rho_0\rangle}{Dk_x}=\sum_{m' n'}(\varepsilon_{m'}-\varepsilon_{n'})f_{0n'}|m'\rangle\langle \partial_y m'|\partial_x n'\rangle\langle n'|,
\label{appendix-D_B-element3}
\end{align}
from which we have
\begin{align}
P^{-1}\frac{D\mathcal{H}_0}{Dk_y}\frac{D\langle \rho_0\rangle}{Dk_x}=-i\hbar\sum_{m' n'}\frac{\varepsilon_{m'}-\varepsilon_{n'}}{\varepsilon_{m'}-\varepsilon_{n'}}f_{0n'}|m'\rangle\langle \partial_y m'|\partial_x n'\rangle\langle n'|=-i\hbar\sum_{m' n'}f_{0n'}|m'\rangle\langle \partial_y m'|\partial_x n'\rangle\langle n'|.
\end{align}
Then a diagonal element of this matrix is obtained as
$\langle m| P^{-1}\frac{D\mathcal{H}_0}{Dk_y}\frac{D\langle \rho_0\rangle}{Dk_x}| m\rangle=-i\hbar f_{0m}\langle \partial_y m|\partial_x m\rangle$.
Similarly we have $\langle m| P^{-1}\frac{D\langle \rho_0\rangle}{Dk_x}\frac{D\mathcal{H}_0}{Dk_y}| m\rangle=+i\hbar f_{0m}\langle \partial_x m|\partial_y m\rangle$.
In the end, we see that
\begin{align}
\langle m| P^{-1}D_B(\langle \rho_0\rangle)| m\rangle=(e/\hbar)f_{0m}B_z\Omega^m_z,
\end{align}
where $\Omega^m_z=i\langle \partial_x m|\partial_y m\rangle-i\langle \partial_y m|\partial_x m\rangle$ is the $z$ component of the Berry curvature of band $m$.
Note that only the terms of the form of (\ref{appendix-D_B-element3}) give rise to nonzero contributions to the diagonal component of $P^{-1}D_B(\langle \rho_0\rangle)$.

\section{General expression for $\mathrm{Tr}\, [D_B(\langle S_E\rangle)]$ in parallel electric and magnetic fields \label{dN/dt-analytical}}
In this Appendix, we derive an explicit expression for the diagonal matrix element $[D_B(\langle S_E\rangle)]^{mm}$ in a general system with $\mathcal{H}_0=\sum_{m}\varepsilon_m |m\rangle\langle m|$ and $\langle\rho_0\rangle=\sum_m f_{0m} |m\rangle\langle m|$ with $m$ a band index and $f_{0m}$ a Fermi-Dirac distribution function, where we have omitted the wavevector $\bm{k}$ dependences in these equations to simplify the notation.
For concreteness, we consider the case of $\bm{E}=(0,0,E_z)$ and $\bm{B}=(0,0,B_z)$.
An off-diagonal component of the electric driving term (\ref{driving-term-E}) reads
\begin{align}
\langle n|D_E(\langle\rho_0\rangle)|n'\rangle=eE_z\sum_{m'}f_{0m'}\langle n|\bigl[ |\partial_z m'\rangle\langle m'|+|m'\rangle\langle\partial_z  m'|\bigr]|n'\rangle=eE_z(f_{0n'}-f_{0n})\langle n|\partial_z n'\rangle
\end{align}
with $\partial_a=\partial/\partial k_a$, from which we obtain the off-diagonal part of the density matrix induced by the electric field
\begin{align}
\langle S_E\rangle=(-i)eE_z\sum_{nn'}\frac{f_{0n'}-f_{0n}}{\varepsilon_n-\varepsilon_{n'}}|n\rangle\langle n|\partial_z n'\rangle\langle n'|,
\end{align}
where $n\neq n'$. On the other hand, the magnetic driving term (\ref{driving-term-B}) is written as
\begin{align}
D_B(\langle S_E\rangle)=\frac{e}{2}\left\{\left(\frac{D \mathcal{H}_0}{D\bm{k}}\times\bm{B}\right)\cdot\frac{D\langle S_E\rangle}{D\bm{k}}\right\}=\frac{eB_z}{2}\left[\left\{\frac{D\mathcal{H}_0}{Dk_y},\frac{D\langle S_E\rangle}{Dk_x}\right\}-\left\{\frac{D\mathcal{H}_0}{Dk_x},\frac{D\langle S_E\rangle}{Dk_y}\right\}\right].
\end{align}
Since we are focusing on the Fermi surface response, we consider only the terms proportional to $\partial_x f_{0}$ in $D\langle S_E\rangle/Dk_x$:
\begin{align}
\frac{D\langle S_E\rangle}{Dk_x}=(-i)eE_z\sum_{nn'}\frac{\partial_x f_{0n'}-\partial_x f_{0n}}{\varepsilon_n-\varepsilon_{n'}}|n\rangle\langle n|\partial_z n'\rangle\langle n'|.
\end{align}
We also have the relevant term
\begin{align}
\frac{D\mathcal{H}_0}{Dk_y}=\sum_{m'}(\varepsilon_{m'}-\varepsilon_{n'})\bigl[|\partial_y m'\rangle\langle m'|+|m'\rangle\langle \partial_y m'|\bigr],
\end{align}
where we have used $\partial_a(\sum_{m'}|m'\rangle\langle m'|)=0$.
Note that the terms proportional to $\partial_y\varepsilon$ in $D\mathcal{H}_0/Dk_y$ do not contribute to the diagonal part of $D_B(\langle S_E\rangle)$.
Then we obtain
\begin{align}
\langle m|\frac{D\mathcal{H}_0}{Dk_y}\frac{D\langle S_E\rangle}{Dk_x}|m\rangle&=(-i)eE_z\sum_{nn'}\sum_{m'}(\varepsilon_{m'}-\varepsilon_{n'})\frac{\partial_x f_{0n'}-\partial_x f_{0n}}{\varepsilon_n-\varepsilon_{n'}}\delta_{n'm}\bigl[\delta_{m'n}\langle m|\partial_y n\rangle\langle n|\partial_z m\rangle+\delta_{mm'}\langle \partial_y m|n\rangle\langle n|\partial_z m\rangle\bigr] \nonumber\\
&=(-i)eE_z\sum_{n}(\partial_x f_{0m}-\partial_x f_{0n})\langle m|\partial_y n\rangle\langle n|\partial_z m\rangle \nonumber\\
&=ieE_z\partial_x f_{0m}\langle \partial_y m|\partial_z m\rangle-ieE_z\sum_n\partial_x f_{0n}\langle\partial_z  n|m\rangle\langle m|\partial_y n\rangle.
\end{align}
Similarly we have
\begin{align}
\langle m|\frac{D\langle S_E\rangle}{Dk_x}\frac{D\mathcal{H}_0}{Dk_y}|m\rangle=-ieE_z\partial_x f_{0m}\langle \partial_z m|\partial_y m\rangle+ieE_z\sum_n\partial_x f_{0n}\langle\partial_y  n|m\rangle\langle m|\partial_z n\rangle.
\end{align}
Finally we arrive at a general expression for the rate of pumping from the Fermi surface contribution:
\begin{align}
\frac{\partial N}{\partial t}=\mathrm{Tr}\left[D_B(\langle S_E\rangle)\right]=e^2E_zB_z\sum_{m,\bm{k}}\left[\partial_x f_{0m}\Omega_{x}^m+\partial_y f_{0m}\Omega_{y}^m\right],
\end{align}
where $\Omega_{a}^m=\epsilon^{abc}\, i\langle \partial_{b} m|\partial_{c} m\rangle$ is the Berry curvature of band $m$.

\section{Evaluation of the anomalous Hall conductivity $\sigma_{xy}^{\mathrm{II}}$ \label{Appendix-AHE-extrinsic-contribution}}
In this Appendix, we calculate explicitly the anomalous Hall conductivity $\sigma_{xy}^{\mathrm{II}}$ which results from the extrinsic (i.e., Fermi surface) contribution, and show that it vanishes when the chemical potential $\mu$ lies sufficiently close to the Weyl points.
We have considered the case where an electric field is applied along the $y$ direction as $\bm{E}=E_y\bm{e}_y$.
In this case, the diagonal part of the density matrix induced by the electric field $\langle n_E\rangle=\mathrm{diag}[n^+_{E\bm{k}},n^-_{E\bm{k}}]$ is given by
\begin{align}
n^+_{E\bm{k}}=-e\tau^+\bm{E}\cdot\frac{\partial \varepsilon^+_{\bm{k}}}{\partial \bm{k}}\frac{\partial f_0(\varepsilon^+_{\bm{k}})}{\partial \varepsilon^+_{\bm{k}}}
=-e\tau^+E_yv_F^2\frac{k_\perp}{\varepsilon_{\bm{k}}}\sin\theta\,\delta(\varepsilon_{\bm{k}}-\mu)\ \ \ \ \ \mathrm{and}\ \ \ \ \ n^-_{E\bm{k}}=0,
\end{align}
where we have used $\varepsilon^\pm_{\bm{k}}=\pm\varepsilon_{\bm{k}}=\pm\sqrt{v_F^2(k_x^2+k_y^2)+m^2}$ and considered the case of $\mu>0$.
Then from Eq.~(\ref{J^+-}) we obtain the imaginary part of $[J(\langle n_E\rangle)]^{+-}_{\bm{k}}$ as
\begin{align}
\mathrm{Im}\left\{[J(\langle n_E\rangle)]_{\bm{k}}^{+-}\right\}&=\pi\frac{n_{\mathrm{imp}}U_0^2}{2}\sum_{\bm{k}'}[\sin\theta'\cos\theta-\cos\theta'\sin\theta]\frac{k_\perp'}{\varepsilon_{\bm{k}'}}\left[n^+_{E\bm{k}}\delta(\varepsilon^+_{\bm{k}}-\varepsilon^+_{\bm{k}'})-n^+_{E\bm{k}'}\delta(\varepsilon^+_{\bm{k}}-\varepsilon^+_{\bm{k}'})\right] \nonumber\\
&=\pi\frac{n_{\mathrm{imp}}U_0^2}{2}e\tau^+E_y v_F^2\cos\theta\sum_{\bm{k}'}\sin^2\theta'\frac{k_\perp'}{\varepsilon_{\bm{k}'}}\delta(\varepsilon_{\bm{k}'}-\mu)\delta(\varepsilon_{\bm{k}}-\varepsilon_{\bm{k}'}),
\label{Im-J-explicit}
\end{align}
where we have used that the contribution from $n_{E\bm{k}}^+$ vanishes since $\sin\theta'=k_y'/k_\perp'$ and $\cos\theta'=k_x'/k_\perp'$ are odd functions.
Similarly, we obtain the real part of $[J(\langle n_E\rangle)]^{+-}_{\bm{k}}$ as
\begin{align}
\mathrm{Re}\left\{[J(\langle n_E\rangle)]_{\bm{k}}^{+-}\right\}&=\pi\frac{n_{\mathrm{imp}}U_0^2}{2}\sum_{\bm{k}'}\left[\frac{k_\perp}{\varepsilon_{\bm{k}}}\frac{m'}{\varepsilon_{\bm{k}'}}-\cos(\theta'-\theta)\frac{m}{\varepsilon_{\bm{k}}}\frac{k_\perp'}{\varepsilon_{\bm{k}'}}\right]\left[n^+_{E\bm{k}}\delta(\varepsilon^+_{\bm{k}}-\varepsilon^+_{\bm{k}'})-n^+_{E\bm{k}'}\delta(\varepsilon^+_{\bm{k}}-\varepsilon^+_{\bm{k}'})\right] \nonumber\\
&=-\pi\frac{n_{\mathrm{imp}}U_0^2}{2}e\tau^+E_y v_F^2\sin\theta\left[\frac{k_\perp}{\varepsilon_{\bm{k}}}\delta(\varepsilon_{\bm{k}}-\mu)\sum_{\bm{k}'}\frac{m'}{\varepsilon_{\bm{k}'}}\delta(\varepsilon_{\bm{k}}-\varepsilon_{\bm{k}'})+\frac{m}{\varepsilon_{\bm{k}}}\sum_{\bm{k}'}\sin^2\theta'\frac{k_\perp'}{\varepsilon_{\bm{k}'}}\delta(\varepsilon_{\bm{k}'}-\mu)\delta(\varepsilon_{\bm{k}}-\varepsilon_{\bm{k}'})\right].
\label{Re-J-explicit}
\end{align}
Finally, the extrinsic contribution to the anomalous Hall conductivity $\sigma^{\mathrm{II}}_{xy}$ is calculated from the definition $\sigma^{\mathrm{II}}_{xy}=\mathrm{Tr}[(-e) v_x \langle S'_E\rangle]/E_y$.
Using Eqs.~(\ref{v_x}) and (\ref{rho-from-J}), in the case of small chemical potential $\mu$ we obtain
\begin{align}
\sigma^{\mathrm{II}}_{xy}=&2e\sum_{t=\pm}\sum_{\bm{k}}\left\{\frac{1}{2\varepsilon_{t\bm{k}}}\mathrm{Re}\left\{[J(\langle n_E\rangle)]_{\bm{k}}^{+-}\right\}\sin\theta-\frac{m_t}{2\varepsilon_{t\bm{k}}^2}\mathrm{Im}\left\{[J(\langle n_E\rangle)]_{\bm{k}}^{+-}\right\}\cos\theta\right\} \nonumber\\
=&-\pi e^2\frac{n_{\mathrm{imp}}U_0^2}{2}\tau^+ v_F^2\sum_{\bm{k},\bm{k}'}\left[\frac{m_-}{\varepsilon_{\bm{k}}^2}\sin^2\theta'\frac{k_\perp'}{\varepsilon_{\bm{k}'}}\delta(\varepsilon_{\bm{k}'}-\mu)+\frac{k_\perp}{\varepsilon_{\bm{k}}^2}\sin^2\theta\frac{m_-'}{\varepsilon_{\bm{k}'}}\delta(\varepsilon_{\bm{k}}-\mu)\right]\delta(\varepsilon_{\bm{k}}-\varepsilon_{\bm{k}'}) \nonumber\\
\approx&-\pi e^2\frac{n_{\mathrm{imp}}U_0^2}{2}\tau^+ v_F^4\int_{-\delta}^\delta\frac{d^3q}{(2\pi)^3}\int_{-\delta}^\delta\frac{d^3q'}{(2\pi)^3}\delta(\varepsilon_{\bm{q}}-\varepsilon_{\bm{q}'}) \nonumber\\
&\times\left[\frac{\frac{k_0}{b}q_z-\frac{k_0}{b}q_z}{\varepsilon_{\bm{q}}^2}\sin^2\theta'\frac{q_\perp'}{\varepsilon_{\bm{q}'}}\delta(\varepsilon_{\bm{q}'}-\mu)+\frac{q_\perp}{\varepsilon_{\bm{q}}^2}\sin^2\theta\frac{\frac{k_0}{b}q_z'-\frac{k_0}{b}q_z'}{\varepsilon_{\bm{q}'}}\delta(\varepsilon_{\bm{q}}-\mu)\right] \nonumber\\
=&0,
\end{align}
where we have omitted the subscript $-$ in $\varepsilon_{-\bm{k}}$ in the third through last lines. 
Note that the contributions only from $t=-$ survive since the chemical potential $\mu$ lies at $\varepsilon_{-\bm{k}}^+$.
Also, we have used the fact that $m_-(k_z)\approx\mp v_F^2(k_0/b)q_z$ around the Weyl nodes, where $\bm{q}=\bm{k}-W_\pm=(k_x,k_y,k_z\mp k_0)$ with $k_0=\sqrt{b^2-\Delta^2}/v_F$ and $\bm{q}^2\ll 1$.

\section{Explicit matrix forms of the magnetic driving term $D_B$ \label{Appendix-D_B-2}}
In this Appendix, we show the explicit matrix forms of the magnetic driving term (\ref{driving-term-B}) obtained from arbitrary diagonal and off-diagonal density matrices.
For concreteness, let us consider the case of a two-band model with the two eigenvalues $\varepsilon^\pm$ and the density matrix $\langle \rho\rangle$.
We define the diagonal and off-diagonal parts of $\langle \rho\rangle$ as $\langle n\rangle$ and $\langle S\rangle$, respectively.
The total density matrix of the system is given by $\langle \rho\rangle=\langle n\rangle+\langle S\rangle$.
In the eigenstate basis, we have
\begin{align}
\mathcal{H}_0=
\begin{bmatrix}
\varepsilon_{\bm{k}}^+-\mu && 0\\
0 && \varepsilon_{\bm{k}}^--\mu
\end{bmatrix},\ \ \ \ \ \ 
\langle n\rangle\equiv
\begin{bmatrix}
n^+_{\bm{k}} && 0\\
0 && n^-_{\bm{k}}
\end{bmatrix},\ \ \ \ \ \ 
\langle  S\rangle\equiv
\begin{bmatrix}
0 && a_{\bm{k}}\\
b_{\bm{k}} && 0
\end{bmatrix},
\end{align}
where $\mu$ is a chemical potential.
We immediately get
\begin{align}
&[\mathcal{R}_\alpha,\mathcal{H}_0]=
\begin{bmatrix}
0 && \mathcal{R}_\alpha^{+-}(\varepsilon^-_{\bm{k}}-\varepsilon^+_{\bm{k}})\\
-\mathcal{R}_\alpha^{-+}(\varepsilon^-_{\bm{k}}-\varepsilon^+_{\bm{k}}) && 0
\end{bmatrix},\ \ \ \ \ \ 
[\mathcal{R}_\alpha,\langle n\rangle]=
\begin{bmatrix}
0 && \mathcal{R}_\alpha^{+-}(n^-_{\bm{k}}-n^+_{\bm{k}})\\
-\mathcal{R}_\alpha^{-+}(n^-_{\bm{k}}-n^+_{\bm{k}}) && 0
\end{bmatrix},\\
&[\mathcal{R}_\alpha,\langle  S\rangle]
=
\begin{bmatrix}
\mathcal{R}_\alpha^{+-}b_{\bm{k}}-\mathcal{R}_\alpha^{-+}a_{\bm{k}} && (\mathcal{R}_\alpha^{++}-\mathcal{R}_\alpha^{--})a_{\bm{k}}\\
-(\mathcal{R}_\alpha^{++}-\mathcal{R}_\alpha^{--})b_{\bm{k}} && -(\mathcal{R}_\alpha^{+-}b_{\bm{k}}-\mathcal{R}_\alpha^{-+}a_{\bm{k}})
\end{bmatrix},
\end{align}
where $\bm{\mathcal{R}}=\sum_{\alpha=x,y,z}\mathcal{R}_\alpha\bm{e}_\alpha$ is the Berry connection vector.

First, let us calculate explicitly the magnetic driving term originating from the diagonal part of the density matrix $\langle n\rangle$.
For concreteness, we consider the case of $\bm{B}=(0,0,B_z)$.
Then the driving term (\ref{driving-term-B}) is written as
\begin{align}
D_B(\langle n\rangle)=\frac{1}{2}e\left\{\left(\frac{D \mathcal{H}_0}{D\bm{k}}\times\bm{B}\right)\cdot\frac{D\langle n\rangle}{D\bm{k}}\right\}=\frac{1}{2}eB_z\left[\left\{\frac{D\mathcal{H}_0}{Dk_y},\frac{D\langle n\rangle}{Dk_x}\right\}-\left\{\frac{D\mathcal{H}_0}{Dk_x},\frac{D\langle n\rangle}{Dk_y}\right\}\right].
\label{D_B-Appendix}
\end{align}
After a straightforward calculation, we obtain
\begin{align}
\frac{D\mathcal{H}_0}{Dk_y}\frac{D\langle n\rangle}{Dk_x}&=
\begin{bmatrix}
\frac{\partial \varepsilon_{\bm{k}}^+}{\partial k_y} && -i\mathcal{R}_y^{+-}(\varepsilon^-_{\bm{k}}-\varepsilon^+_{\bm{k}})\\
i\mathcal{R}_y^{-+}(\varepsilon^-_{\bm{k}}-\varepsilon^+_{\bm{k}}) && \frac{\partial \varepsilon_{\bm{k}}^-}{\partial k_y}
\end{bmatrix}
\begin{bmatrix}
\frac{\partial n_{\bm{k}}^+}{\partial k_x} && -i\mathcal{R}_x^{+-}(n_{\bm{k}}^- - n_{\bm{k}}^+)\\
i\mathcal{R}_x^{-+}(n_{\bm{k}}^- - n_{\bm{k}}^+) && \frac{\partial n_{\bm{k}}^-}{\partial k_x}
\end{bmatrix} \nonumber\\
&=
\begin{bmatrix}
\frac{\partial \varepsilon_{\bm{k}}^+}{\partial k_y}\frac{\partial n_{\bm{k}}^+}{\partial k_x}+\mathcal{R}_y^{+-}\mathcal{R}_x^{-+}(\varepsilon^-_{\bm{k}}-\varepsilon^+_{\bm{k}})(n_{\bm{k}}^- - n_{\bm{k}}^+) && -i\frac{\partial \varepsilon_{\bm{k}}^+}{\partial k_y}\mathcal{R}_x^{+-}(n_{\bm{k}}^- - n_{\bm{k}}^+)-i\frac{\partial n_{\bm{k}}^-}{\partial k_x}\mathcal{R}_y^{+-}(\varepsilon^-_{\bm{k}}-\varepsilon^+_{\bm{k}})\\
i\frac{\partial n_{\bm{k}}^+}{\partial k_x}\mathcal{R}_y^{-+}(\varepsilon^-_{\bm{k}}-\varepsilon^+_{\bm{k}})+i\frac{\partial \varepsilon_{\bm{k}}^-}{\partial k_y}\mathcal{R}_x^{-+}(n_{\bm{k}}^- - n_{\bm{k}}^+) && \frac{\partial \varepsilon_{\bm{k}}^-}{\partial k_y}\frac{\partial n_{\bm{k}}^-}{\partial k_x}+\mathcal{R}_y^{-+}\mathcal{R}_x^{+-}(\varepsilon^-_{\bm{k}}-\varepsilon^+_{\bm{k}})(n_{\bm{k}}^- - n_{\bm{k}}^+)
\end{bmatrix},
\end{align}
\begin{align}
\frac{D\langle n\rangle}{Dk_x}\frac{D\mathcal{H}_0}{Dk_y}&=
\begin{bmatrix}
\frac{\partial n_{\bm{k}}^+}{\partial k_x} && -i\mathcal{R}_x^{+-}(n_{\bm{k}}^- - n_{\bm{k}}^+)\\
i\mathcal{R}_x^{-+}(n_{\bm{k}}^- - n_{\bm{k}}^+) && \frac{\partial n_{\bm{k}}^-}{\partial k_x}
\end{bmatrix}
\begin{bmatrix}
\frac{\partial \varepsilon_{\bm{k}}^+}{\partial k_y} && -i\mathcal{R}_y^{+-}(\varepsilon^-_{\bm{k}}-\varepsilon^+_{\bm{k}})\\
i\mathcal{R}_y^{-+}(\varepsilon^-_{\bm{k}}-\varepsilon^+_{\bm{k}}) && \frac{\partial \varepsilon_{\bm{k}}^-}{\partial k_y}
\end{bmatrix} \nonumber\\
&=
\begin{bmatrix}
\frac{\partial \varepsilon_{\bm{k}}^+}{\partial k_y}\frac{\partial n_{\bm{k}}^+}{\partial k_x}+\mathcal{R}_y^{-+}\mathcal{R}_x^{+-}(\varepsilon^-_{\bm{k}}-\varepsilon^+_{\bm{k}})(n_{\bm{k}}^- - n_{\bm{k}}^+) && -i\frac{\partial \varepsilon_{\bm{k}}^-}{\partial k_y}\mathcal{R}_x^{+-}(n_{\bm{k}}^- - n_{\bm{k}}^+)-i\frac{\partial n_{\bm{k}}^+}{\partial k_x}\mathcal{R}_y^{+-}(\varepsilon^-_{\bm{k}}-\varepsilon^+_{\bm{k}})\\
i\frac{\partial n_{\bm{k}}^-}{\partial k_x}\mathcal{R}_y^{-+}(\varepsilon^-_{\bm{k}}-\varepsilon^+_{\bm{k}})+i\frac{\partial \varepsilon_{\bm{k}}^+}{\partial k_y}\mathcal{R}_x^{-+}(n_{\bm{k}}^- - n_{\bm{k}}^+) && \frac{\partial \varepsilon_{\bm{k}}^-}{\partial k_y}\frac{\partial n_{\bm{k}}^-}{\partial k_x}+\mathcal{R}_y^{+-}\mathcal{R}_x^{-+}(\varepsilon^-_{\bm{k}}-\varepsilon^+_{\bm{k}})(n_{\bm{k}}^- - n_{\bm{k}}^+)
\end{bmatrix},
\end{align}
which results in
\begin{align}
\left\{\frac{D\mathcal{H}_0}{Dk_y},\frac{D\langle n\rangle}{Dk_x}\right\}&=
\begin{bmatrix}
2\frac{\partial \varepsilon_{\bm{k}}^+}{\partial k_y}\frac{\partial n_{\bm{k}}^+}{\partial k_x}+g^{++}_{xy} && -i\mathcal{R}_{y}^{+-}(\varepsilon^-_{\bm{k}}-\varepsilon^+_{\bm{k}})\frac{\partial}{\partial k_x}(n_{\bm{k}}^+ + n_{\bm{k}}^-)+g^{+-}_{xy}\\
i\mathcal{R}_{y}^{-+}(\varepsilon^-_{\bm{k}}-\varepsilon^+_{\bm{k}})\frac{\partial}{\partial k_x}(n_{\bm{k}}^+ + n_{\bm{k}}^-)+g^{-+}_{xy} && 2\frac{\partial \varepsilon_{\bm{k}}^-}{\partial k_y}\frac{\partial n_{\bm{k}}^-}{\partial k_x}+g^{--}_{xy}
\end{bmatrix}, \nonumber\\
\left\{\frac{D\mathcal{H}_0}{Dk_x},\frac{D\langle n\rangle}{Dk_y}\right\}&=
\begin{bmatrix}
2\frac{\partial \varepsilon_{\bm{k}}^+}{\partial k_x}\frac{\partial n_{\bm{k}}^+}{\partial k_y}+g^{++}_{yx} && -i\mathcal{R}_{x}^{+-}(\varepsilon^-_{\bm{k}}-\varepsilon^+_{\bm{k}})\frac{\partial}{\partial k_y}(n_{\bm{k}}^+ + n_{\bm{k}}^-)+g^{+-}_{yx}\\
i\mathcal{R}_{x}^{-+}(\varepsilon^-_{\bm{k}}-\varepsilon^+_{\bm{k}})\frac{\partial}{\partial k_y}(n_{\bm{k}}^+ + n_{\bm{k}}^-)+g^{-+}_{yx} && 2\frac{\partial \varepsilon_{\bm{k}}^-}{\partial k_x}\frac{\partial n_{\bm{k}}^-}{\partial k_y}+g^{--}_{yx}
\end{bmatrix},
\label{D_B-diagonal-component}
\end{align}
where
\begin{align}
g^{++}_{xy}&=\mathcal{R}_y^{+-}\mathcal{R}_x^{-+}(\varepsilon^-_{\bm{k}}-\varepsilon^+_{\bm{k}})(n_{\bm{k}}^- - n_{\bm{k}}^+)+\mathcal{R}_y^{-+}\mathcal{R}_x^{+-}(\varepsilon^-_{\bm{k}}-\varepsilon^+_{\bm{k}})(n_{\bm{k}}^- - n_{\bm{k}}^+), \nonumber\\
g^{--}_{xy}&=\mathcal{R}_y^{-+}\mathcal{R}_x^{+-}(\varepsilon^-_{\bm{k}}-\varepsilon^+_{\bm{k}})(n_{\bm{k}}^- - n_{\bm{k}}^+)+\mathcal{R}_y^{+-}\mathcal{R}_x^{-+}(\varepsilon^-_{\bm{k}}-\varepsilon^+_{\bm{k}})(n_{\bm{k}}^- - n_{\bm{k}}^+), \nonumber\\
g^{+-}_{xy}&=-i\mathcal{R}_{x}^{+-}(n^-_{\bm{k}}-n^+_{\bm{k}})\frac{\partial}{\partial k_y}(\varepsilon_{\bm{k}}^+ + \varepsilon_{\bm{k}}^-), \nonumber\\
g^{-+}_{xy}&=i\mathcal{R}_{x}^{-+}(n^-_{\bm{k}}-n^+_{\bm{k}})\frac{\partial}{\partial k_y}(\varepsilon_{\bm{k}}^+ + \varepsilon_{\bm{k}}^-).
\end{align}
Here, note that $g^{++}_{xy}=g^{++}_{yx}$ and $g^{--}_{xy}=g^{--}_{yx}$.
In the case of $n_{\bm{k}}^\pm=f_0(\varepsilon_{\bm{k}}^\pm)$ where $f_0(\varepsilon_{\bm{k}}^\pm)$ is the Fermi-Dirac distribution function, we have $\frac{\partial \varepsilon_{\bm{k}}^\pm}{\partial k_y}\frac{\partial f_0(\varepsilon_{\bm{k}}^\pm)}{\partial k_x}-\frac{\partial \varepsilon_{\bm{k}}^\pm}{\partial k_x}\frac{\partial f_0(\varepsilon_{\bm{k}}^\pm)}{\partial k_y}=0$.
Namely, the magnetic driving term (\ref{D_B-Appendix}) is purely off-diagonal in this case.
Furthermore, in the case of $\varepsilon_{\bm{k}}^\pm=\pm\varepsilon_{\bm{k}}$, which applies to Weyl semimetals, we have $g^{+-}=g^{-+}=0$.
Then in such a case the magnetic driving term is simplified to be
\begin{align}
D_B(\langle n\rangle)=eB_z
\begin{bmatrix}
0 && -i\varepsilon_{\bm{k}}(\mathcal{R}_{x}^{+-}\frac{\partial}{\partial k_y}-\mathcal{R}_{y}^{+-}\frac{\partial}{\partial k_x})[f_0(\varepsilon_{\bm{k}}^+) + f_0(\varepsilon_{\bm{k}}^-)]\\
i\varepsilon_{\bm{k}}(\mathcal{R}_{x}^{-+}\frac{\partial}{\partial k_y}-\mathcal{R}_{y}^{-+}\frac{\partial}{\partial k_x})[f_0(\varepsilon_{\bm{k}}^+) + f_0(\varepsilon_{\bm{k}}^-)] && 0
\end{bmatrix}.
\label{D_B-from-diagonal-density}
\end{align}

Next, let us calculate explicitly the magnetic driving term originating from the off-diagonal part of the density matrix $\langle S\rangle$.
For concreteness, we consider the case of $\bm{B}=(0,0,B_z)$.
Then the driving term (\ref{driving-term-B}) is written as
\begin{align}
D_B(\langle S\rangle)=\frac{1}{2}e\left\{\left(\frac{D \mathcal{H}_0}{D\bm{k}}\times\bm{B}\right)\cdot\frac{D\langle S\rangle}{D\bm{k}}\right\}=\frac{1}{2}eB_z\left[\left\{\frac{D\mathcal{H}_0}{Dk_y},\frac{D\langle S\rangle}{Dk_x}\right\}-\left\{\frac{D\mathcal{H}_0}{Dk_x},\frac{D\langle S\rangle}{Dk_y}\right\}\right].
\end{align}
After a straightforward calculation, we obtain
\begin{align}
\frac{D\mathcal{H}_0}{Dk_y}\frac{D\langle S\rangle}{Dk_x}&=
\begin{bmatrix}
\frac{\partial \varepsilon_{\bm{k}}^+}{\partial k_y} && -i\mathcal{R}_y^{+-}(\varepsilon^-_{\bm{k}}-\varepsilon^+_{\bm{k}})\\
i\mathcal{R}_y^{-+}(\varepsilon^-_{\bm{k}}-\varepsilon^+_{\bm{k}}) && \frac{\partial \varepsilon_{\bm{k}}^-}{\partial k_y}
\end{bmatrix}
\begin{bmatrix}
-iF && -i\Delta a+\frac{\partial a}{\partial k_x}\\
i\Delta b+\frac{\partial b}{\partial k_x} && iF
\end{bmatrix} \nonumber\\
&=
\begin{bmatrix}
-i\frac{\partial \varepsilon_{\bm{k}}^+}{\partial k_y}F-i\mathcal{R}_y^{+-}(\varepsilon^-_{\bm{k}}-\varepsilon^+_{\bm{k}})[i\Delta b+\frac{\partial b}{\partial k_x}] && \frac{\partial \varepsilon_{\bm{k}}^+}{\partial k_y}[-i\Delta a+\frac{\partial a}{\partial k_x}]+\mathcal{R}_y^{+-}(\varepsilon^-_{\bm{k}}-\varepsilon^+_{\bm{k}})F \\
\mathcal{R}_y^{-+}(\varepsilon^-_{\bm{k}}-\varepsilon^+_{\bm{k}})F+\frac{\partial \varepsilon_{\bm{k}}^-}{\partial k_y}[i\Delta b+\frac{\partial b}{\partial k_x}] && i\mathcal{R}_y^{-+}(\varepsilon^-_{\bm{k}}-\varepsilon^+_{\bm{k}})[-i\Delta a+\frac{\partial a}{\partial k_x}]+i\frac{\partial \varepsilon_{\bm{k}}^-}{\partial k_y}F
\end{bmatrix},
\end{align}
\begin{align}
\frac{D\langle S\rangle}{Dk_x}\frac{D\mathcal{H}_0}{Dk_y}&=
\begin{bmatrix}
-iF && -i\Delta a+\frac{\partial a}{\partial k_x}\\
i\Delta b+\frac{\partial b}{\partial k_x} && iF
\end{bmatrix}
\begin{bmatrix}
\frac{\partial \varepsilon_{\bm{k}}^+}{\partial k_y} && -i\mathcal{R}_y^{+-}(\varepsilon^-_{\bm{k}}-\varepsilon^+_{\bm{k}})\\
i\mathcal{R}_y^{-+}(\varepsilon^-_{\bm{k}}-\varepsilon^+_{\bm{k}}) && \frac{\partial \varepsilon_{\bm{k}}^-}{\partial k_y}
\end{bmatrix} \nonumber\\
&=
\begin{bmatrix}
-i\frac{\partial \varepsilon_{\bm{k}}^+}{\partial k_y}F+i\mathcal{R}_y^{-+}(\varepsilon^-_{\bm{k}}-\varepsilon^+_{\bm{k}})[-i\Delta a+\frac{\partial a}{\partial k_x}] && \frac{\partial \varepsilon_{\bm{k}}^-}{\partial k_y}[-i\Delta a+\frac{\partial a}{\partial k_x}]-\mathcal{R}_y^{+-}(\varepsilon^-_{\bm{k}}-\varepsilon^+_{\bm{k}})F \\
-\mathcal{R}_y^{-+}(\varepsilon^-_{\bm{k}}-\varepsilon^+_{\bm{k}})F+\frac{\partial \varepsilon_{\bm{k}}^+}{\partial k_y}[i\Delta b+\frac{\partial b}{\partial k_x}] && -i\mathcal{R}_y^{+-}(\varepsilon^-_{\bm{k}}-\varepsilon^+_{\bm{k}})[i\Delta b+\frac{\partial b}{\partial k_x}]+i\frac{\partial \varepsilon_{\bm{k}}^-}{\partial k_y}F\end{bmatrix},
\end{align}
where $\Delta=\mathcal{R}_x^{++}-\mathcal{R}_x^{--}$ and $F=(\mathcal{R}_x^{+-}b-\mathcal{R}_x^{-+}a)$.
In the case of $\varepsilon_{\bm{k}}^\pm=\pm\varepsilon_{\bm{k}}$, which applies to Weyl semimetals, we have
\begin{align}
\frac{1}{2}\left\{\frac{D\mathcal{H}_0}{Dk_y},\frac{D\langle S\rangle}{Dk_x}\right\}=
\begin{bmatrix}
\mathcal{A}_{\bm{k}} && 0\\
0 && \mathcal{A}_{\bm{k}}
\end{bmatrix},\ \ \ \ \ \ \ \ 
\frac{1}{2}\left\{\frac{D\mathcal{H}_0}{Dk_x},\frac{D\langle S\rangle}{Dk_y}\right\}=
\begin{bmatrix}
\mathcal{B}_{\bm{k}} && 0\\
0 && \mathcal{B}_{\bm{k}}
\end{bmatrix}
\end{align}
with
\begin{align}
\mathcal{A}_{\bm{k}}&=-i\frac{\partial \varepsilon_{\bm{k}}}{\partial k_y}(\mathcal{R}_x^{+-}b_{\bm{k}}-\mathcal{R}_x^{-+}a_{\bm{k}})+i\mathcal{R}_y^{+-}\varepsilon_{\bm{k}}\left[i(\mathcal{R}_x^{++}-\mathcal{R}_x^{--})b_{\bm{k}}+\frac{\partial b_{\bm{k}}}{\partial k_x}\right] -i\mathcal{R}_y^{-+}\varepsilon_{\bm{k}}\left[-i(\mathcal{R}_x^{++}-\mathcal{R}_x^{--})a_{\bm{k}}+\frac{\partial a_{\bm{k}}}{\partial k_x}\right], \nonumber\\
\mathcal{B}_{\bm{k}}&=-i\frac{\partial \varepsilon_{\bm{k}}}{\partial k_x}(\mathcal{R}_y^{+-}b_{\bm{k}}-\mathcal{R}_y^{-+}a_{\bm{k}})+i\mathcal{R}_x^{+-}\varepsilon_{\bm{k}}\left[i(\mathcal{R}_y^{++}-\mathcal{R}_y^{--})b_{\bm{k}}+\frac{\partial b_{\bm{k}}}{\partial k_y}\right] -i\mathcal{R}_x^{-+}\varepsilon_{\bm{k}}\left[-i(\mathcal{R}_y^{++}-\mathcal{R}_y^{--})a_{\bm{k}}+\frac{\partial a_{\bm{k}}}{\partial k_y}\right].
\end{align}
Especially in the case of $a_{\bm{k}}=-ic_{\bm{k}}$ and $b_{\bm{k}}=ic_{\bm{k}}$, i.e., $\langle S\rangle_{\bm{k}}=c_{\bm{k}}\sigma_y$, we have
\begin{align}
\mathcal{A}_{\bm{k}}&=(\mathcal{R}_x^{+-}+\mathcal{R}_x^{-+})\frac{\partial \varepsilon_{\bm{k}}}{\partial k_y}c_{\bm{k}}-i(\mathcal{R}_y^{+-}-\mathcal{R}_y^{-+})(\mathcal{R}_x^{++}-\mathcal{R}_x^{--})\varepsilon_{\bm{k}}c_{\bm{k}}-(\mathcal{R}_y^{+-}+\mathcal{R}_y^{-+})\varepsilon_{\bm{k}}\frac{\partial c_{\bm{k}}}{\partial k_x}, \nonumber\\
\mathcal{B}_{\bm{k}}&=(\mathcal{R}_y^{+-}+\mathcal{R}_y^{-+})\frac{\partial \varepsilon_{\bm{k}}}{\partial k_x}c_{\bm{k}}-i(\mathcal{R}_x^{+-}-\mathcal{R}_x^{-+})(\mathcal{R}_y^{++}-\mathcal{R}_y^{--})\varepsilon_{\bm{k}}c_{\bm{k}}-(\mathcal{R}_x^{+-}+\mathcal{R}_x^{-+})\varepsilon_{\bm{k}}\frac{\partial c_{\bm{k}}}{\partial k_y}
.
\label{Expression-of-AB}
\end{align}
Finally, we obtain the magnetic driving term which is purely diagonal,
\begin{align}
D_B(\langle S\rangle)=\frac{1}{2}eB_z\left[\left\{\frac{D\mathcal{H}_0}{Dk_y},\frac{D\langle S\rangle}{Dk_x}\right\}-\left\{\frac{D\mathcal{H}_0}{Dk_x},\frac{D\langle S\rangle}{Dk_y}\right\}\right]=eB_z
\begin{bmatrix}
\mathcal{F}_{\bm{k}} && 0\\
0 && \mathcal{F}_{\bm{k}}
\end{bmatrix}
\label{D_B-from-offdiagonal-density}
\end{align}
with $\mathcal{F}_{\bm{k}}=\mathcal{A}_{\bm{k}}-\mathcal{B}_{\bm{k}}$.

\end{widetext}

\end{document}